\documentclass[letterpaper,twocolumn,10pt]{article}
\usepackage{usenix}

\usepackage{tikz}
\usepackage{amsmath,amssymb,amsfonts}
\usepackage{cite}
\usepackage{algorithmic}
\usepackage{graphicx}
\usepackage{textcomp}
\usepackage{xcolor}
\usepackage[hyphens]{xurl}

\usepackage{amsthm}
\usepackage[linesnumbered, lined, boxed]{algorithm2e}

\newtheorem{definition}{Definition}
\newtheorem{theorem}{Theorem}
\newtheorem{lemma}{Lemma}

\newtheorem{corollary}{Corollary}

\SetKwInput{KwData}{Input}
\SetKwInput{KwResult}{Output}

\newcommand{\nats}{\mathbb{N}}
\newcommand{\ints}{\mathbb{Z}}
\newcommand{\nnints}{\mathbb{Z}_{\ge0}}
\newcommand{\reals}{\mathbb{R}}
\newcommand{\nnreals}{\mathbb{R}_{\ge0}}

\renewcommand{\epsilon}{\varepsilon}

\newcommand{\calD}{\mathcal{D}}

\newcommand{\calI}{\mathcal{I}}

\newcommand{\calM}{\mathcal{M}}

\newcommand{\calR}{\mathcal{R}}
\newcommand{\calS}{\mathcal{S}}
\newcommand{\calT}{\mathcal{T}}

\newcommand{\calZ}{\mathcal{Z}}

\newcommand{\E}{\operatorname{\mathbb{E}}}
\newcommand{\V}{\operatorname{\mathbb{V}}}

\newcommand{\bmc}{\mathbf{c}}

\newcommand{\bmf}{\mathbf{f}}

\newcommand{\bmh}{\mathbf{h}}

\newcommand{\bmx}{\mathbf{x}}
\newcommand{\bmy}{\mathbf{y}}
\newcommand{\bmz}{\mathbf{z}}

\newcommand{\hf}{\hat{f}}

\newcommand{\tn}{\tilde{n}}
\newcommand{\tx}{\tilde{x}}

\newcommand{\tO}{\tilde{O}}
\newcommand{\bmtx}{\mathbf{\tx}}
\newcommand{\bmhf}{\mathbf{\hf}}

\newcommand{\bn}{\bar{n}}

\newcommand{\FME}{\textsf{FME}}

\def\FOUD{\textsf{FOUD}}
\def\FOLNF{\textsf{FOLNF}}
\def\FOLNFast{\textsf{FOLNF$^*$}}
\def\Largesf{\textsf{Large}}
\def\FOUDLarge{\textsf{FOUD-L}}
\def\FOLNFLarge{\textsf{FOLNF-L}}
\def\FOLNFastLarge{\textsf{FOLNF$^*$-L}}

\def\GGK{\textsf{GGK+21}}

\def\IPUMS{IPUMS}
\def\Localization{Localization}
\def\Foursquare{Foursquare}
\def\AOL{AOL}

\def\GRR{\textsf{GRR}}
\def\OUE{\textsf{OUE}}
\def\OLH{\textsf{OLH}}

\def\BC{\textsf{BC20}}
\def\CM{\textsf{CM22}}
\def\LWY{\textsf{LWY22}}
\def\LWYL{\textsf{LWY22-L}}

\def\AGeo{\textsf{AGeo}}
\def\OGeo{\textsf{1Geo}}

\newcommand{\colorB}[1]{\textcolor{black}{#1}}

\newif\ifconferenceon\conferenceonfalse
\ifconferenceon
\newcommand{\conference}[1]{#1}
\newcommand{\arxiv}[1]{}
\else
\newcommand{\conference}[1]{}
\newcommand{\arxiv}[1]{#1}
\usepackage{balance}
\fi

\begin{document}

\date{}

\title{\Large \bf Fully Oblivious Differential Privacy for Frequency Estimation in the Augmented Shuffle Model with Trusted Processors}

\author{
{\rm Takao Murakami}\\
ISM/ROIS/AIST/RIKEN AIP
\and
{\rm Yuichi Sei}\\
UEC
\and
{\rm Reo Eriguchi}\\
AIST
}

\maketitle

\begin{abstract}
In the shuffle model of DP (Differential Privacy), a shuffler randomly permutes users' data to achieve high accuracy and privacy. 
Recent studies show that most existing shuffle protocols are vulnerable to collusion attacks by the data collector and users. 
They address this issue by introducing the augmented shuffle model 
\colorB{that} 
incorporates random sampling and dummy data addition into the shuffler. 
However, it remains open how to ensure the shuffler follows the protocol and does not collude with the data collector in this model. 

We address this trust issue by thoroughly exploring the augmented shuffle model with TEEs (Trusted Execution Environments). 
We first introduce a new privacy notion, \textit{FODP (Fully Oblivious DP)}, which strengthens DP  
to prevent various TEE side-channel attacks based on external/internal memory access patterns and control flows. 
We propose a general framework for FODP algorithms based on memory-size obfuscation 
and three concrete algorithms within it. 
We also improve the efficiency of our algorithms 
by using the count-min sketch and optimizing the number of hashes. 
We evaluate our algorithms on Intel SGX and demonstrate 
their effectiveness through comparisons with nine baselines.
\end{abstract}

\section{Introduction}
\label{sec:intro}
DP (Differential Privacy) \cite{DP} enables data analysis with 
strong privacy, especially when a parameter called the privacy budget $\epsilon$ is small. 
Depending on its underlying architecture, DP can be divided into some 
models, such as the central~\cite{DP}, local~\cite{Kasiviswanathan_FOCS08}, and shuffle models~\cite{Erlingsson_SODA19}. 
In the central model, a single data collector holds all users' raw data and adds noise 
to their statistics (e.g., histograms, heavy hitters). 
A drawback of this model is that all data can be leaked from the data collector through data breaches or cyberattacks~\cite{data_breach_2025}. 
The local model prevents data leakage by 
adding noise 
to a user's data before sending it to the data collector. 
However, it suffers from low accuracy due to large local noise. 
The shuffle model addresses this issue by introducing a \textit{shuffler}, which does not collude with the data collector. 
In this model, each user sends noisy data to the shuffler. 
Then, the shuffler randomly shuffles the noisy data and sends the shuffled data to the data collector. 
The random shuffling amplifies privacy; i.e., it significantly reduces the value of $\epsilon$. 
Thus, the shuffle model significantly improves the local model's accuracy at the same value of $\epsilon$. 

However, most existing protocols 
\colorB{assume a \textit{pure shuffle model}, where the shuffler only shuffles noisy values sent from users, and} 
are vulnerable to collusion attacks by the data collector and some malicious users (abbreviated as \textit{collusion with users})~\cite{Wang_PVLDB20,Murakami_SP25}. 
Specifically, the data collector may obtain noisy data from some users by colluding with them or compromising their accounts. 
\colorB{As a result, the pure shuffle model suffers from an increase of $\epsilon$ (or suboptimal accuracy when attempting to prevent it); see Section~\ref{sub:collusion_attacks} for details.}

To prevent collusion with users, some recent studies~\cite{Wang_PVLDB20,Murakami_SP25} introduce an \textit{augmented shuffle model}, which incorporates additional operations (e.g., random sampling, adding dummies) into the shuffler. 
For example, \cite{Wang_PVLDB20} proposes the UD (Uniform Dummies) protocol, in which the shuffler adds dummies uniformly randomly sampled from 
the space of noisy data. 
\cite{Wang_PVLDB20} shows that the uniform dummies significantly improve the robustness against collusion with users. 
\cite{Murakami_SP25} proposes the LNF (Local-Noise-Free) protocol, in which users do not add noise and the shuffler performs random sampling and dummy data addition. 
This protocol achieves ideal robustness against collusion with users in that the value of $\epsilon$ is not increased by the collusion. 
\cite{Murakami_SP25} also shows that the LNF protocol provides much higher accuracy than other shuffle DP protocols. 

\smallskip{}
\noindent{\textbf{Trust Issue.}}~~The story does not end there, as one important trust issue remains in the augmented shuffle model. 
Specifically, we must ensure that the shuffler follows the protocol and does not collude with the data collector. 
For example, we must ensure that the shuffler randomly shuffles data sent from users and does not leak non-shuffled data to the data collector. 

One promising solution to achieve this is to use a \textit{TEE (Trusted Execution Environment)}~\cite{TEE_book}, 
which protects data and the running code by memory encryption. 
The TEE guarantees data confidentiality by preventing entities outside the TEE from reading raw data. 
It also guarantees integrity by 
providing a cryptographic proof that the running code is genuine \cite{Anati_HASP13}. 
Therefore, we can prevent the shuffler from deviating from the protocol or leaking non-shuffled data by implementing it within the TEE~\cite{Allen_NeurIPS19,Bittau_SOSP17}. 
In this case, 
users communicate with a single server; i.e., the TEE inside the server serves as a shuffler, and the data collector is outside the TEE. 

\smallskip{}
\noindent{\textbf{Side-Channel Attacks.}}~~However, recent studies show that TEEs are vulnerable to various side-channel attacks based on external/internal memory access patterns~\cite{Allen_NeurIPS19,Bulck_USENIX17,Lee_USENIX20} 
and control flows (e.g., if statements, loops)~\cite{Lee_USENIX17,Moghimi_USENIX20}. 
Note that although external memory-based attacks can be prevented by oblivious shuffling~\cite{Allen_NeurIPS19,Bittau_SOSP17,Ohrimenko_ICALP14}, adversaries can also obtain internal memory access patterns and control flows through state-of-the-art attacks; see~\cite{Sasy_CCS22} for surveys of these attacks. 
In addition, the shuffler performs not only shuffling but also random sampling and dummy data addition in the augmented shuffle model. 
Thus, we arrive at the following question: 
\textit{how can we prevent all of the side-channel attacks explained above in the augmented shuffle model with TEEs}? 

\subsection{Our Approaches}
\label{sub:technical_overview}
We comprehensively study the above question in frequency estimation~\cite{Wang_USENIX17,Wang_PVLDB20,Luo_CCS22,Murakami_SP25}, a fundamental data analysis task that estimates the frequency of each item. 

We first introduce a new privacy notion called \textit{FODP (Fully Oblivious 
DP)}. 
FODP strengthens DP based on the notion of \textit{full obliviousness}~\cite{Sasy_CCS22,Sasy_CCS23}, 
which guarantees that no information about secret data is leaked from external/internal memory access patterns and control flows. 
Specifically, FODP provides DP guarantees for 
all of the 
algorithm's output, external/internal memory access patterns, and control flows. 
We use FODP as a privacy notion for \textit{internal adversaries} (i.e., server administrators) who can obtain side-channel information, and DP for \textit{external adversaries} (i.e., everyone) who can access the published frequency distribution. 

Then, we propose a general framework for FODP algorithms. 
At a high level, our framework \textit{obfuscates the memory size to be allocated by adding bots ``$\bot$'' for each item}. 
It also makes control flows dependent only on the obfuscated memory size. 
Our key insight is that the memory access patterns and control flows in our framework depend on users' data only through \textit{bot counts}, i.e., the number of $\bot$ for each item (Lemma~\ref{lem:general_privacy}). 
This enables us to reduce FODP of our framework to the privacy of shuffled data and bot counts. 

Based on this insight, we propose three concrete algorithms in our framework: \FOUD{}, \FOLNF{}, and \FOLNFast{}. 
\FOUD{} and \FOLNF{} are fully oblivious versions of the UD and LNF protocols, respectively. 
In \FOUD{}, we modify the UD protocol so that the value of $\epsilon$ is not increased by collusion with users. 
Both \FOUD{} and \FOLNF{} set bot counts to constants and therefore provide FODP. 
\FOLNFast{} is a variant of \FOLNF{} that generates bot counts under DP. 
\FOLNFast{} 
improves the efficiency (runtime) 
of \FOLNF{} at the same privacy budget in DP by increasing the privacy budget in FODP, i.e., by relaxing the privacy against internal adversaries. 
It also 
achieves \textit{pure} DP ($\delta=0$) when using random sampling. 

For large-domain data, we further 
improve the efficiency 
of our algorithms by using the count-min sketch \cite{Cormode_JA05}. 
For our improved algorithms, we show an accuracy bound much tighter than \cite{Cheu_SP22,Ghazi_EUROCRYPT21} and optimize the number of hashes in the count-min sketch based on our bound. 

\smallskip{}
\noindent{\textbf{Our Contributions.}}~~Our contributions are as follows: 
\begin{itemize}
\item We present a new privacy notion called FODP to provide privacy guarantees against adversaries who can obtain 
all of the randomized algorithm's output, external/internal memory access patterns, and control flows. 
\item We propose a general framework for FODP algorithms and three instantiations: \FOUD{}, \FOLNF{}, and \FOLNFast{}. 
We prove that they achieve DP, FODP, and robustness against collusion with users. 
We also analyze the accuracy and runtime of these algorithms. 
\item For large-domain data, we improve the efficiency of our algorithms using the count-min sketch. 
We show an accuracy bound much tighter than \cite{Cheu_SP22,Ghazi_EUROCRYPT21} and propose a technique to optimize the number of hashes based on it. 
\item We evaluate our algorithms on Intel SGX \cite{Anati_HASP13}, 
a popular trusted processor (TEE implementation). 
We show 
the effectiveness of our algorithms 
through comparisons with nine baselines 
(eight state-of-the-art shuffle algorithms and 
\colorB{a central DP algorithm explained later}). 
\end{itemize}

The proofs of all statements 
are in \conference{\colorB{our full paper~\cite{Murakami_arXiv26}}}\arxiv{Appendix~\ref{sec:proofs}}. 

\smallskip{}
\noindent{\textbf{Technical Novelty.}}~~In this paper, we introduce several new theoretical and algorithmic ideas, as explained below. 

First, we introduce not only a privacy notion called FODP but also two proof techniques for FODP. 
Specifically, we connect FODP to DP and full obliviousness for each user's data 
(Theorem~\ref{thm:DP_FO_FODP}). 
Based on this, we introduce two proof techniques: (i) \textit{showing FODP via full obliviousness} and (ii) \textit{directly showing FODP}. 
We prove FODP of \FOUD{} and \FOLNF{} by (i) and that of \FOLNFast{} by (ii). 

Second, among our three algorithms, \FOLNFast{} presents the most novel ideas. 
Specifically, 
in \FOLNFast{}, 
the numbers of dummies and bots for each item follow a dummy-count distribution $\calD$ and a bot-count distribution $\calD'$, respectively. 
For 
$(\calD,\calD')$, we propose a novel \textit{joint asymmetric geometric distribution}, which is a pair of the \textit{positively skewed} geometric distribution $\calD$ and the \textit{negatively skewed} geometric distribution $\calD'$. 
\colorB{Although they are instantiations of the GDL (General Discrete Laplace distribution)~\cite{Lekshmi_IJMSI14}, the existing studies that apply the GDL to DP \cite{Athanasiou_PoPETs25,Harrison_FORC25} consider only a symmetric GDL. 
Our joint distribution is new in that it is a pair of positively and negatively skewed asymmetric GDLs and that the left/right curves are adjusted based on the sampling probability.} 

Finally, 
although we use the count-min sketch 
in the same way as \cite{Cheu_SP22,Ghazi_EUROCRYPT21}, we present a new accuracy bound \colorB{much tighter than \cite{Cheu_SP22,Ghazi_EUROCRYPT21}} and an optimization technique using our bound.

\smallskip{}
\noindent{\textbf{\colorB{Comparison with Other Models.}}}~~\colorB{Below, we emphasize the motivation for using the augmented shuffle model within TEEs by comparing it with other models.} 

\colorB{First, the pure shuffle model suffers from collusion with users, as explained above. 
Moreover, this model suffers from \textit{poisoning attacks} from users~\cite{Cao_USENIX21,Cheu_SP21}. 
This issue is especially serious when $\epsilon$ is small, as honest users need to add a large amount of noise to their data (and malicious users do not). 
In contrast, our algorithms are robust against poisoning attacks, as noise is added on the shuffler side (see Appendix~\ref{sub:robustness_poisoning}).} 

\colorB{Second, it is possible to run a central DP algorithm (e.g., basic DP histogram~\cite{Balcer_JPC19,Dwork_TCC06}, stability histogram~\cite{Bun_JMLR19,Lebeda_FORC25}) within the TEE. 
However, it suffers from prohibitively large runtime, as it must access the entire memory to add each user's data to the histogram~\cite{Allen_NeurIPS19}. 
For example, when there are $10^8$ users and $10^8$ items, 
it requires about $270$ days, as shown in our experiments. 
In contrast, our algorithms require only $16$ hours or less. 
We also note that sorting results in higher complexity than shuffling, as it must be fully oblivious~\cite{Sasy_CCS23}.\footnote{\colorB{Additionally, if we can perform sorting, we can also perform shuffling by assigning a random value to each input and then sorting the random values. Thus, sorting \textit{cannot} be more efficient than shuffling.}}}

\colorB{Third, the MPC (Multi-Party Computation)-DP model~\cite{Wagh_CACM21} requires multiple servers and communication between them. 
For example, the protocol in \cite{Bell_CCS22} requires two semi-honest 
servers and four rounds of interaction between the two. 
In contrast, our approaches need only a \textit{single server}. 
Although it might be possible to simulate the MPC-DP model using the TEE, it is not obvious how to do that, as the size of the internal memory is typically limited (e.g., $93.5$ MB~\cite{Lee_EECS22}). 
Our algorithms need a small amount of memory and are efficient.} 

\colorB{We also note that there are advantages of the pure shuffle, central, and MPC-DP models. 
For example, the pure shuffle model can be implemented using the mix-net~\cite{Bittau_SOSP17}. 
The central DP algorithm is more efficient than our algorithms when $n$ (\#users) and $d$ (\#items) are small, e.g., $n=d\leq 10^3$ in our experiments. 
When multiple servers are available, the MPC-DP model can be used to achieve accuracy comparable to the central model for general tasks. 
Therefore, we should use an appropriate model depending on the application.} 

\section{Related Work}
\label{sec:related}
\noindent{\textbf{Shuffle Model.}}~~Starting from Google's Prochlo~\cite{Bittau_SOSP17}, the shuffle model has been widely studied 
for various tasks, such as frequency estimation~\cite{Balcer_ITC20,Cheu_SP22,Luo_CCS22,Murakami_SP25,Wang_PVLDB20}, 
graph analysis~\cite{Fang_CCS25,Imola_CCS22}, 
and federated learning~\cite{Girgis_AISTATS21,Liu_AAAI21}. 
Most existing protocols assume the pure shuffle model 
\colorB{and} 
are vulnerable to collusion with users\colorB{, as described in Section~\ref{sec:intro}}. 

Some recent studies address this issue by introducing the augmented shuffle model. 
Protocols in this model include the UD~\cite{Wang_PVLDB20} and LNF~\cite{Murakami_SP25} protocols. 
We propose a general framework that includes fully oblivious versions of these protocols. 
A recent study \cite{Murakami_NDSS26} proposes the FME (Filtering-with-Multiple-Encryption) protocol, which improves the efficiency of the LNF protocol for large-domain data. 
However, it (still) requires high communication costs due to the use of multiple encryptions. 
We show that our algorithms with the count-min sketch are much more efficient than the FME protocol, while achieving higher (or comparable) accuracy. 
Another important difference is that the FME protocol requires two rounds of communication between the shuffler and the data collector, whereas our algorithms need only one round. 

Some studies~\cite{Allen_NeurIPS19,Bittau_SOSP17} consider the shuffle model with TEEs and use oblivious shuffling to prevent external memory-based side-channel attacks. 
However, their approaches cannot be applied to our setting, 
as explained in Section~\ref{sec:intro}.

\smallskip{}
\noindent{\textbf{Oblivious Algorithms.}}~~Oblivious RAM \cite{Asharov_JACM22,Stefanov_JACM18} has been widely studied to hide memory access patterns. 
The combination of DP and obliviousness for memory access patterns has also been introduced 
\colorB{(e.g., ODP \cite{Allen_NeurIPS19}, DO \cite{Chan_SODA19})}. 
However, they are not designed to prevent control flow-based attacks~\cite{Lee_USENIX17,Moghimi_USENIX20}. 

Sasy \textit{et al.}~\cite{Sasy_CCS22,Sasy_CCS23} introduce the notion of full obliviousness to prevent all of the attacks based on external/internal memory access patterns and control flows. 
They propose ORShuffle~\cite{Sasy_CCS22} and WaksShuffle~\cite{Sasy_CCS23} as fully oblivious shuffle algorithms. 
However, they do not consider random sampling, dummy data addition, or providing DP to the output. 
In this work, we use their algorithms as building blocks and propose algorithms that provide FODP for the entire process, including random sampling and dummy data addition.

\section{Preliminaries}
\label{sec:preliminaries}

\subsection{Basic Notations}
\label{sub:notations}
Below, we introduce our notations. 
Let $\reals$, $\nnreals$, $\nats$, $\ints$, $\nnints$ be the sets of real numbers, non-negative real numbers, natural numbers, integers, and non-negative integers, respectively. 
For a set $\calZ$, let $\calZ^*$ be the set of all finite sequences of elements in $\calZ$. 
For $a\in\nats$, we denote the set $\{1,2,\ldots,a\}$ by $[a]$. 

Let $n \in \nats$ be the number of users, and $u_i$ be the $i$-th user. 
Let $d \in \nats$ be the number of items. 
Each item is represented as an index from $1$ to $d$. 
For $i \in [n]$, let $x_i \in [d]$ be the input value of user $u_i$. 
Let $\bmx = (x_1,\ldots,x_n)$ be the database. 
For $i \in [d]$, let $f_i \in [0,1]$ be the relative frequency of item $i$ in $\bmx$ such that $\sum_{i=1}^d f_i = 1$. 
Let $\hf_i \in \reals$ be the estimate of $f_i$. 
Let $\bmf = (f_1,\ldots,f_d)$ and $\bmhf = (\hf_1,\ldots,\hf_d)$. 
Frequency estimation is the task of calculating $\bmhf$ as close to $\bmf$ as possible. 

\subsection{DP (Differential Privacy)}
\label{sub:DP}
\noindent{\textbf{DP and LDP.}}~~DP is 
formally defined as follows: 

\begin{definition}[$(\epsilon,\delta)$-DP~\cite{DP}] \label{def:DP}
Let $\epsilon \in \nnreals$ and $\delta \in [0,1]$. 
Let $\calR$ be a randomized algorithm with domain $[d]^n$. 
Then, $\calR$ 
provides \emph{$(\epsilon,\delta)$-DP} (or \emph{$\epsilon$-DP} when $\delta=0$) if for any neighboring databases 
\colorB{$\bmx = (x_1,\ldots,x_i,\ldots,x_n)$ and $\bmx' = (x_1,\ldots,x'_i,\ldots,x_n)$ that differ in one entry $i\in[n]$} 
and any 
$O \subseteq \mathrm{Range}(\calR)$, 
\begin{align}
\Pr[\calR(\bmx) \in O] \leq e^\epsilon \Pr[\calR(\bmx') \in O] + \delta. 
\label{eq:DP_inequality}
\end{align}
\end{definition}
The parameters $\epsilon$ and $\delta$ should be small. 
For example, $\epsilon$ should ideally satisfy $\epsilon \leq 1$ and should not exceed $5$ or $10$~\cite{DP_Li}. 
$\delta$ must satisfy $\delta \ll 1/n$~\cite{DP}. 

Most shuffle protocols use an LDP (Local DP) algorithm, such as 
the GRR~\cite{Kairouz_ICML16,Wang_PVLDB20}, OUE~\cite{Wang_USENIX17}, and OLH~\cite{Wang_USENIX17}. 
LDP is a variant of DP in the local model and is defined as follows: 
\begin{definition}[$\epsilon$-LDP~\cite{Kasiviswanathan_FOCS08,Duchi_FOCS13}] \label{def:LDP}
Let $\epsilon \in \nnreals$. 
Let $\calR_L$ be a randomized algorithm (local randomizer) with domain $[d]$. 
$\calR_L$ provides \emph{$\epsilon$-LDP} if for any $x,x' \in [d]$ and any $O \subseteq \mathrm{Range}(\calR_L)$, 
\begin{align*}
\Pr[\calR_L(\bmx) \in O] \leq e^\epsilon \Pr[\calR_L(\bmx') \in O]. 
\end{align*}
\end{definition}

\smallskip
\noindent{\textbf{Pure Shuffle DP Protocols.}}~~A shuffle DP protocol 
consists of a \textit{shuffler part} (also called a \textit{shuffle algorithm}) that, given input values, outputs shuffled values and an \textit{analyzer part} that, given shuffled values, outputs analysis results. 
Below, we explain pure shuffle protocols using LDP algorithms. 

Assume that each user $u_i$ ($i \in [n]$) 
applies a local randomizer $\calR_L$ providing $\epsilon_L$-LDP to her input value $x_i$ 
and sends the noisy value $\calR_L(x_i)$ to the shuffler. 
The shuffler randomly shuffles noisy values $\calR_L(x_1),\ldots,\calR_L(x_n)$ and sends the shuffled values to the data collector. 
Then, the random shuffling serves as anonymization and amplifies privacy. 
Specifically, the shuffled values provide $(\epsilon,\delta)$-DP, where 
$\epsilon = g(n,\epsilon_L,\delta)$, 
and $g(n,\epsilon_L,\delta)$ is monotonically decreasing (resp.~increasing) with respect to $n$ and $\delta$ (resp.~$\epsilon_L$). 
$\epsilon$ is generally much smaller than $\epsilon_L$. 
See Appendix~\ref{sub:pure_shuffle_collusion} for more details.

\subsection{Robustness against Collusion with Users}
\label{sub:collusion_attacks}
\noindent{\textbf{Collusion with Users.}}~~A major drawback of the pure shuffle DP protocols is that they are vulnerable to collusion attacks between users and the data collector. 
Let $\Omega \subset [n]$ be indices of colluding users; i.e., users $\{u_i | i \in \Omega\}$ 
share their noisy values $\{\calR_L(x_i) | i \in \Omega \}$ with the data collector. 
In this case, the value of $\epsilon$ after shuffling is increased from $g(n,\epsilon_L,\delta)$ to $g(n-|\Omega|,\epsilon_L,\delta)$, as the data collector can remove $|\Omega|$ noisy values from $n$ shuffled values. 
\colorB{This results in a great increase in $\epsilon$ in some cases. 
For example, suppose that $n=6 \times 10^5$, $\delta=10^{-12}$, and $|\Omega| = 6 \times 10^4$, i.e., $10\%$ of users are malicious. 
Then, $\epsilon$ can be increased from $1$ to $7.2$ (resp.~from $4$ to $20$) after collusion attacks in the pure shuffle protocols using LDP mechanisms (resp.~the protocol in \cite{Cheu_SP22}), as shown in \cite{Murakami_SP25}.}

\colorB{We can prevent the increase in $\epsilon$ by estimating the number $|\Omega|$ of colluding users in advance. 
However, it is difficult to estimate the exact value of $|\Omega|$ in practice. 
For example, in X (formerly Twitter), 
the estimates of the proportion of fake users range from $11\%$ to $20\%$~\cite{Twitter_fake2,Ng_SR25} 
and can be even $43\%$ during elections~\cite{Ng_SR25}. 
Although we can conservatively estimate $|\Omega|$, this results in suboptimal accuracy. 
For example, when we estimate $|\Omega|$ to be $50\%$, the squared error can increase by a factor of four~\cite{Murakami_SP25}. 
Note that this is a general issue, not limited to specific protocols or privacy parameters.}

\smallskip
\noindent{\textbf{Robustness.}}~~Shuffle DP protocols should be robust against collusion with users; i.e., 
$\epsilon$ and $\delta$ 
after shuffling should not be increased even if some users share their data sent to the shuffler with the data collector. 
Following~\cite{Murakami_NDSS26}, we say that a shuffle DP protocol is \textit{robust against collusion with users} if the privacy parameters $(\epsilon, \delta)$ are not increased by collusion with users; we give its formal definition in Appendix~\ref{sub:robustness_collusion_formal}. 
 
Note that pure shuffle DP protocols \textit{cannot} provide this robustness for the following reason. 
First, they must add noise on the user side, because otherwise it merely shuffles input values and cannot provide $(\epsilon,\delta)$-DP. 
Then, they are not robust against collusion with users, 
because the adversary can increase $\epsilon$ from $g(n,\epsilon_L,\delta)$ to $g(n-|\Omega|,\epsilon_L,\delta)$ by removing $|\Omega|$ noisy values from $n$ shuffled values, as explained above. 

\subsection{Augmented Shuffle DP Protocols}
\label{sub:augmented_shuffle}
The robustness against collusion with users 
can be achieved by introducing the augmented shuffle model, which allows the shuffler to perform additional operations: random sampling and adding dummies. 
The additional operations act as \textit{noise addition on the shuffler side}, improving the robustness against collusion with users. 
Below, we briefly explain two existing augmented shuffle protocols: 
UD~\cite{Wang_PVLDB20} and LNF~\cite{Murakami_SP25}. 

\smallskip
\noindent{\textbf{UD (Uniform Dummies) Protocol~\cite{Wang_PVLDB20}}.}~~In the UD protocol, each user $u_i$ ($i \in [n]$) 
sends a noisy value $\calR_L(x_i)$ to the shuffler, where $\calR_L$ is a local randomizer. 
Then, the shuffler adds $\lambda \in \nnints$ dummy values $y_1,\ldots,y_\lambda \in \mathrm{Range}(\calR_L)$ uniformly randomly sampled from the range of $\calR_L$. 
Finally, the shuffler randomly shuffles the values $\calR_L(x_1),\ldots,\calR_L(x_n),y_1,\ldots,y_\lambda$ and sends the shuffled values to the data collector. 

It is proved in \cite{Wang_PVLDB20} that $\epsilon$ of the UD protocol in the presence of collusion with users is much smaller than that of the pure shuffle protocol. 
However, 
$\epsilon$ of the UD protocol still increases slightly~\cite{Murakami_SP25}, 
as it adds local noise on the user side. 
In this paper, we modify this protocol to achieve robustness against collusion with users 
by not adding local noise. 

\smallskip
\noindent{\textbf{LNF (Local-Noise-Free) Protocol~\cite{Murakami_SP25}}.}~~In the LNF protocol, each user $u_i$ sends her input value $x_i$ to the shuffler without adding noise. 
Then, the shuffler performs three operations: random sampling, adding dummies, and shuffling. 
Specifically, the shuffler randomly samples each input value $x_i$ ($i \in [n]$) with probability $\beta \in [0,1]$. 
In other words, it discards $x_i$ with probability $1-\beta$. 
Then, for each item $i\in[d]$, the shuffler samples $z_i$ from a dummy-count distribution $\calD$ over $\nnints$ ($z_i \sim \calD$) 
and adds the following dummy values: 
\begin{center} $\underbrace{1,\ldots,1}_{z_1},\underbrace{2,\ldots,2}_{z_2},\ldots,\underbrace{d,\ldots,d}_{z_d}$.
\end{center} 
Finally, 
it shuffles the sampled values and dummy values. 

It is proved in \cite{Murakami_SP25} that the LNF protocol provides DP and is robust against collusion with users 
if $\calD$ provides DP. 
As the dummy-count distribution $\calD$, \cite{Murakami_SP25} introduces the \textit{asymmetric geometric distribution}, 
whose left tail shrinks 
as $\beta$ decreases. 
The LNF protocol with this distribution provides 
much higher accuracy than other existing shuffle DP protocols 
when $\beta=1$. 
When $\beta = 1-e^{-\epsilon/2}$, the distribution becomes a one-sided geometric distribution (i.e., the left tail disappears). 
In this case, the LNF protocol achieves pure DP ($\delta=0$) at the cost of accuracy. 

\subsection{Full Obliviousness}
\label{sub:FO}

\noindent{\textbf{Definition.}}~~Full obliviousness \cite{Sasy_CCS22,Sasy_CCS23} 
prevents 
side-channel attacks based on external/internal memory access patterns and control flows 
by making 
the memory access patterns 
and the sequence of executed instructions independent of secret data. 

Formally, an \textit{instruction trace} of the randomized algorithm $\calR$ is a sequence of indices $(i_1,i_2,\ldots) \in \nats^*$, where each index points to an instruction (e.g., line number) in $\calR$. 
A \textit{memory trace} of $\calR$ is a sequence of tuples $(t,b,l)$, where $t\in\nats$ represents that the memory is accessed during the $t$-th instruction execution, $b\in\{0,1\}$ 
indicates a read (0) or write (1), 
and $l \in \nats$ represents the location of memory accessed. 

Assume that $\calR$ leaks the instruction and memory traces along with its output. 
Throughout this paper, 
we denote by $\calR^\calM$ (resp.~$\calR^\calI$) a randomized algorithm that outputs memory (resp.~instruction) traces $\calR$ leaks. 
Then, 
full obliviousness is defined as follows: 

\begin{definition}[Full obliviousness~\cite{Sasy_CCS22,Sasy_CCS23}] \label{def:FO}
Let $\Lambda \subseteq [n]$. 
Let $\calR$ be a randomized algorithm with domain $[d]^n$. 
Then, $\calR$ is \emph{fully oblivious 
to 
$\{x_i\}_{i\in\Lambda}$} if there exists a simulator $\calS$ such that for any database $\bmx=(x_1,\ldots,x_n)$ 
and any $(o,(o_M,o_I)) \in \mathrm{Range}(\calR \times (\calR^\calM, \calR^\calI))$, 
\begin{align}
&\Pr\left[\left(\calR(\bmx),\left(\calR^\calM(\bmx),\calR^\calI(\bmx)\right)\right) = (o,(o_M,o_I))\right] \nonumber\\
&= \Pr\left[\left(\calR(\bmx),\calS \left((x_i)_{i\in [n] \setminus \Lambda}, (|x_i|)_{i \in \Lambda}\right)\right) \hspace{-0.5mm}=\hspace{-0.5mm} (o,(o_M,o_I))\right]. 
\label{eq:FO_equality}
\end{align}
\end{definition}
Definition~\ref{def:FO} regards the input values $(x_i)_{i \in \Lambda}$ as secrets. 
Full obliviousness guarantees that the memory traces $\calR^\calM(\bmx)$ and instruction traces $\calR^\calI(\bmx)$ 
are independent of $(x_i)_{i \in \Lambda}$ and do not reveal any information about it 
beyond what is implied by non-secrets $(x_i)_{i\in [n] \setminus \Lambda}$ given to the simulator $\calS$. 
Note that the size $|x_i|$ is public for any input value $i\in[n]$. 
This applies to our scenario, as the number $d$ of items is public (i.e., $|x_i| = \log_2 d$). 

Following the convention in \cite{Sasy_CCS22,Sasy_CCS23}, we also say that for any data $A$, 
an algorithm is fully oblivious to $A$ if its memory and instruction traces are independent of $A$. 

\smallskip{}
\noindent{\textbf{Fully Oblivious Primitives.}}~~Following \cite{Sasy_CCS22,Sasy_CCS23}, we make the following assumptions on fully oblivious primitives. 
We assume that arithmetic and bitwise operations are fully oblivious to the input values. 
We also assume that random bits can be generated fully obliviously by simply reading them from a special memory region.  
Moreover, we 
assume that a comparison of two real values $y_1\in\reals$ and $y_2\in\reals$ is fully oblivious to $y_1$ and $y_2$. 
Let $[y_1 \geq y_2]$ be a fully oblivious comparison that outputs $1$ if $y_1 \geq y_2$ and $0$ otherwise. 

Other basic primitives include 
fully oblivious swapping, random number generators, and shuffling~\cite{Sasy_CCS22,Sasy_CCS23}. 
We 
define these primitives. 
For $m\in\nats$, let $C = (c_1,\ldots,c_m)$ be an array of length $m$. 
Let $\zeta \in \nnreals$ be the size of each element in $C$. 

We denote a fully oblivious swap function by $\texttt{OSWAP}(c_i,c_j,\allowbreak b)$, where $i \neq j$ and $b \in \{0,1\}$. 
$\texttt{OSWAP}(c_i,c_j,b)$ swaps the values of $c_i$ and $c_j$ if $b=1$ and does nothing otherwise. 
$\texttt{OSWAP}(c_i,c_j,b)$ is fully oblivious to the values $c_i$, $c_j$, and $b$, but not to the indices $i$ and $j$. 
$\texttt{OSWAP}(c_i,c_j,b)$ can be implemented using CMOV or XOR operations~\cite{Ngai_SP24}. 

We denote fully oblivious random generators for real and natural values by $\texttt{ORAND\_REAL}()$ and $\texttt{ORAND\_NATS}(x)$, respectively, where $x \in \nats$. 
$\texttt{ORAND\_REAL}()$ randomly generates a real value $r \in [0,1)$. 
This can be implemented 
by 
generating 
$\gamma$ ($\gamma \in \nats$) random bits, regarding them as a binary integer $r_0 \in [0,2^\gamma)$, and dividing it by $2^{\gamma}$. 
$\texttt{ORAND\_NATS}(x)$ randomly generates a natural value $r \in [x]$ and can be implemented by generating a 
$\lceil \log_2 x \rceil$-bit binary integer $r_0$ until $r_0 < x$ and outputting $r = r_0 + 1$. 
Both $\texttt{ORAND\_REAL}()$ and $\texttt{ORAND\_NATS}(x)$ are fully oblivious to the random value $r$. 

Finally, we denote a fully oblivious shuffle algorithm by $\texttt{OSHUFFLE}\allowbreak(c_1,\ldots,c_m)$. 
$\texttt{OSHUFFLE}\allowbreak(c_1,\ldots,c_m)$ randomly shuffles $c_1,\ldots,c_m$ and is fully oblivious to $C$. 
Examples of this algorithm include ORShuffle~\cite{Sasy_CCS22} and WaksShuffle~\cite{Sasy_CCS23}. 
The total runtime of ORShuffle and WaksShuffle is $O(\zeta m \log^2 m)$ and 
$O(m \log^3 m + \zeta m \log m)$, respectively.  

\section{Fully Oblivious Differential Privacy}
\label{sec:proposal}
In this section, we introduce 
FODP, a fully oblivious version of DP. 
We first clarify our 
system model and threat model 
(Section~\ref{sub:threat_model}). 
Then, we define FODP (Section~\ref{sub:FODP}) and show 
its relationship with full obliviousness (Section~\ref{sub:FODP_theoretical}). 

\subsection{Assumptions}
\label{sub:threat_model}

\noindent{\textbf{System Model.}}~~We consider the augmented shuffle DP model with TEEs. 
In this model, 
the TEE inside a server serves as a shuffler, and the data collector is outside the TEE. 
Thus, each user sends data to the TEE, which then performs augmented shuffling (i.e., random sampling, adding dummies, and shuffling) and places the shuffled data outside the TEE. 
Finally, the server (outside the TEE) calculates an estimate $\bmhf$ of the frequency distribution $\bmf$ 
and publishes $\bmhf$. 

The TEE provides data confidentiality and integrity. 
For example, Intel SGX \cite{Anati_HASP13} 
provides data confidentiality by allocating 
a hardware-protected container 
called an \textit{enclave}. 
Data in the internal memory (i.e., the memory inside the enclave) is encrypted, and keys used for encryption/decryption are not compromised. 
Thus, the data and code within the enclave are isolated from outside the enclave, including the main operating system. 
Note that the internal memory is often small (e.g., $93.5$ MB~\cite{Lee_EECS22}).  
Therefore, the code may store encrypted and integrity-protected data in a large external memory. 

Intel SGX also provides integrity of the running code through an attestation mechanism \cite{Anati_HASP13}, which provides a cryptographic proof that the code 
is genuine. 
After the verification is complete, users can establish a secure channel with the enclave and communicate with it using cryptographic keys. 

\smallskip{}
\noindent{\textbf{Threat Model.}}~~In this paper, anyone except a single user $u_i$ (referred to as a victim) can be an adversary, e.g., other users, server administrators, or individuals who have obtained the published estimate $\bmhf$. 
We refer to adversaries comprising the server administrators as \textit{internal adversaries} and to other adversaries as \textit{external adversaries}. 
Following the conventions of the DP literature~\cite{DP,Dwork_TCC06}, we consider the worst case for background knowledge of the adversaries; i.e., they may know all input values 
$\{x_j | j \ne i \}$ 
other than the victim's $x_i$. 

We can ensure 
that the shuffler follows the protocol and does not collude with the data collector by using the TEE. 
However, the data collector can collude with some users or inject some fake user accounts~\cite{Twitter_fake1,Twitter_fake2,Facebook_fake,Thomas_USENIX13} to obtain their (plaintext) data sent to the shuffler. 

In addition, the internal adversaries (i.e., server administrators) can 
obtain external memory access patterns, as it is outside the TEE's control~\cite{Allen_NeurIPS19}. 
They may also obtain internal memory access patterns and control flows by launching state-of-the-art attacks, such as \cite{Bulck_USENIX17,Lee_USENIX20,Lee_USENIX17,Moghimi_USENIX20}.

\smallskip{}
\noindent{\textbf{Remark.}}~~We exclude timing attacks~\cite{Haeberlen_USENIX11,Jin_SP22}, which infer secret values based on the time to run the code, from our scope in the same way as~\cite{Sasy_CCS22,Sasy_CCS23}. 
Our algorithms can easily be modified to be secure against timing attacks by making the runtime constant~\cite{Haeberlen_USENIX11,Jin_SP22}; we have also confirmed that the variance in the runtime of our algorithms is small.  

\colorB{We also note that \cite{Wang_TIFS25} proposes side-channel attacks for shuffle protocols with multiple rounds, multiple messages (per user), or variable-length messages. 
Our algorithms are secure against all of these attacks because they are single-round protocols with a single, fixed-length message.}

\subsection{FODP (Fully Oblivious DP)}
\label{sub:FODP}

\noindent{\textbf{Definition.}}~~To provide strong privacy guarantees against the internal adversary, we introduce FODP: 
\begin{definition}[$(\epsilon,\delta)$-FODP] \label{def:FODP}
Let $\epsilon \in \nnreals$ and $\delta \in [0,1]$. 
Let $\calR$ be a randomized algorithm with domain $[d]^n$. 
Then, $\calR$ provides \emph{$(\epsilon,\delta)$-FODP} if for any neighboring databases 
\colorB{$\bmx = (x_1,\ldots,x_i,\ldots,x_n)$ and $\bmx' = (x_1,\ldots,x'_i,\ldots,x_n)$ that differ in one entry $i\in[n]$} 
and any $(O,O_M,O_I) \subseteq \mathrm{Range}((\calR, \calR^\calM, \calR^\calI))$, 
\begin{align}
&\Pr[(\calR(\bmx),\calR^\calM(\bmx),\calR^\calI(\bmx)) \in (O,O_M,O_I)] \nonumber\\
&\leq e^\epsilon \Pr[(\calR(\bmx'),\calR^\calM(\bmx'),\calR^\calI(\bmx')) \in (O,O_M,O_I)] + \delta. 
\label{eq:FODP_inequality}
\end{align}
\end{definition}
\colorB{Since FODP is defined using (\ref{eq:FODP_inequality}), it also holds under composition~\cite{DP}.} 
Chan \textit{et al.}~\cite{Chan_SODA19} introduce a notion of DO (Differential Obliviousness), which provides DP guarantees for only memory traces $\calR^\calM(\bmx)$. 
Later, Allen \textit{et al.}~\cite{Allen_NeurIPS19} introduce 
ODP (Oblivious DP), which provides DP for the output $\calR(\bmx)$ and the external memory traces. 
FODP 
is a generalization of 
DO and ODP in that it provides DP for 
all of the output $\calR(\bmx)$, 
memory traces $\calR^\calM(\bmx)$, and 
instruction traces $\calR^\calI(\bmx)$. 

\smallskip{}
\noindent{\textbf{FODP and DP.}}~~It is clear that 
FODP is stronger than DP; i.e., if an algorithm $\calR$ running within the TEE provides $(\epsilon,\delta)$-FODP, then it also provides $(\epsilon,\delta)$-DP. 
Then, by the post-processing invariance~\cite{DP}, the estimate $\bmhf$ exposed to the external adversary also provides $(\epsilon,\delta)$-DP. 

\begin{figure}[t]
  \centering
  \includegraphics[width=0.99\linewidth]{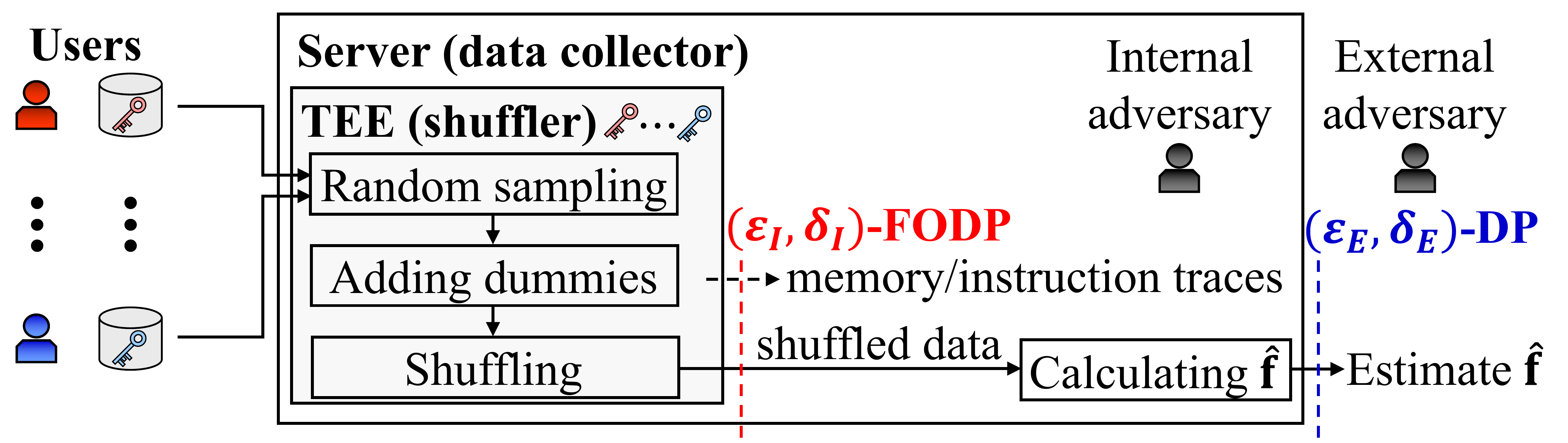}
  \caption{FODP/DP in our system model ($\epsilon_E \leq \epsilon_I$, $\delta_E \leq \delta_I$).} 
  \label{fig:FODP}
\end{figure}

Figure~\ref{fig:FODP} shows the roles of FODP and DP in our system model. 
We use $(\epsilon_I,\delta_I)$-FODP as a privacy notion against the internal adversary and $(\epsilon_E,\delta_E)$-DP against the external adversary, where $\epsilon_E \leq \epsilon_I$ and $\delta_E \leq \delta_I$. 
Note that it is reasonable to set smaller privacy parameters against the external adversary (e.g., $\epsilon_E = 0.1$, $\epsilon_I = 1$, $\delta_E = 0$, $0 < \delta_I \ll 1/n$) because \textit{everyone} can be an external adversary after publishing $\bmhf$. 

\subsection{Relationship with Full Obliviousness}
\label{sub:FODP_theoretical}

At first glance, the relationship between 
FODP and full obliviousness (Definition~\ref{def:FO}) 
may appear unclear, as 
the latter 
is defined using a simulator $\calS$. 
However, we can show that FODP is connected to DP and full obliviousness as follows: 
\begin{definition}[Full obliviousness with respect to a single secret] \label{def:FO_single}
We say that a randomized algorithm $\calR$ with domain $[d]^n$ is \emph{fully oblivious 
to a single secret} if $\calR$ is fully oblivious 
to 
$\{x_1\},\{x_2\},\ldots,$ and $\{x_n\}$. 
\end{definition}

\begin{theorem} \label{thm:DP_FO_FODP}
Let $\calR$ be a randomized algorithm with domain $[d]^n$. 
If $\calR$ provides $(\epsilon,\delta)$-DP and is fully oblivious 
to a single secret, then $\calR$ provides $(\epsilon,\delta)$-FODP. 
\end{theorem}

Theorem~\ref{thm:DP_FO_FODP} states that any $(\epsilon,\delta)$-DP algorithm can be modified to satisfy $(\epsilon,\delta)$-FODP by making it fully oblivious to a single secret. 
We make use of Theorem~\ref{thm:DP_FO_FODP} to prove that our 
\FOUD{} and \FOLNF{} 
provide FODP. 
Meanwhile, it is important to note that providing DP and full obliviousness with respect to a single secret is \textit{not} a necessary condition (but a sufficient condition) for providing FODP. 
In fact, we prove FODP of our 
\FOLNFast{} 
by directly showing the inequality (\ref{eq:FODP_inequality}). 

\section{Our FODP Algorithms}
\label{sec:FODP_algorithm}

In this section, we propose a general framework and three concrete algorithms for FODP in the augmented shuffle model. 
We first explain our motivation and the overview of our algorithms (Section~\ref{sub:motivation}). 
We then present our general framework (Section~\ref{sub:general_framework}) and three concrete algorithms (Sections~\ref{sub:FOUD}-\ref{sub:FOLNF_DP}). 
Finally, we compare our three algorithms (Section~\ref{sub:comparison_three_algorithms}). 

\subsection{Overview}
\label{sub:motivation}
\begin{figure}[t]
  \centering
  \includegraphics[width=0.99\linewidth]{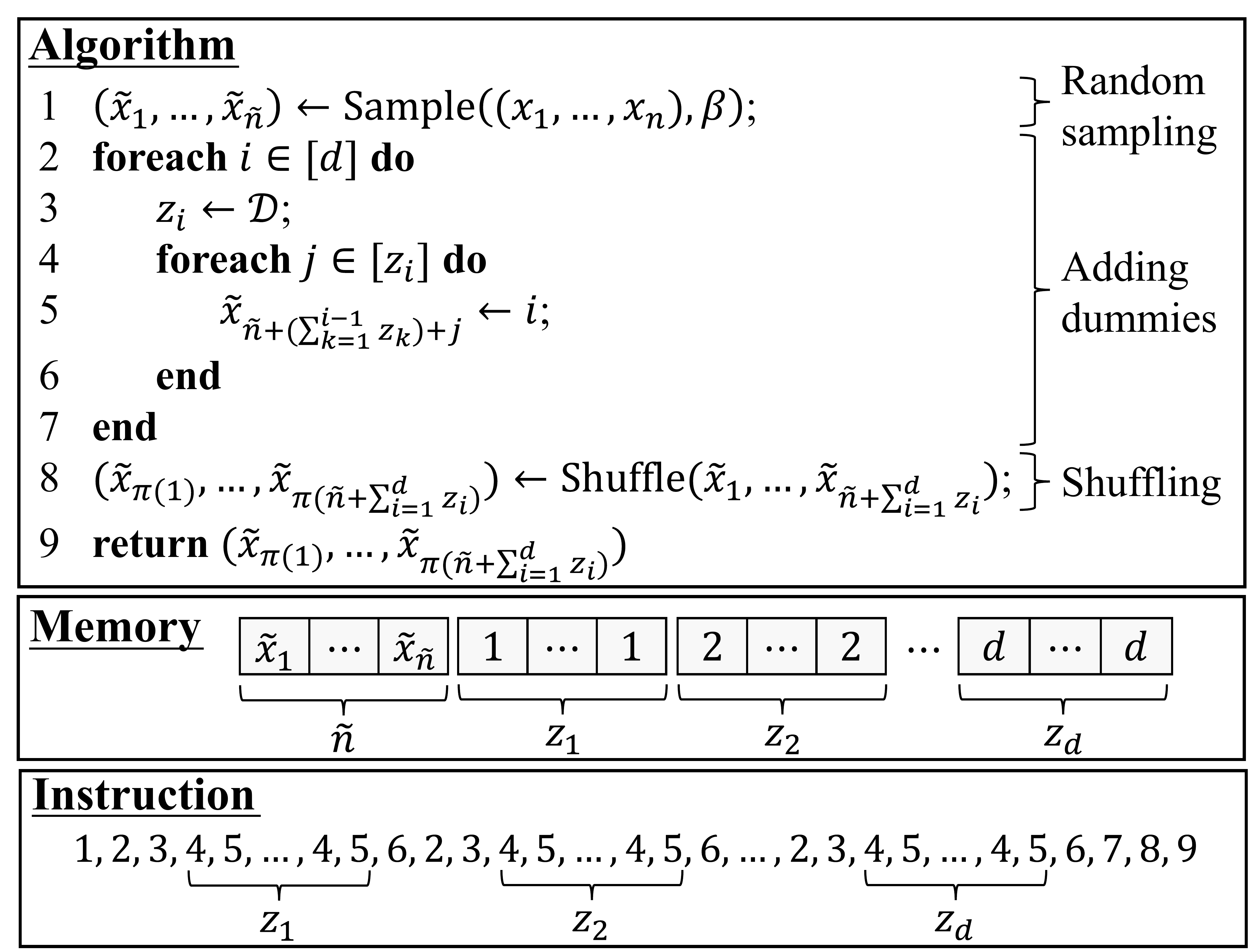}
  \caption{Direct implementation of the LNF protocol~\cite{Murakami_SP25}.} 
  \label{fig:LNF_memory_instruction}
\end{figure}

\noindent{\textbf{Technical Motivation.}}~~The goal of this work is to design 
FODP algorithms for frequency estimation in the augmented shuffle model with TEEs. 
However, directly implementing augmented shuffle algorithms within TEEs can result in the leakage of substantial side-channel information, thereby violating FODP. 
To explain this issue, we show pseudocode for the shuffler part of the LNF protocol~\cite{Murakami_SP25} and a visualization of its memory usage and instruction traces in Figure~\ref{fig:LNF_memory_instruction}. 

As shown in the top panel of Figure~\ref{fig:LNF_memory_instruction}, the shuffler first samples input values $x_1,\ldots,x_n$ with probability $\beta$ and obtains selected values $\tx_1,\ldots,\tx_{\tn}$, where $\tn \leq n$ (line 1). 
Then, for each item $i \in [d]$, the shuffler adds $z_i$ dummy values, where $z_i$ is generated from the dummy-count distribution $\calD$ (lines 2-7). 
Finally, the shuffler randomly shuffles the input and dummy values and outputs $(\tx_{\pi(1)},\ldots,\tx_{\pi(\tn + \sum_{i=1}^d z_i)})$, where $\pi$ is a random permutation over $[\tn + \sum_{i=1}^d z_i]$ (lines 8-9). 
\colorB{The key insight here is that sampling and adding dummies, followed by shuffling, are equivalent to adding discrete noise to each histogram bin~\cite{Murakami_SP25}. Thus, the output provides DP.}

In this 
\colorB{algorithm}, 
the shuffler allocates 
memory of size $\tn$ for selected input values and of size $z_i$ ($i\in[d]$) for dummy values $i$, as shown in the middle panel of Figure~\ref{fig:LNF_memory_instruction}. 
Thus, the values $\tn, z_1, \ldots, z_d$ can be leaked from the size of the allocated memory. 
In addition, the shuffler repeats the second ``foreach'' loop (lines 4-5) $z_i$ times. 
Thus, the instruction trace can be represented as a sequence of line numbers in the bottom panel of Figure~\ref{fig:LNF_memory_instruction}, from which $z_1, \ldots, z_d$ are also leaked.

In summary, 
the values $\tn, z_1, \ldots, z_d$ can be leaked from the memory and instruction traces. 
This leads to a privacy violation because the adversary who knows $z_1, \ldots, z_d$ can remove dummy values from the shuffled values. 
Note that the number $\tn$ of selected input values must also be kept secret to obtain privacy amplification by sampling~\cite{Li_AsiaCCS12,Balle_NeurIPS18,Murakami_SP25}. 
Thus, the leakage of $\tn$ also reduces privacy. 

\begin{figure}[t]
  \centering
  \includegraphics[width=0.99\linewidth]{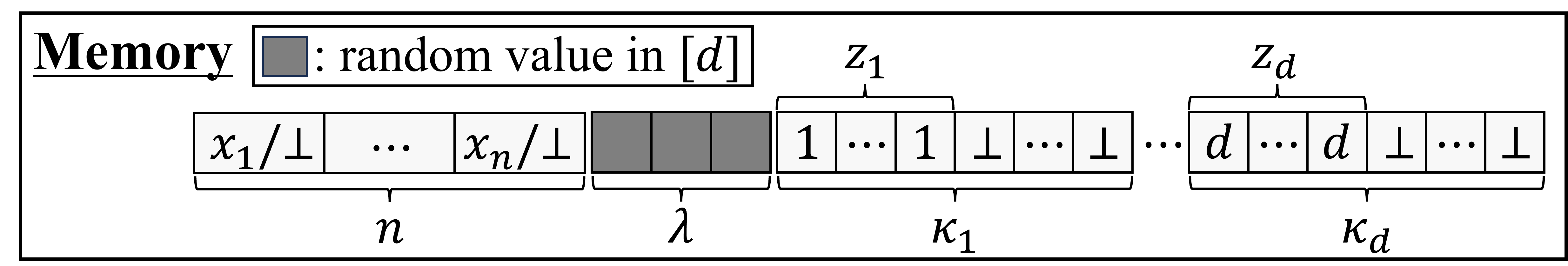}
  \caption{Allocated memory in our general framework.} 
  \label{fig:proposal_memory}
\end{figure}

\smallskip{}
\noindent{\textbf{Algorithm Overview.}}~~We propose a general framework for FODP that \colorB{includes the UD~\cite{Wang_PVLDB20} and LNF~\cite{Murakami_SP25} protocols and} 
addresses the leakage issue explained above. 
Our framework prevents 
the leakage of side-channel information 
by \textit{obfuscating the allocated memory size}. 
Figure~\ref{fig:proposal_memory} shows the allocated memory in our framework. 
Specifically, we allocate memory of size $n$ for input values, of size $\lambda$ for uniform dummies, 
and of size $\kappa_i$ ($\geq z_i$) for dummy values $i$. 
We call $\kappa_i$ the \textit{bot count}, as we initialize $\kappa_i$ memory locations with $\bot$. 
We transform each input value into $\bot$ with probability $1-\beta$ and store $\lambda$ uniform dummies and $z_i$ dummy values $i$ ($i\in[d]$) in the memory using fully oblivious primitives in Section~\ref{sub:FO}. 

Our key insight is that 
an algorithm outputting memory and instruction traces in our framework can be expressed as an \textit{algorithm outputting bot counts $(\kappa_1,\ldots,\kappa_d)$ followed by a randomized post-processing algorithm} (Lemma~\ref{lem:general_privacy}). 
This lemma enables us to reduce FODP of our framework to the privacy of shuffled values and bot counts $(\kappa_1,\ldots,\kappa_d)$. 

Based on this, we propose three instances: 
\FOUD{}, \FOLNF{}, and \FOLNFast{}. 
\FOUD{} and \FOLNF{} output the same shuffled values as the UD and LNF protocols, respectively (after removing $\bot$). 
Both \FOUD{} and \FOLNF{} set $\kappa_i$ to a constant, 
thus providing full obliviousness with respect to a single secret. 
Then, by Theorem~\ref{thm:DP_FO_FODP}, \FOUD{} and \FOLNF{} provide FODP. 

\FOLNFast{} is a variant of \FOLNF{} that generates bot counts $(\kappa_1,\ldots,\kappa_d)$ under DP. 
Specifically, it calculates $\kappa_i$ as $\kappa_i = z_i + \omega_i$, where $\omega_i$ is generated from a \textit{bot-count distribution $\calD'$}. 
For \FOLNFast{}, we introduce 
\colorB{a novel \textit{joint asymmetric geometric distribution}. 
We prove that \FOLNFast{} with this distribution provides FODP by directly showing (\ref{eq:FODP_inequality}).}

We compare \FOUD{}, \FOLNF{}, and \FOLNFast{} in Section~\ref{sub:comparison_three_algorithms}.

\subsection{Our General Framework}
\label{sub:general_framework}
\noindent{\textbf{Algorithm.}}~~Algorithm~\ref{alg:proposal} shows a general framework for our algorithms, denoted by $\calR_{\calD,\calD',\beta,\lambda}$. 
It 
has four parameters: 
a dummy-count distribution $\calD$ over $\nnints$ (mean: $\mu$, variance: $\sigma^2$), 
a bot-count distribution $\calD'$ over $\nnints$, 
a sampling probability $\beta \in [0,1]$, and 
the number $\lambda \in \nnints$ of uniform dummies. 
These parameters are public. 
In our framework, each user $u_i$ ($i\in[n]$) sends her input value $x_i$ to the TEE without adding noise. 
Then, the TEE runs Algorithm~\ref{alg:proposal} as follows. 

\setlength{\algomargin}{5mm}
\begin{algorithm}[t]
  \SetAlgoLined
  \KwData{Database $\bmx = (x_1, \ldots, x_n)$, 
  \#items $d \in \nats$. 
  }
  \KwResult{Shuffled values $\bmtx = (\tx_{\pi(1)}, \ldots, \tx_{\pi(n + \lambda + \sum_{i=1}^d \kappa_i)})$.}
  \tcc{Random sampling}
  \ForEach{$i \in [n]$}{
    $a_1 \leftarrow x_i$; $a_2 \leftarrow \bot$\;
    $r \leftarrow \texttt{ORAND\_REAL}()$\;
    $\texttt{OSWAP}$($a_1$, $a_2$, $[r \geq \beta]$)\;
    $\tx_i \leftarrow a_1$\;
  }
  \tcc{Dummy data addition}
  \ForEach{$i \in [\lambda]$}{
    $r \leftarrow \texttt{ORAND\_NATS}(d)$\;
    $\tx_{n + i} \leftarrow r$\;
  }
  \ForEach{$i \in [d]$}{
    $z_i \leftarrow \texttt{DummyCountGeneration}(\calD)$\;
    $\kappa_i \leftarrow \texttt{BotCountGeneration}(z_i, \calD')$\;
    $(a_1,\ldots,a_{\kappa_i}) \leftarrow (\bot,\ldots,\bot)$\;
    \ForEach{$k \in [\kappa_i]$}{
      $\texttt{OSWAP}$($a_k$, $i$, $[z_i \geq k]$)\;
    }
    $(\tx_{n + \lambda + \sum_{j=1}^{i-1} \kappa_j +1},\ldots,\tx_{n + \lambda + \sum_{j=1}^{i} \kappa_j}) \leftarrow (a_1,\ldots,a_{\kappa_i})$\;
  }
  \tcc{Shuffling}
  $\tx_{\pi(1)},\ldots,\tx_{\pi(n + \lambda + \sum_{i=1}^d \kappa_i)}$ $\leftarrow$ $\texttt{OSHUFFLE}$($\tx_1,\ldots,\tx_{n + \lambda + \sum_{i=1}^d \kappa_i}$)\;
  \KwRet{$\bmtx = (\tx_{\pi(1)}, \ldots, \tx_{\pi(n + \lambda + \sum_{i=1}^d \kappa_i)})$}
  \caption{Our general algorithm $\calR_{\calD,\calD',\beta,\lambda}$ 
  ($\calD$: dummy-count distribution with mean $\mu$ and variance $\sigma^2$; 
  $\calD'$: bot-count distribution; 
  $\beta \in [0,1]$: sampling probability; 
  $\lambda \in \nnints$: \#uniform dummies).}
  \label{alg:proposal}
\end{algorithm}

First, for each input value $x_i$ ($i\in[n]$), we perform 
random sampling. 
Specifically, we randomly generate a real value $r \in [0,1)$ by $\texttt{ORAND\_REAL}()$ and perform 
swapping $\texttt{OSWAP}(a_1,a_2,[r \geq \beta])$, where $a_1 = x_i$ and $a_2 = \bot$ (lines 2-4). 
Then, we copy $a_1$ to $\tx_i$ (line 5). 
These processes select $x_i$ (i.e., $\tx_i = x_i$) with probability $\beta$ and do not select $x_i$ (i.e., $\tx_i = \bot$) with probability $1 - \beta$. 

Then, 
we randomly generate $\lambda$ 
uniform dummies 
from $[d]$. 
Specifically, for $i\in[\lambda]$, we generate a natural value $r \in [d]$ by $\texttt{ORAND\_NATS}(d)$ and copy $r$ to $\tx_{n + i}$ (lines 7-10). 
As a result, $\lambda$ uniform dummies are copied to $(\tx_{n + 1},\ldots,\tx_{n + \lambda})$. 

We also generate dummies for each item based on the dummy-count distribution $\calD$. 
Specifically, 
for each item $i\in[d]$, 
we call the function $\texttt{DummyCountGeneration}$, which generates a random sample $z_i$ from $\calD$ (line 12). 
This function can be implemented via a CDF (Cumulative Distribution Function) \cite{Karmakar_TC18} or arithmetic operations and is fully oblivious to $z_i$; 
see Appendix~\ref{sub:dummy-count_distribution} for details. 
Hereafter, we assume that this function is implemented via arithmetic operations. 
$z_i$ 
is used as the number of dummies for item $i$. 

After generating the number $z_i$ of dummies, we generate $(a_1,\ldots,a_{\kappa_i}) = (\bot,\ldots,\bot)$, where $\kappa_i \in \nnints$ is generated by the function $\texttt{BotCountGeneration}$ (lines 13-14). 
This function takes $z_i$ and the bot-count distribution $\calD'$ as input and outputs $\kappa_i$. 
$\texttt{BotCountGeneration}$ consists of generating a random sample from $\calD'$ and arithmetic operations. 
Thus, it is also fully oblivious to $z_i$. 
We introduce three instances of 
this function 
in Sections~\ref{sub:FOUD}-\ref{sub:FOLNF_DP}. 
Then, we perform 
swapping $\texttt{OSWAP}(a_k,i,[z_i \geq k])$ for each $k\in[\kappa_i]$ and copy $(a_1,\ldots,a_{\kappa_i})$ to 
$(\tx_{n + \lambda + \sum_{j=1}^{i-1} \kappa_j +1},\ldots,\tx_{n + \lambda + \sum_{j=1}^{i} \kappa_j})$ (lines 15-18). 
Note that running $\texttt{OSWAP}(a_k,i,[z_i \geq k])$ for each $k\in[\kappa_i]$ results in swapping between $\bot$ and $i$ for $z_i$ times. 
Thus, 
$(\tx_{n + \lambda + \sum_{j=1}^{i-1} \kappa_j +1},\ldots,\allowbreak\tx_{n + \lambda + \sum_{j=1}^{i} \kappa_j})$ includes $z_i$ dummies for item $i$. 

Finally, we perform fully oblivious shuffling $\texttt{OSHUFFLE}$($\tx_1,\allowbreak \ldots,\allowbreak\tx_{n + \lambda + \sum_{i=1}^d \kappa_i}$) 
and output 
$\bmtx = (\tx_{\pi(1)}, \ldots, \allowbreak \tx_{\pi(n + \lambda + \sum_{i=1}^d \kappa_i)})$, 
where $\pi$ is a random permutation over 
$[n + \lambda + \sum_{i=1}^d \kappa_i]$ 
(lines 20-21). 
The output consists of shuffled values, where each input value is sampled with probability $\beta$, $\lambda$ uniform dummies are added, and $z_i$ dummies are added to each item $i\in[d]$. 

\smallskip{}
\noindent{\textbf{Calculation of $\bmhf$.}}~~After our general algorithm $\calR_{\calD,\calD',\beta,\lambda}$ outputs the shuffled values $\bmtx = (\tx_{\pi(1)}, \ldots, \allowbreak \tx_{\pi(n + \lambda + \sum_{i=1}^d \kappa_i)})$, the server (outside the TEE) discards $\bot$ and counts the number $c_i \in \nnints$ of each item $i$ in the shuffled values. 
Then, the server calculates an unbiased estimate $\bmhf$ of the frequency $\bmf$ 
as follows: 
\begin{align}
\textstyle{\hf_i = \frac{1}{\beta n}(c_i - \frac{\lambda}{d} - \mu) ~~ (i\in[d]),}
\label{eq:hf_i_general}
\end{align}
where $\mu$ 
is the mean of the dummy-count distribution $\calD$. 

\smallskip{}
\noindent{\textbf{Privacy and Robustness.}}~~Our general algorithm $\calR_{\calD,\calD',\beta,\lambda}$ is a combination of the UD and LNF protocols in that it samples each input value with probability $\beta$, adds $\lambda$ uniform dummies over $[d]$, adds $z_i$ dummies for each item $i\in[d]$, and then shuffles them. 
Thus, $\calR_{\calD,\calD',\beta,\lambda}$ provides DP. 
The remaining issue is side-channel information 
of $\calR_{\calD,\calD',\beta,\lambda}$. 

As shown in Figure~\ref{fig:proposal_memory} and Algorithm~\ref{alg:proposal}, our general algorithm $\calR_{\calD,\calD',\beta,\lambda}$ obfuscates the allocated memory size with $\bot$ and makes control flows (i.e., ``foreach'' loops) dependent only on the memory size. 
Based on this, 
we show our key insight that memory and instruction traces of $\calR_{\calD,\calD',\beta,\lambda}$ depend on the database $\bmx$ only through bot counts $(\kappa_1,\ldots,\kappa_d)$: 
\begin{lemma} \label{lem:general_privacy}
Let $\calR^\bot_{\calD,\calD',\beta,\lambda}$ be a randomized algorithm that, given a database $\bmx=(x_1,\ldots,x_n)$, outputs $(\kappa_1, \ldots, \kappa_d)$ in $\calR_{\calD,\calD',\beta,\lambda}$ (Algorithm~\ref{alg:proposal}). 
Then, there exists 
a randomized post-processing algorithm $\phi$ 
such that 
for any $\bmx \in [d]^n$, 
\begin{align}
(\calR^\calM_{\calD,\calD',\beta,\lambda}(\bmx), \calR^\calI_{\calD,\calD',\beta,\lambda}(\bmx)) = \phi(\calR^\bot_{\calD,\calD',\beta,\lambda}(\bmx)). 
\label{eq:lemma_post_processing}
\end{align}
\end{lemma}
This is a key lemma that reduces FODP of our general algorithm to the privacy of \textit{shuffled values} output by $\calR_{\calD,\calD',\beta,\lambda}$ and \textit{bot counts $(\kappa_1, \ldots, \kappa_d)$} output by $\calR^\bot_{\calD,\calD',\beta,\lambda}$. 
We call $\calR^\bot_{\calD,\calD',\beta,\lambda}$ the \textit{bot-count algorithm}. 
Our 
\FOUD{} and \FOLNF{} set $\kappa_i$ to a constant, while 
\FOLNFast{} generates $\kappa_i$ under DP. 
In Theorems~\ref{thm:FOUD_FODP}-\ref{thm:FOLNFast_FODP}, we prove FODP of these algorithms using Lemma~\ref{lem:general_privacy}. 

Our general algorithm $\calR_{\calD,\calD',\beta,\lambda}$ is also robust against collusion with users, as users do not add noise: 
\begin{lemma} \label{lem:general_robustness}
Let $\epsilon_E,\epsilon_I \in \nnreals$ and $\delta_E, \delta_I \in [0,1]$. 
If $\calR_{\calD,\calD',\beta,\lambda}$ (Algorithm~\ref{alg:proposal}) provides $(\epsilon_E,\delta_E)$-DP and $(\epsilon_I,\delta_I)$-FODP, then it is robust against collusion with users; i.e., the values of $(\epsilon_E,\delta_E)$ and $(\epsilon_I,\delta_I)$ are not increased by collusion with users 
(see Appendix~\ref{sub:robustness_collusion_formal} for the formal definition of the robustness). 
\end{lemma}

\colorB{Furthermore, in Appendix~\ref{sub:robustness_poisoning}, we show that our general algorithm is robust against poisoning attacks~\cite{Cao_USENIX21,Cheu_SP21}. 
Specifically, it prevents \textit{output poisoning attacks}~\cite{Li_USENIX23}, as users can change only their input values. 
See Appendix~\ref{sub:robustness_poisoning} for details.}

\smallskip{}
\noindent{\textbf{Accuracy.}}~~We also show the accuracy of the estimate $\bmhf$. 
Following \cite{Kairouz_ICML16,Wang_PVLDB20,Murakami_SP25}, we use the expected $l_2$ loss 
as a metric: 
\begin{theorem}\label{thm:general_accuracy}
The estimate $\bmhf$ in (\ref{eq:hf_i_general}) 
is unbiased (i.e., $\E[\hf_i] = f_i$ for any $i\in[d]$) and achieves the following expected $l_2$ loss: 
\begin{align*}
\textstyle{\E[\sum_{i=1}^d (\hf_i - f_i)^2] = \frac{1-\beta}{\beta n} + \frac{\lambda(d-1)}{\beta^2 n^2 d} + \frac{\sigma^2 d}{\beta^2 n^2},}
\end{align*}
where 
$\sigma^2$ 
is the variance of the dummy-count distribution $\calD$. 
\end{theorem}
\FOUD{} sets $\sigma^2 = 0$ and achieves $\E[\sum_{i=1}^d (\hf_i - f_i)^2] = \frac{1-\beta}{\beta n} + \frac{\lambda(d-1)}{\beta^2 n^2 d}$. 
\FOLNF{} and \FOLNFast{} set $\lambda = 0$ and achieve $\E[\sum_{i=1}^d (\hf_i - f_i)^2] = \frac{1-\beta}{\beta n} + \frac{\sigma^2 d}{\beta^2 n^2}$, which is the same as \cite{Murakami_SP25}.

\smallskip{}
\noindent{\textbf{Remark.}}~~Although our general algorithm $\calR_{\calD,\calD',\beta,\lambda}$ adds both uniform dummies (Algorithm~\ref{alg:proposal}, lines 7-10) and dummies for each item 
(lines 11-19), our three concrete algorithms add either of the two. 
This is because $\lambda$ uniform dummies over $[d]$ can be approximated by binomial dummies for each item (i.e., $z_i \sim Bin(\lambda,\frac{1}{d})$ dummies for each item $i$). 
We add uniform dummies in $\calR_{\calD,\calD',\beta,\lambda}$ because they do not need any $\bot$ and make \FOUD{} more efficient than \FOLNF{} with $\calD = Bin(\lambda,\frac{1}{d})$. 

\subsection{FOUD (Fully Oblivious UD)}
\label{sub:FOUD}
\noindent{\textbf{Algorithm.}}~~\FOUD{} 
is an instance of our general algorithm, where $\beta = 1$ and $z_i = \kappa_i = 0$ for any $i\in[d]$; i.e., $\calD$ is a degenerate distribution at $0$, and $\texttt{BotCountGeneration}$ (line 13 in Algorithm~\ref{alg:proposal}) always outputs $0$. 
In other words, \FOUD{} performs adding uniform dummies (lines 7-10) and shuffling (line 20). 
\FOUD{} outputs the same data as the UD protocol~\cite{Wang_PVLDB20} without LDP noise. 
We denote this algorithm by $\calR_\lambda^{\FOUD}$.

\smallskip{}
\noindent{\textbf{Privacy and Robustness.}}~~$\calR_\lambda^{\FOUD}$ 
provides the following privacy and robustness guarantees: 

\begin{theorem} \label{thm:FOUD_FODP}
For any $\theta_1 \in \nnreals$ and $\theta_2 \in [0,1)$, $\calR_\lambda^{\FOUD}$ provides 
$(\epsilon,\delta)$-DP and $(\epsilon,\delta)$-FODP, where 
$\epsilon = \ln \frac{d + (1+\theta_1)\lambda}{(1-\theta_2)\lambda}$ and $\delta = e^{- \theta_1^2 \lambda / ((2 + \theta_1)d)} + e^{- \theta_2^2 \lambda / (2d)}$. 
It is also robust against collusion with users. 
\end{theorem}
We prove FODP 
in Theorem~\ref{thm:FOUD_FODP} 
via full obliviousness (Theorem~\ref{thm:DP_FO_FODP}). 
In Appendix~\ref{sub:comparison_bounds_FOUD}, we also show that the bound on $(\epsilon,\delta)$ in Theorem~\ref{thm:FOUD_FODP} is tighter than the bound in \cite{Wang_PVLDB20}. 

\subsection{FOLNF (Fully Oblivious LNF)}
\label{sub:FOLNF_fixed}
\noindent{\textbf{Algorithm.}}~~\FOLNF{} is an instance of our general algorithm, 
where $\lambda = 0$ and $\texttt{BotCountGeneration}$ 
always outputs a predetermined value $\kappa \in \nats$, i.e., 
$\kappa_i = \kappa$ for any $i\in[d]$. 
After discarding $\bot$, the output of \FOLNF{} is the same as that of the LNF protocol~\cite{Murakami_SP25}. 
We denote this algorithm by $\calR_{\calD,\beta}^{\FOLNF}$. 

\smallskip{}
\noindent{\textbf{Privacy and Robustness.}}~~To show the privacy and robustness of $\calR_{\calD,\beta}^{\FOLNF}$, we introduce the binary input mechanism in~\cite{Murakami_SP25}: 
\begin{definition}[Binary input mechanism~\cite{Murakami_SP25}] \label{def:binary_input}
Let 
$Ber(\beta)$ be the Bernoulli distribution with parameter $\beta$. 
A \emph{binary input mechanism $\calM_{\calD,\beta}$} 
takes binary data $x\in\{0,1\}$ as input and outputs 
$\calM_{\calD,\beta}(x) = \alpha x + z$, 
where 
$\alpha \sim Ber(\beta)$ and $z \sim \calD$. 
\end{definition}
$\calM_{\calD,\beta}$ is a binary input version of the LNF protocol; i.e., it samples $x$ with probability $\beta$ and adds $z$ dummies. 
If $\calM_{\calD,\beta}$ provides $(\frac{\epsilon}{2},\frac{\delta}{2})$-DP, then the LNF protocol provides $(\epsilon,\delta)$-DP~\cite{Murakami_SP25}. 
Note that $(\epsilon,\delta)$ is doubled because neighboring databases $\bmx$ and $\bmx'$ differ by $1$ in two bins on their histograms. 

We show the privacy and robustness of $\calR_{\calD,\beta}^{\FOLNF}$ using the binary input mechanism: 
\begin{theorem} \label{thm:FOLNF_FODP}
Let $F: \nnints \rightarrow [0,1]$ be a CDF for the dummy-count distribution $\calD$. 
If the binary input mechanism $\calM_{\calD,\beta}$ provides $(\frac{\epsilon}{2},\frac{\delta}{2})$-DP, then $\calR_{\calD,\beta}^{\FOLNF}$ 
provides 
$(\epsilon,\delta + \delta_0)$-DP and $(\epsilon,\delta + \delta_0)$-FODP, where 
$\delta_0 = 2(1 - F(\kappa - 1))$, 
and is robust against collusion with users. 
\end{theorem}
We prove FODP in Theorem~\ref{thm:FOLNF_FODP} via full obliviousness. 
Note that \FOLNF{} introduces an additional term $\delta_0$ in DP, as it truncates the number $z_i$ of dummy values to $\kappa$ when $z_i > \kappa$. 
\colorB{The parameter $\kappa$ controls the trade-off between $\delta_0$ and the number of bots. 
In our experiments, we set $\delta = \delta_0$ and then set $\kappa$ so that $\delta + \delta_0$ $(=2\delta)$ does not exceed a required value.}

For $\calD$, we use the asymmetric geometric distribution: 

\begin{definition}[Asymmetric Geometric Distribution~\cite{Murakami_SP25}]\label{def:ageo}
Let $\nu \in \nnints$, $q_l \in [0,1]$, and $q_r \in [0,1]$. 
Let 
$\eta = q_l(1-q_l^\nu)/(1-q_l) + 1/(1-q_r)$. 
Let \AGeo$(\nu, q_l, q_r)$ be the \emph{asymmetric geometric distribution}, and $X$ be a random variable that follows \AGeo$(\nu, q_l, q_r)$. 
Then, the probability mass function at $X = k$ is given by
\begin{align*}
\Pr[X = k] = 
\begin{cases}
\frac{1}{\eta} q_l^{\nu - k}   &   \text{(if $k = 0, 1, \ldots, \nu-1$)}\\
\frac{1}{\eta} q_r^{k - \nu}   &   \text{(if $k = \nu, \nu+1, \ldots $)}.
\end{cases}
\end{align*}
\end{definition}
The binary input mechanism with $\calD = \AGeo(\nu, q_l, q_r)$, $q_l = \frac{e^{-\epsilon/2} - 1 + \beta}{\beta}$, and $q_r = \frac{\beta}{e^{\epsilon/2} - 1 + \beta}$ 
provides $(\frac{\epsilon}{2},\frac{\delta}{2})$-DP, where 
\begin{align}
\delta = 
\begin{cases}
\frac{2}{\eta} q_l^\nu (1 - e^{\epsilon/2} + \beta e^{\epsilon/2})    &   \text{(if $\beta > 1 - e^{-\epsilon/2}$)}\\
0   &   \text{(if $\beta = 1 - e^{-\epsilon/2}$)}
\end{cases}
\label{eq:AGeo_delta}
\end{align}
(see \cite{Murakami_SP25} for the proof). 
Thus, by Theorem~\ref{thm:FOLNF_FODP}, 
\FOLNF{} 
with this distribution provides 
$(\epsilon,\delta + \colorB{\delta_0})$-DP and $(\epsilon,\delta + \colorB{\delta_0})$-FODP. 

\subsection{FOLNF$^*$ (Fully Oblivious LNF with Differentially Private Bot Counts)}
\label{sub:FOLNF_DP}
\noindent{\textbf{Algorithm.}}~~\FOLNFast{} is a modification of \FOLNF{} that generates 
bot counts $(\kappa_1,\ldots,\kappa_d)$ under DP. 
Specifically, \FOLNFast{} is an instance of our general algorithm, where $\lambda = 0$ and $\texttt{BotCountGeneration}$ 
generates a sample $\omega_i$ from a bot-count distribution $\calD'$ over $\nnints$ and outputs 
$\kappa_i = z_i + \omega_i$ ($\omega_i \sim \calD'$). 
We denote this algorithm by $\calR_{\calD,\calD',\beta}^{\FOLNFast}$.

\smallskip{}
\noindent{\textbf{Privacy and Robustness.}}~~Privacy analysis of 
$\calR_{\calD,\calD',\beta}^{\FOLNFast}$ 
is non-trivial because $\kappa_i$ ($= z_i + \omega_i$) depends on the number $z_i$ of dummies for item $i$. 
In other words, bot counts $(\kappa_1,\ldots,\kappa_d)$ output by the bot-count algorithm 
depend on shuffled values output by $\calR_{\calD,\calD',\beta}^{\FOLNFast}$. 
To enable privacy analysis of $\calR_{\calD,\calD',\beta}^{\FOLNFast}$, 
we introduce 
\colorB{a mechanism called} 
the \textit{joint binary input mechanism:} 

\begin{definition}[Joint binary input mechanism] \label{def:joint_binary_input}
Let $\beta \in [0,1]$. 
The \emph{joint binary input mechanism $\calM^*_{\calD,\calD',\beta} = (\calM_{\calD,\beta}, \calM^-_{\calD',\beta})$} is a pair of two mechanisms 
$(\calM_{\calD,\beta},\calM^-_{\calD',\beta})$ that 
takes binary data $x\in\{0,1\}$ as input and outputs 
\begin{align}
(\calM_{\calD,\beta}(x),\calM^-_{\calD',\beta}(x)) = (\alpha x + z, -\alpha x + \omega),
\label{eq:M_M_minus}
\end{align}
where 
$\alpha \sim Ber(\beta)$, $z \sim \calD$, and $\omega \sim \calD'$. 
\end{definition}
Note that outputting $(\ref{eq:M_M_minus})$ is equivalent to outputting $(\calM_{\calD,\beta}(x), \calM_{\calD,\beta}(x) + \calM^-_{\calD',\beta}(x)) = (\alpha x + z, z + \omega)$ and that $z+\omega$ follows the same distribution as $\kappa_i$ ($= z_i + \omega_i$). 
Thus, this joint mechanism can be regarded as a binary input version of $\calR_{\calD,\calD',\beta}^{\FOLNFast}$ (\conference{\colorB{see~\cite{Murakami_arXiv26}}}\arxiv{see Appendix~\ref{sub:proof_thm_FONLFast_FODP}} for details). 
Based on this, we reduce FODP of $\calR_{\calD,\calD',\beta}^{\FOLNFast}$ to DP of this joint mechanism: 
\begin{theorem} \label{thm:FOLNFast_FODP}
Let $\calM^*_{\calD,\calD',\beta} = (\calM_{\calD,\beta}, \calM^-_{\calD',\beta})$ be a joint binary input mechanism. 
If $\calM_{\calD,\beta}$ provides $(\frac{\epsilon}{2},\frac{\delta}{2})$-DP and $\calM^*_{\calD,\calD',\beta}$ provides $(\frac{\epsilon^*}{2},\frac{\delta^*}{2})$-DP, then $\calR_{\calD,\calD',\beta}^{\FOLNFast}$ provides $(\epsilon,\delta)$-DP and $(\epsilon^*,\delta^*)$-FODP and is robust against collusion with users. 
\end{theorem}
We prove FODP in Theorem~\ref{thm:FOLNFast_FODP} by directly showing (\ref{eq:FODP_inequality}).

\smallskip{}
\noindent{\textbf{Joint Asymmetric Geometric Distribution.}}~~The remaining issue is how to design $(\calD,\calD')$ so that the joint binary input mechanism $\calM^*_{\calD,\calD',\beta}$ provides DP. 
To address this issue, we propose the 
\textit{joint asymmetric geometric distribution:}

\begin{definition}[Joint asymmetric geometric distribution] \label{def:joint_ageo}
Let $\epsilon \in \nnreals$, $\epsilon^* \in [\epsilon,\infty)$, $\beta \in [1 - e^{-\epsilon/2},1]$, $\nu, \nu' \in \nnints$, $L(\epsilon) = \frac{e^{-\epsilon/2} - 1 + \beta}{\beta}$, and $R(\epsilon) = \frac{\beta}{e^{\epsilon/2} - 1 + \beta}$. 
The \emph{joint asymmetric geometric distribution} is a pair of two asymmetric geometric distributions $($\emph{\AGeo}$(\nu, q_l, q_r)$, \emph{\AGeo}$(\nu', q'_l, q'_r))$, where 
$q_l = L(\epsilon)$, $q_r = R(\epsilon)$, $q'_l = \frac{1}{q_r}R(\epsilon^*)$, and $q'_r = \frac{1}{q_l}L(\epsilon^*)$ if $\beta > 1 - e^{-\epsilon/2}$ and $q'_r = 0$ otherwise. 
\end{definition}

\begin{figure}[t]
  \centering
  \includegraphics[width=0.99\linewidth]{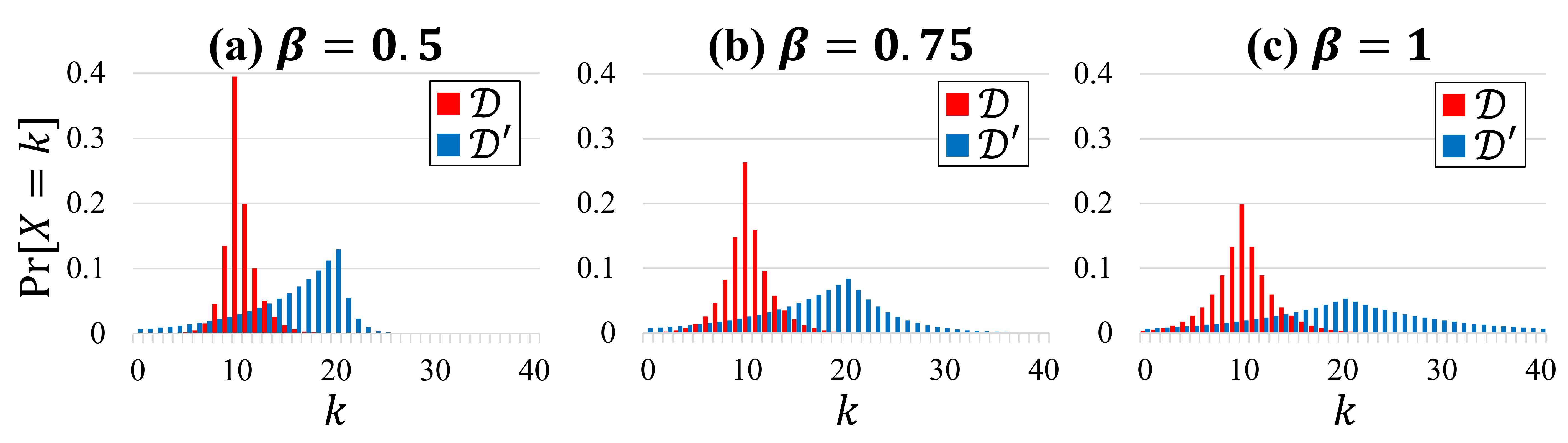}
  \caption{Examples of the joint asymmetric geometric distribution $(\calD,\calD') = (\AGeo(\nu, q_l, q_r), \AGeo(\nu', {q'}_l, {q'}_r))$ 
  ($\epsilon=0.8$, $\epsilon^*=1$, $\nu=10$, $\nu'=20$).} 
  \label{fig:joint_ageo}
\end{figure}
We use $\AGeo(\nu, q_l, q_r)$ as $\calD$ and $\AGeo(\nu', {q'}_l, {q'}_r)$ as $\calD'$. 
Figure~\ref{fig:joint_ageo} shows their examples. 
In this case, 
$\calD$ is identical to the asymmetric geometric distribution in \cite{Murakami_SP25}, whose \textit{left tail} shrinks as $\beta$ decreases. 
An interesting feature of 
$\calD'$ is that the \textit{right tail} shrinks as $\beta$ decreases. 
In other words, 
$\calD$ is \textit{positively skewed}, whereas $\calD'$ is \textit{negatively skewed}. 
This corresponds to the fact that (\ref{eq:M_M_minus}) adds samples from $\calD$ and $\calD'$ to a positive value $\alpha x$ and a negative value $- \alpha x$, respectively. 

We now prove that the joint binary input mechanism $\calM^*_{\calD,\calD',\beta}$ with this joint distribution provides DP: 
\begin{theorem} \label{thm:joint_ageo_DP}
Let $\calM^*_{\calD,\calD',\beta} = (\calM_{\calD,\beta}, \calM^-_{\calD',\beta})$ be a joint binary input mechanism, where $(\calD,\calD')$ is given by 
$(\AGeo(\nu, q_l, q_r), \allowbreak \AGeo(\nu', {q'}_l, {q'}_r))$ in Definition~\ref{def:joint_ageo}. 
Then, $\calM_{\calD,\beta}$ provides $(\frac{\epsilon}{2},\frac{\delta}{2})$-DP and $\calM^*_{\calD,\calD',\beta}$ 
provides $(\frac{\epsilon^*}{2},\frac{\delta^*}{2})$-DP,
where $\delta$ is given by (\ref{eq:AGeo_delta}), 
$\delta^* = \max\{\delta, 2 \beta {q'}_l^{\nu'} / \eta' \}$, 
and 
$\eta' = q'_l(1-{q'}_l^{\nu'})/(1-q'_l) + 1/(1-q'_r)$. 
\end{theorem}

By Theorems~\ref{thm:FOLNFast_FODP} and \ref{thm:joint_ageo_DP}, 
\FOLNFast{} 
with 
the joint asymmetric geometric distribution 
provides $(\epsilon,\delta)$-DP and $(\epsilon^*,\delta^*)$-FODP, where $\epsilon \leq \epsilon^*$ and $\delta \leq \delta^*$. 
We can control the privacy parameters $(\epsilon^*,\delta^*)$ in FODP by changing the parameters in $\calD'$. 
\colorB{The parameters $\nu$ and $\nu'$ are uniquely determined as minimum values such that $\delta^*$ does not exceed a required value.} 

\smallskip{}
\noindent{\textbf{\colorB{Comparison with~\cite{Ghazi_ICML20}}.}}~~\colorB{The asymmetric geometric distribution $\calD=\AGeo(\nu, q_l, q_r)$ becomes a \textit{symmetric} geometric distribution when $\beta=1$, which is near-optimal in the central model~\cite{Ghosh_SICOMP12}. 
Ghazi \textit{et al.}~\cite{Ghazi_ICML20} propose a multi-message shuffle protocol that achieves accuracy close to the symmetric geometric mechanism. 
However, the protocol in \cite{Ghazi_ICML20} requires many dummies generated from the negative binomial distribution to achieve almost the same accuracy as the geometric mechanism; the number of dummies even diverges to infinity. 
Thus, our \FOLNF{}/\FOLNFast{} with $\calD=\AGeo(\nu, q_l, q_r)$ significantly outperforms \cite{Ghazi_ICML20} in terms of accuracy and efficiency.}

\subsection{Comparisons of Our Three Algorithms}
\label{sub:comparison_three_algorithms}

\noindent{\textbf{\FOUD{} vs. \FOLNF{}.}}~~We first compare our \FOUD{} with \FOLNF{}. 
The advantage of \FOUD{} is that it uses only \texttt{ORAND\_NATS} and \texttt{OSHUFFLE} as primitives and is easy to implement. 
However, \FOLNF{} is more flexible in that it can use any distribution providing DP as $\calD$. 
In particular, \FOLNF{} with 
the asymmetric geometric distribution $\calD = \AGeo(\nu, q_l, q_r)$ provides higher accuracy and efficiency than \FOUD{}, as shown in Section~\ref{sec:experiments}. 

\smallskip{}
\noindent{\textbf{\FOLNF{} vs. \FOLNFast{}.}}~~Next, we compare our \FOLNF{} with \FOLNFast{}. 
\FOLNF{} achieves DP and FODP with $(\epsilon_E,\delta_E) = (\epsilon_I,\delta_I) = (\epsilon,\delta+\delta_0)$. 
In contrast, \FOLNFast{} with the same dummy-count distribution $\calD$ provides DP and FODP with $(\epsilon_E,\delta_E) = (\epsilon,\delta)$, $(\epsilon_I,\delta_I) = (\epsilon^*,\delta^*)$, $\epsilon \leq \epsilon^*$, and $\delta \leq \delta^*$. 
Thus, 
these algorithms differ in the following two aspects. 

First, 
\FOLNFast{} achieves FODP with a larger privacy budget $\epsilon_I$ than \FOLNF{}. 
The resulting benefit is \textit{higher efficiency}. 
Specifically, 
\FOLNFast{} can reduce the number of bots $\bot$ (thereby improving efficiency) by increasing $\epsilon_I$ in FODP.

Second, 
\FOLNF{} introduces an additional term $\delta_0$ in DP, whereas \FOLNFast{} does not introduce $\delta_0$. 
In particular, 
when $\beta = 1 - e^{-\epsilon_E/2}$, \FOLNFast{} provides \textit{pure} DP (i.e., $\delta_E = 0$). 

\section{Efficiency of Our FODP Algorithms}
\label{sec:FODP_algorithm_large}

In this section, we analyze the runtime of our algorithms (Section~\ref{sub:comparison_direct_histogram}) and improve it using the count-min (Section~\ref{sub:count_min}). 

\subsection{Efficiency Analysis}
\label{sub:comparison_direct_histogram}
We first show the runtime of our general algorithm $\calR_{\calD,\calD',\beta,\lambda}$: 
\begin{theorem} \label{thm:Proposal_runtime}
The runtime of random sampling and dummy data addition in $\calR_{\calD,\calD',\beta,\lambda}$ (Algorithm~\ref{alg:proposal}) is $O(n \log d)$ and $O((\lambda + \sum_{i=1}^d \kappa_i) \log d)$, respectively. 
The total runtime of $\calR_{\calD,\calD',\beta,\lambda}$ is $O(\bn \log d + \rho)$, where $\bn = n + \lambda + \sum_{i=1}^d \kappa_i$ is the number of shuffled values and $\rho$ is the runtime of $\texttt{OSHUFFLE}$. 
\end{theorem}
If we use ORShuffle~\cite{Sasy_CCS22} (resp.~WaksShuffle~\cite{Sasy_CCS23}) as $\texttt{OSHUFFLE}$, then 
$\rho = O((\bn \log^2 \bn) \log d)$ 
(resp.~$O(\bn \log^3 \bn + \bn \log \bn \log d)$). 
Since $\lambda=O(d)$, 
the runtime of $\calR_{\calD,\calD',\beta,\lambda}$ can be expressed as $\tO(n + d)$ by ignoring the logarithmic factor.

\colorB{It is also possible to run a central DP algorithm~\cite{Dwork_TCC06,Bun_JMLR19,Lebeda_FORC25} within the TEE. 
However,} 
it requires prohibitively large runtime $\tO(nd)$, as it must access the entire memory to provide obliviousness when adding each input value to the histogram~\cite{Allen_NeurIPS19}. 
\colorB{See Appendix~\ref{sub:direct_histogram} for details.} 

\subsection{Improving the Efficiency}
\label{sub:count_min}

\noindent{\textbf{Overview.}}~~Our algorithms achieve a runtime of $\tO(n+d)$. 
Note that we can reduce $n$ by sampling users in advance. 
However, we cannot reduce $d$ by sampling items, as we need to estimate the frequency for each item. 
Thus, our algorithms are still inefficient for large-domain data with large $d$. 

To address this issue, 
we improve the efficiency of our general algorithm 
using the count-min sketch \cite{Cormode_JA05}. 
The count-min sketch compresses each input or dummy value using $\tau \in \nats$ hash functions. 
Then, it estimates the frequency of each item by taking the minimum of $\tau$ counts 
to mitigate the impact of hash collisions. 
We show a bound on the accuracy much tighter than \cite{Cheu_SP22,Ghazi_EUROCRYPT21} and optimize $\tau$ based on our bound. 

\smallskip{}
\noindent{\textbf{Algorithm.}}~~Algorithm~\ref{alg:proposal_large} shows our general framework for large-domain data, denoted by $\calR_{\calD,\calD',\beta,\lambda}^{\Largesf}$. 
This algorithm uses $\tau$ hash functions $h_1,\ldots,h_\tau: [d] \rightarrow [b]$, where $b < d$. 

For each $t \in [\tau]$, we transform input values $(x_1, \ldots, x_n)$ into their hash values $\bmx_t = (h_t(x_1), \ldots, h_t(x_n))$ 
(line 2). 
Then, we run our general algorithm $\calR_{\calD,\calD',\beta,\lambda}$ (Algorithm~\ref{alg:proposal}) with inputs $\bmx_t$ and $b$ to obtain shuffled values $\bmtx_t$ (line 3). 
The output of $\calR_{\calD,\calD',\beta,\lambda}^{\Largesf}$ is the tuple $(\bmtx_1,\ldots,\bmtx_\tau)$ of shuffled values. 

\setlength{\algomargin}{5mm}
\begin{algorithm}[t]
  \SetAlgoLined
  \KwData{Database $\bmx = (x_1, \ldots, x_n)$, \#items $d \in \nats$, hash functions $h_1,\ldots,h_\tau: [d] \rightarrow [b]$. 
  }
  \KwResult{Tuple $(\bmtx_1,\ldots,\bmtx_\tau)$ of shuffled values.}
  \ForEach{$t \in [\tau]$}{
    $\bmx_t \leftarrow (h_t(x_1), \ldots, h_t(x_n))$\;
    $\bmtx_t \leftarrow \calR_{\calD,\calD',\beta,\lambda}(\bmx_t, b)$\;
  }
  \KwRet{$(\bmtx_1,\ldots,\bmtx_\tau)$}
  \caption{Our general algorithm $\calR_{\calD,\calD',\beta,\lambda}^{\Largesf}$ for large-domain data. 
  }
  \label{alg:proposal_large}
\end{algorithm}

\smallskip{}
\noindent{\textbf{Calculation of $\bmhf$.}}~~After $\calR_{\calD,\calD',\beta,\lambda}^{\Largesf}$ outputs $(\bmtx_1,\ldots,\bmtx_\tau)$, the server (outside the TEE) discards $\bot$ from the shuffled data. 
Then, for each $i\in[d]$ and each $t\in[\tau]$, the server counts the number $c_{t,h_t(i)} \in \nnints$ of hash value $h_t(i)$ in $\bmtx_t$. 
Based on the count $c_{t,h_t(i)}$, the server calculates an estimate $\bmhf$ as follows: 
\begin{align}
\textstyle{\hf_i = \frac{1}{n\beta}(\min_{t\in[\tau]} c_{t,h_t(i)} - \frac{\lambda}{b} - \mu)  ~~ (i\in[d]).}
\label{eq:hf_i_count_min}
\end{align}
where $\mu$ is the mean of the dummy-count distribution $\calD$. 

\smallskip{}
\noindent{\textbf{Theoretical Properties.}}~~It is easy to show that $\calR_{\calD,\calD',\beta,\lambda}^{\Largesf}$ provides DP, FODP, and robustness based on the composition theorem~\cite{Kairouz_ICML15}; see 
Appendix~\ref{sub:comparison_bounds_Large}. 
The runtime of $\calR_{\calD,\calD',\beta,\lambda}^{\Largesf}$ is 
$\tO(n+b)$. 
For example, when $n \ll d$, we can reduce the runtime from $\tO(n+d)$ to $\tO(n)$ by setting $b = O(n)$. 

Below, we analyze the accuracy of $\calR_{\calD,\calD',\beta,\lambda}^{\Largesf}$. 
Following~\cite{Ghazi_EUROCRYPT21}, we use 
additive error as a metric: 

\begin{theorem} \label{thm:Proposal_large_accuracy}
Let $\gamma \in (0,\infty)$, $\xi \sim Bin(nf_i, \beta) + Bin(\lambda, \frac{1}{b}) + \calD$, and $\xi_{avg} = nf_i\beta+\frac{\lambda}{b}+\mu$. 
Then, for each item $i \in [d]$, $\calR_{\calD,\calD',\beta,\lambda}^{\Largesf}$ provides the following accuracy guarantee: 
\begin{align}
\textstyle{\Pr[|\hf_i - f_i| \leq \gamma]} 
&\geq \textstyle{1 - \left(\frac{2}{b \gamma} + \Pr[\xi > \xi_{avg} + \frac{n\beta\gamma}{2}]\right)^\tau} \nonumber\\
&\hspace{4mm}\textstyle{- \tau \cdot \Pr[\xi < \xi_{avg} - n\beta\gamma]}. 
\label{eq:Proposal_large_accuracy}
\end{align}
\end{theorem}
Theorem~\ref{thm:Proposal_large_accuracy} decomposes the increase in the additive error $|\hf_i - f_i|$ into two factors: \textit{hash collision} ($\frac{2}{b \gamma}$ in (\ref{eq:Proposal_large_accuracy})) and \textit{DP noise} ($\Pr[\xi > \xi_{avg} + \frac{n\beta\gamma}{2}]$ and $\Pr[\xi < \xi_{avg} - n\beta\gamma]$ in (\ref{eq:Proposal_large_accuracy})). 
The values of $\Pr[\xi > \xi_{avg} + \frac{n\beta\gamma}{2}]$ and $\Pr[\xi < \xi_{avg} - n\beta\gamma]$ depend on the concrete algorithm of $\calR_{\calD,\calD',\beta,\lambda}$. 
In 
Appendix~\ref{sub:comparison_bounds_Large}, 
we analyze these values in \FOUD{}, \FOLNF{}, and \FOLNFast{}. 

\begin{figure}[t]
  \centering
  \includegraphics[width=0.99\linewidth]{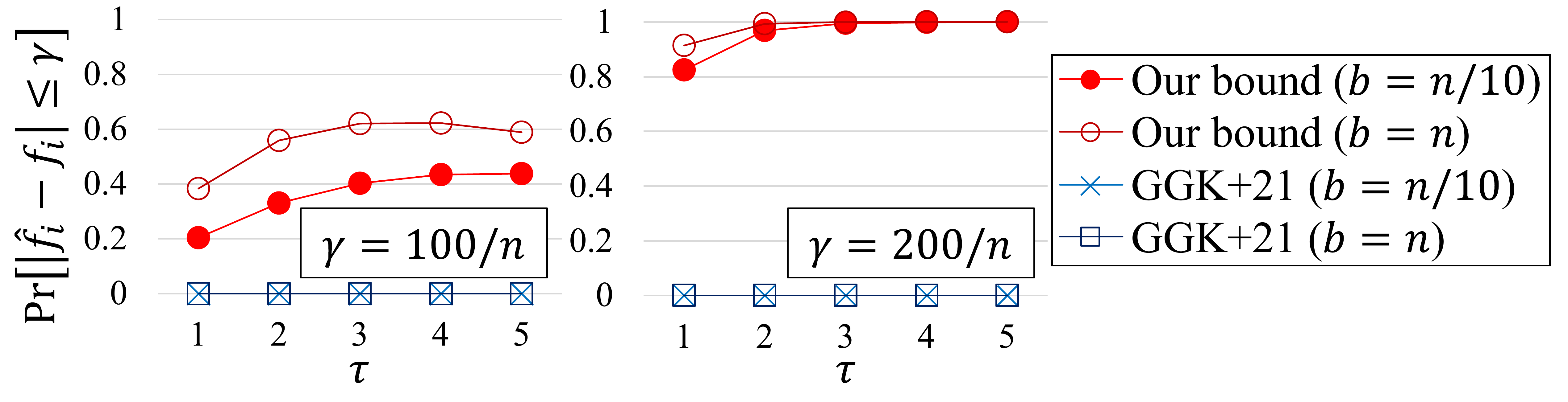}
  \caption{Our bound (Theorem~\ref{thm:Proposal_large_accuracy}) and  the bound in \cite{Ghazi_EUROCRYPT21} (\GGK{}) for \FOLNF{} with $\calD=Bin(n,\varphi)$ ($\varphi \in [0,1]$) and $\beta=1$. 
  We set $(n,\varphi)=(10^4,0.26)$, in which case $(\epsilon_I,\delta_I)=(\epsilon_E,\delta_E)=(1,10^{-12})$ for a single hash function ($\tau=1$).} 
  \label{fig:efficient_bound}
\end{figure}

Figure~\ref{fig:efficient_bound} shows our bound in Theorem~\ref{thm:Proposal_large_accuracy} and the bound in \cite{Ghazi_EUROCRYPT21} (denoted by \GGK{}) when applied to \FOLNF{} with the binomial distribution $\calD=Bin(n,\varphi)$ ($\varphi \in [0,1]$) and $\beta=1$. 
We do not show the bound in \cite{Cheu_SP22}, as it yields the same results as \GGK{}; see Appendix~\ref{sub:comparison_bounds_Large} for details of these bounds. 

We observe that \GGK{} cannot lower bound 
$\Pr[|\hf_i - f_i| \leq \gamma]$. 
This is because \GGK{} calculates the lower bound based on the probability that no hash collision occurs among $n$ input values, which is almost zero when $b \leq n$. 
In contrast, our bound gives a non-trivial lower bound on $\Pr[|\hf_i - f_i| \leq \gamma]$, as it directly evaluates the increase in the additive error caused by the hash collision. 
For example, our bound shows that when $\tau=2$ and $b=n$, the additive error is smaller than $\frac{100}{n}$ (resp.~$\frac{200}{n}$) with probability at least $0.56$ (resp.~$0.99$). 

\smallskip{}
\noindent{\textbf{Optimization of $\tau$.}}~~The probability $\Pr[|\hf_i - f_i| \leq \gamma]$ increases with an increase in $\tau$ because the impact of hash collisions can be mitigated by taking the minimum of $\tau$ counts. 
However, the increase in $\tau$ results in larger privacy parameters $(\epsilon_I,\delta_I)$ and $(\epsilon_E,\delta_E)$ due to the composition. 
Therefore, we 
calculate an optimal value of $\tau$ that maximizes the accuracy at the same privacy parameters based on our bound in Theorem~\ref{thm:Proposal_large_accuracy}. 

Specifically, given $\tau$ and the required values for $(\epsilon_I,\delta_I)$ and $(\epsilon_E,\delta_E)$, we calculate the privacy parameters for each hash function based on the composition theorem~\cite{Kairouz_ICML15}. 
Then, we calculate the value of $\gamma$ such that the right side of (\ref{eq:Proposal_large_accuracy}) is larger than a predetermined threshold $p_{thr} \in [0,1]$. 
We perform this process for various values of $\tau$ and select $\tau$ that minimizes $\gamma$. 
We show that this approach works well in our experiments. 

\section{Experimental Evaluation}
\label{sec:experiments}

\subsection{Experimental Set-up}
\label{sub:setup}

In this section, we evaluate the accuracy and runtime of our algorithms. 
Following \cite{Wang_PVLDB20,Luo_CCS22,Murakami_SP25,Murakami_NDSS26}, we use the following datasets with various $n$, $d$, and data types: 

\smallskip{}
\noindent{\textbf{\IPUMS{}~\cite{ruggles2023ipums}.}}~~U.S. Census data with $n=602156$ users and $d=915$ cities (we sampled $1\%$ of users, as in \cite{Wang_PVLDB20}). 

\smallskip{}
\noindent{\textbf{\Localization{}~\cite{kaluvza2010agent}.}}~~Person activity data with $n=164860$ instances and $d=11$ activity types (e.g., walking, lying). 

\smallskip{}
\noindent{\textbf{\Foursquare{}~\cite{Yang_TIST16}.}}~~Location data containing $n=18201$ check-ins in New York 
divided into $d=10^6$ ($=10^3 \times 10^3$) regions. 

\smallskip{}
\noindent{\textbf{\AOL{}~\cite{Pass_InfoScale06}.}}~~Website access data. 
We used $n=10000$ accesses and $d=16777216$ ($=2^{24}$) domain names comprising the first three characters of the URL, as in \cite{Luo_CCS22}. 

\smallskip{}
\noindent{\IPUMS{}} and \Localization{} are small-domain datasets, whereas \Foursquare{} and \AOL{} are large-domain datasets. 

\smallskip{}
\noindent{\textbf{Algorithms.}}~~We first evaluated our three algorithms (\FOUD{}, \FOLNF{}, and \FOLNFast{}) using two small-domain datasets. 
As $(\calD,\calD')$, we used the joint asymmetric geometric distribution with $\beta = 1$ (denoted by \AGeo{}) or $1 - e^{\epsilon_E/2}$ (denoted by \OGeo{}, as $\calD$ and $\calD'$ are one-sided geometric distributions in this case). 
Note that we used the same distribution $\calD$ for \FOLNF{} and \FOLNFast{}. 
In \FOLNF{}, we set $\delta = \delta_0$ ($= \frac{\delta_E}{2} = \frac{\delta_I}{2}$). 
We compared the accuracy 
of 
our algorithms 
with that of six existing pure shuffle algorithms. 
Specifically, as competitors, we evaluated three shuffle algorithms based on the GRR~\cite{Kairouz_ICML16,Wang_PVLDB20}, OUE~\cite{Wang_USENIX17}, and OLH~\cite{Wang_USENIX17} 
and three multi-message shuffle algorithms in \cite{Balcer_ITC20,Cheu_SP22,Luo_CCS22} (denoted by \GRR{}, \OUE{}, \OLH{}, \BC{}, \CM{}, and \LWY{}, respectively). 

Then, 
we evaluated our three algorithms with the count-min sketch (denoted by \FOUDLarge{}, \FOLNFLarge{}, and \FOLNFastLarge{}) using 
two large-domain datasets. 
We set $p_{thr} = 0.5$ and 
compared the accuracy 
of our algorithms with 
that 
of two existing algorithms: the FME algorithm~\cite{Murakami_NDSS26} (denoted by \FME{}) and \LWY{} for a large domain~\cite{Luo_CCS22} (denoted by \LWYL{}). 
\colorB{For \FME{}, we evaluated an accurate version (``large $l$'' in \cite{Murakami_NDSS26}).  
For the other existing algorithms, we used the same parameters and privacy amplification bounds as~\cite{Murakami_SP25} (see~\cite{Murakami_SP25}).} 

Note that none of the existing algorithms provides FODP. 
Thus, we compared our algorithms with the existing ones at the same 
value of $\epsilon_E$ in DP. 
We evaluated $\epsilon_I$ in FODP only for our algorithms. 
For the delta values, we set $\delta_E, \delta_I \leq 10^{-12}$. 

\smallskip{}
\noindent{\textbf{Accuracy Metrics.}}~~We used the MSE (Mean Squared Error) $\frac{1}{d} \sum_{i=1}^d (\hf_i - f_i)^2$ 
as a metric for 
small-domain datasets. 
For large-domain datasets, 
we used the MSE over the top-50 items with the highest frequencies, as most items have zero frequencies. 
We evaluated the average MSE over $100$ runs. 

\smallskip{}
\noindent{\textbf{Runtime on Intel SGX.}}~~We also measured the runtime of our algorithms 
on a Microsoft Azure Virtual Machine (Standard DC2s v3) \cite{Azure}, a public cloud platform that supports Intel SGX. 
It includes 2 vCPUs and 16 GiB of memory and runs on Ubuntu 20.04.6 LTS (Linux kernel 5.15.0-1089-azure). 
Since the codes of ORShuffle and WaksShuffle in \cite{Sasy_code} could not handle over ten million data points due to stack overflow, we implemented ORShuffle 
without a stack. 
\conference{\colorB{In~\cite{Murakami_arXiv26}}}\arxiv{In Appendix~\ref{sub:runtime_Sasy_code}}, we also 
evaluate the runtime of our algorithms using the codes of ORShuffle and WaksShuffle in \cite{Sasy_code}. 

Note that the runtime depends on $n$ and $d$, but not on input values $(x_1,\ldots,x_n)$. 
Thus, we randomly generated each input value $x_i$ ($i\in[n]$) from $[d]$ when measuring the runtime. 
We also confirmed that the runtime is almost the same when we used real datasets with the same $n$ and $d$. 

\begin{figure}[t]
  \centering
  \includegraphics[width=0.99\linewidth]{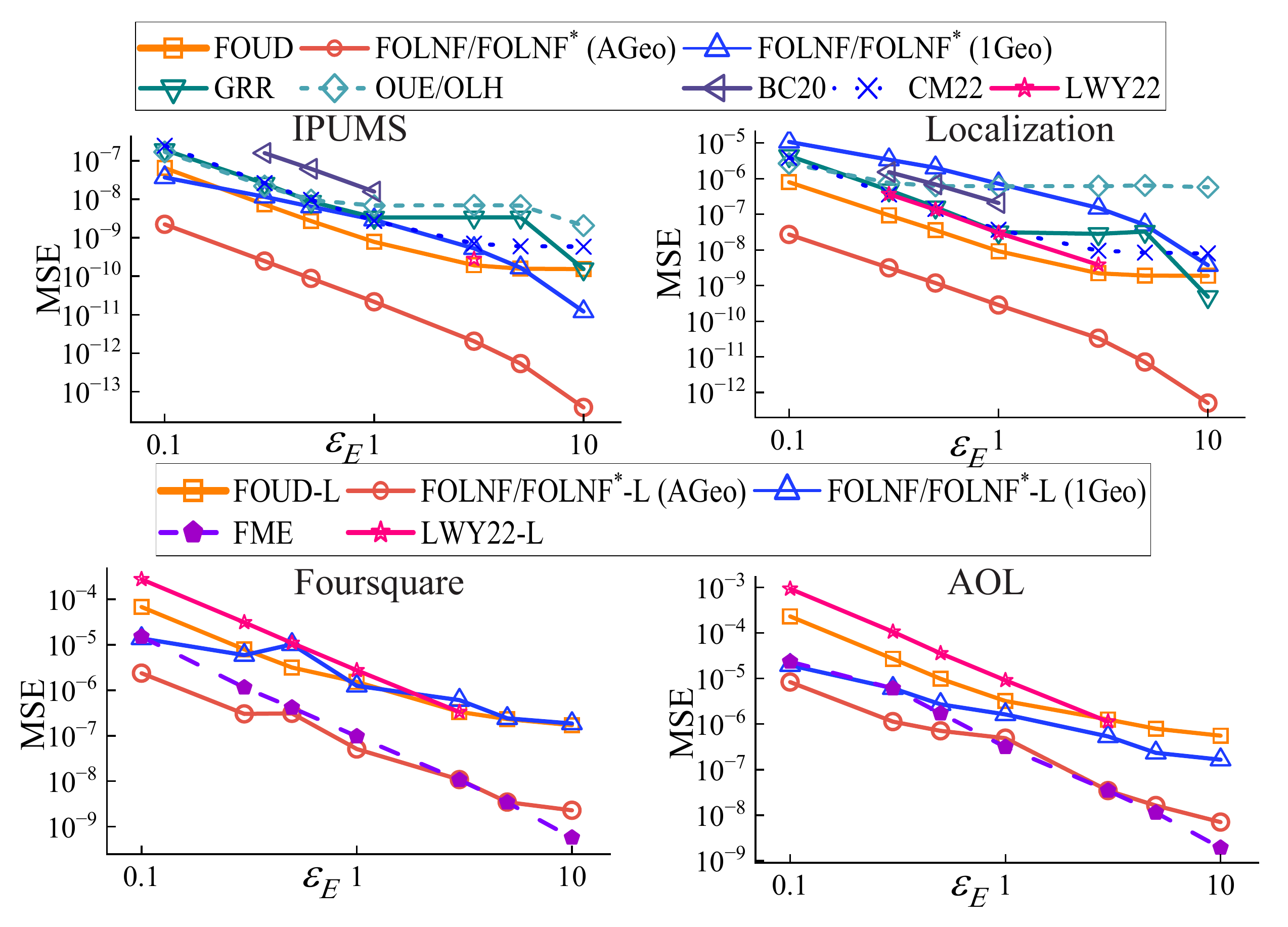}
  \caption{MSE vs. $\epsilon_E$ in DP ($\tau$: optimized, $b=n$).} 
  \label{fig:res_MSE}
\end{figure}
\begin{figure}[t]
  \centering
  \includegraphics[width=0.99\linewidth]{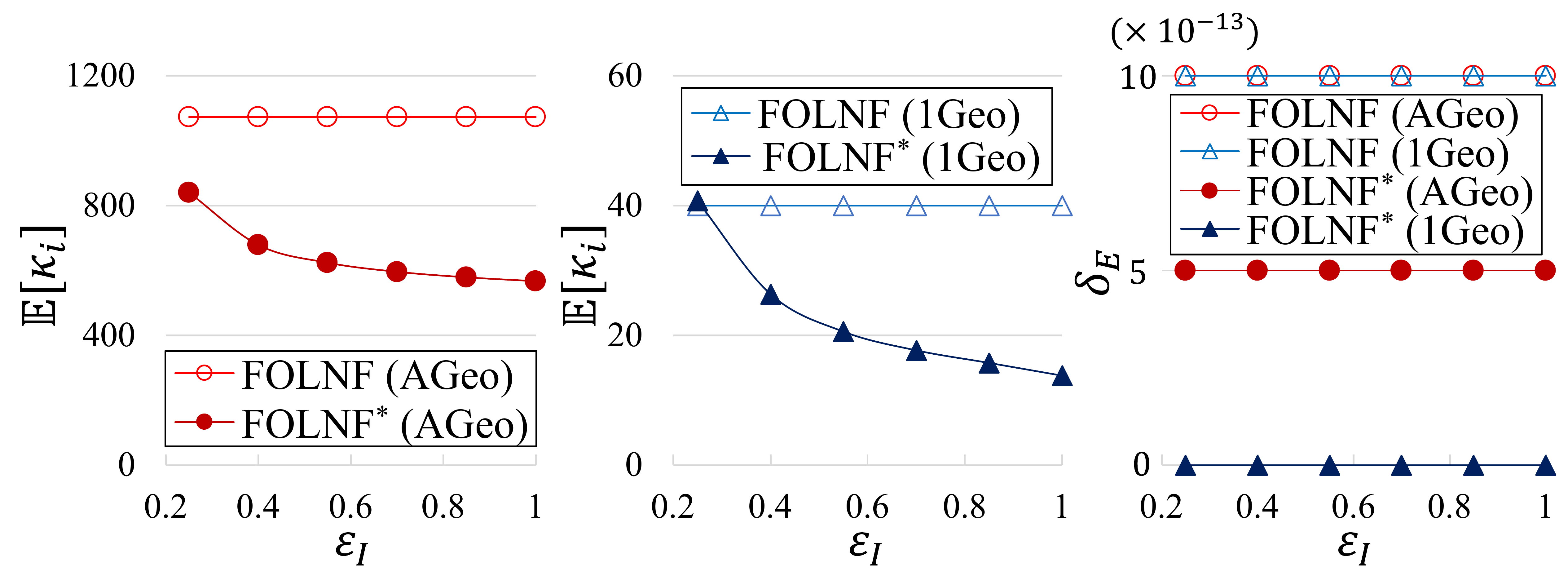}
  \caption{Relationship between the expected number $\E[\kappa_i]$ of bots, $\delta_E$ in DP, and $\epsilon_I$ in FODP ($\epsilon_E = 0.1$, $\delta_I = 10^{-12}$).} 
  \label{fig:res_FOLNF}
\end{figure}

\subsection{Experimental Results}
\label{sub:results}
\noindent{\textbf{Accuracy.}}~~Figure~\ref{fig:res_MSE} shows the MSE over a wide range of $\epsilon_E$ from $0.1$ to $10$. 
In \FOUDLarge{}, \FOLNFLarge{}, and \FOLNFastLarge{}, we set $b=n$ and optimized $\tau$ using our 
method. 
\FOLNF{} and \FOLNFast{} provide the same accuracy, as they use the same $\calD$. 

Figure~\ref{fig:res_MSE} shows that 
our \FOLNF{}/\FOLNFast{} (\AGeo{}) and \FOUD{} achieve the best and second-best accuracy, respectively, which indicates that accuracy can be improved by introducing the augmented shuffle model. 
Although \OGeo{} is less accurate than \AGeo{}, it achieves pure DP and higher efficiency, as shown later. 
For large-domain data, 
our \FOLNFLarge{} and \FOLNFastLarge{} provide the best (or almost the best) accuracy when $\epsilon \leq 5$. 
\conference{\colorB{In~\cite{Murakami_arXiv26}}}\arxiv{In Appendix~\ref{sub:communication_efficiency}}, we also show that \FOLNFLarge{} and \FOLNFastLarge{} are much more communication-efficient than \FME{}.

\smallskip{}
\noindent{\textbf{\FOLNF{} vs. \FOLNFast{}.}}~~Next, we compared \FOLNF{} with \FOLNFast{}. 
Figure~\ref{fig:res_FOLNF} shows the expected number $\E[\kappa_i]$ of bots for each item $i\in[d]$ and the value of $\delta_E$ in DP. 
Here, we set $\epsilon_E = 0.1$\colorB{, which} 
is the most stringent in terms of efficiency; i.e., both $\E[\kappa_i]$ and the runtime are smaller when $\epsilon_E > 0.1$. 

Figure~\ref{fig:res_FOLNF} shows that 
$\E[\kappa_i]$ in \FOLNFast{} decreases with an increase in $\epsilon_I$ in FODP, as  
the variance of the bot-count distribution $\calD'$ decreases in this case. 
For example, when $\epsilon_I = 1$, \FOLNFast{} with \AGeo{} (resp.~\OGeo{}) reduces $\E[\kappa_i]$ of \FOLNF{} by about half (resp.~one-third). 

Figure~\ref{fig:res_FOLNF} also shows that \FOLNFast{} reduces $\delta_E$ in DP. 
Specifically, \FOLNFast{} (\AGeo{}) reduces $\delta_E$ of \FOLNF{} by half, and \FOLNFast{} (\OGeo{}) achieves pure DP ($\delta_E = 0$). 

\begin{figure}[t]
  \centering
  \includegraphics[width=0.99\linewidth]{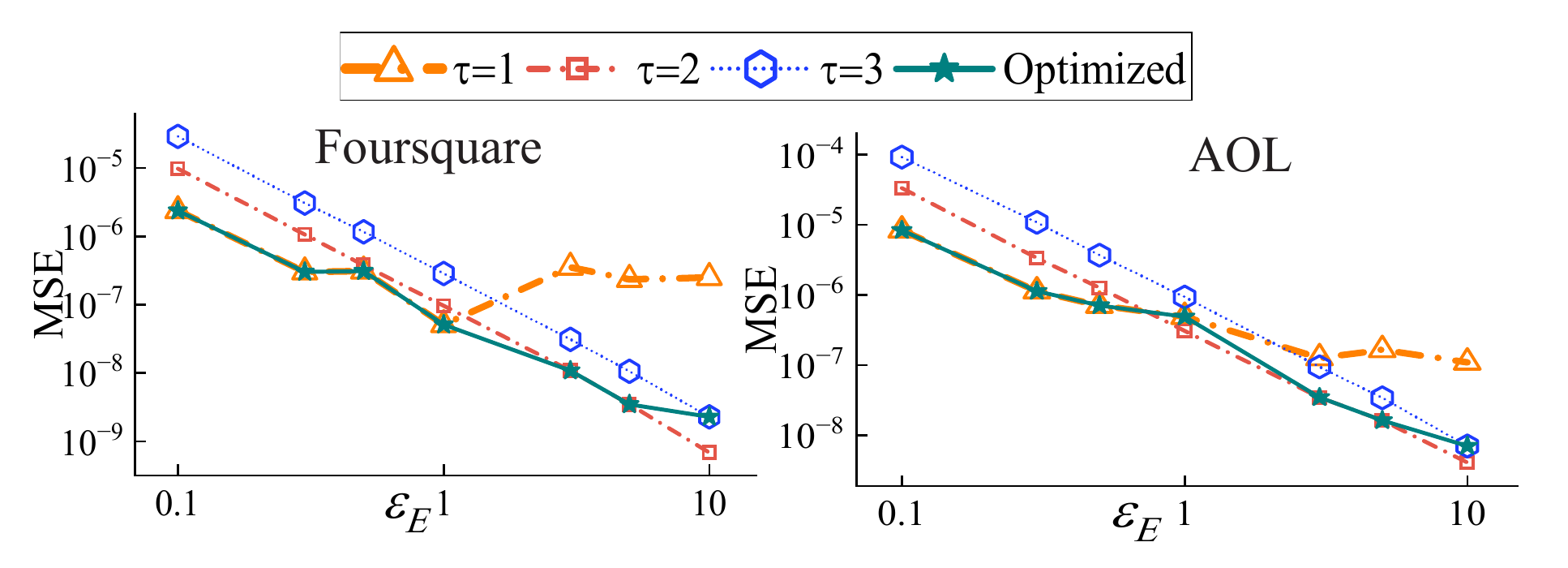}
  \caption{MSE for various values of $\tau$ 
  ($b=n$, \AGeo{}).} 
  \label{fig:res_optimization}
\end{figure}
\begin{figure}[t]
  \centering
  \includegraphics[width=0.99\linewidth]{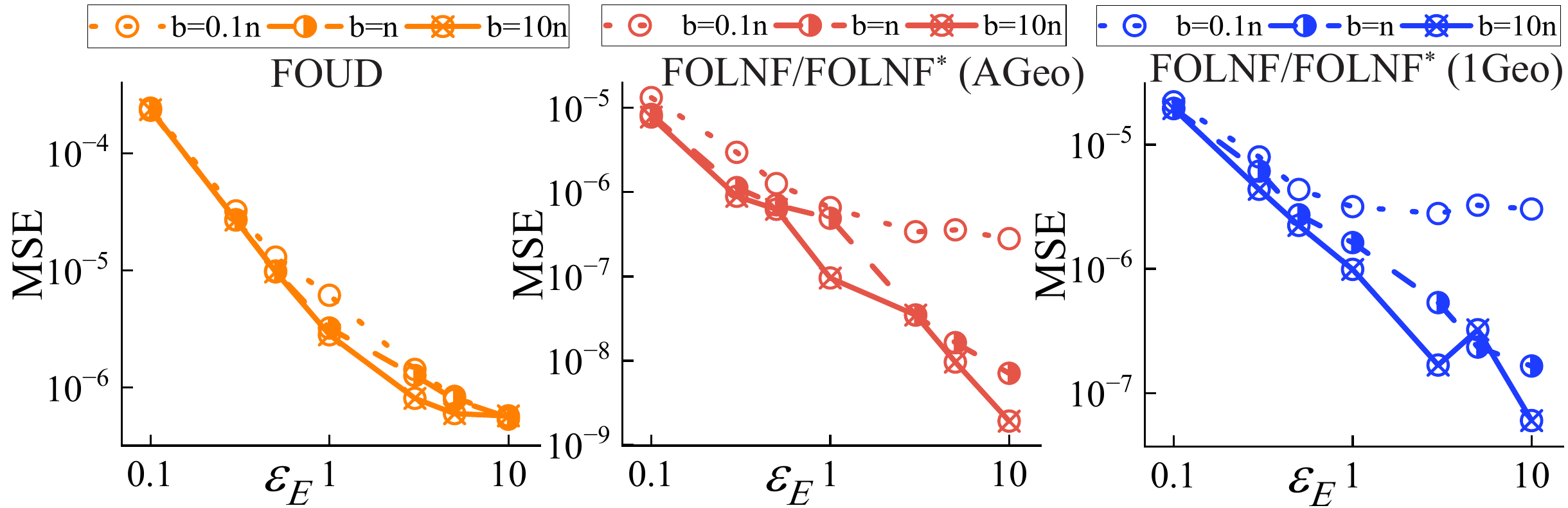}
  \caption{MSE for various values of $b$ ($\tau$: optimized, \AOL{}).} 
  \label{fig:res_hash_range}
\end{figure}

\smallskip{}
\noindent{\textbf{Changing $\tau$ and $b$.}}~~We also evaluated our algorithms for large-domain data while changing 
the number $\tau$ of hashes and the hash range $b$. 
Figures~\ref{fig:res_optimization} and \ref{fig:res_hash_range} show the results. 
\conference{\colorB{In~\cite{Murakami_arXiv26}}}\arxiv{In Appendix~\ref{sub:tau_b}}, we also show that similar results are obtained when varying $\tau$ in \OGeo{} and varying $b$ in \Foursquare{}. 

Figure~\ref{fig:res_optimization} shows that setting $\tau=1$ results in the smallest MSE for small $\epsilon_E$, as it avoids an increase in $\epsilon_E$ due to composition.  
However, setting $\tau \geq 2$ results in a smaller MSE for large $\epsilon_E$, as it mitigates the impact of hash collisions. 
Our optimization method successfully takes the best of both worlds and finds the optimal $\tau$ in most cases. 

Figure~\ref{fig:res_hash_range} shows that the MSE is reduced by increasing $b$. 
However, it comes at the cost of efficiency, as the runtime of our algorithms is $\tO(n+b)$. 
Figure~\ref{fig:res_hash_range} also shows that the accuracy is close between $b=n$ and $b=10n$. 

\begin{figure}[t]
  \centering
  \includegraphics[width=0.99\linewidth]{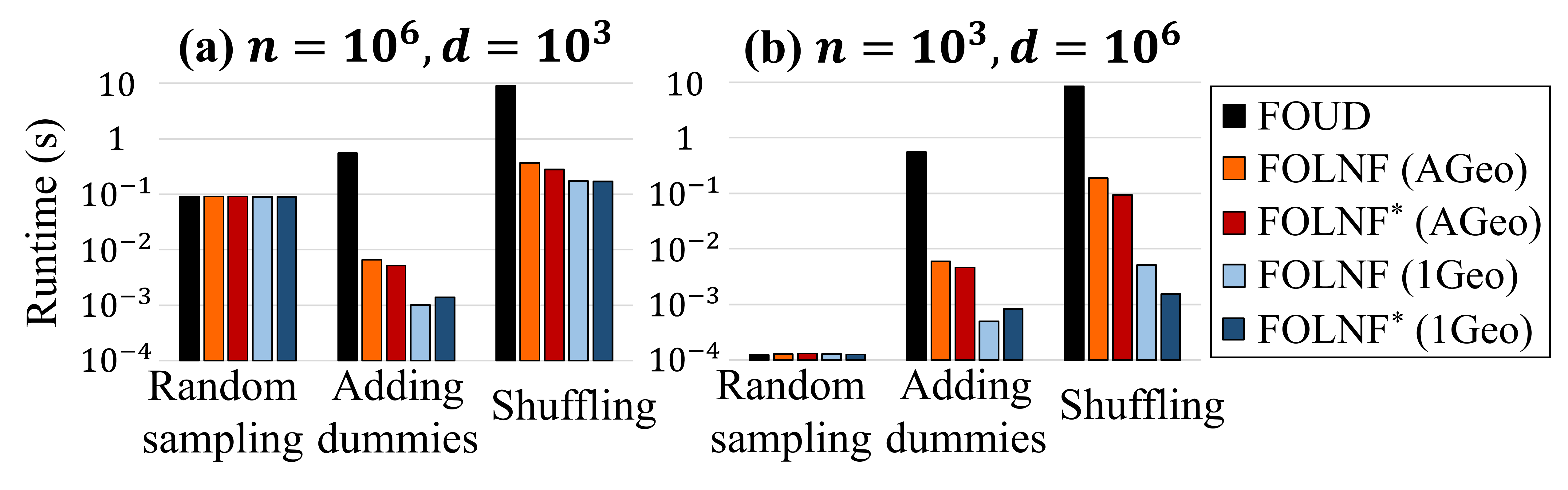}
  \caption{Time for random sampling, dummy data addition, and shuffling in our algorithms ($\epsilon_E = 0.1$, $\epsilon_I = 1$). 
  In (b), we used the count-min sketch ($b=n$).} 
  \label{fig:res_each_runtime}
\end{figure}

\begin{figure}[t]
  \centering
  \includegraphics[width=0.99\linewidth]{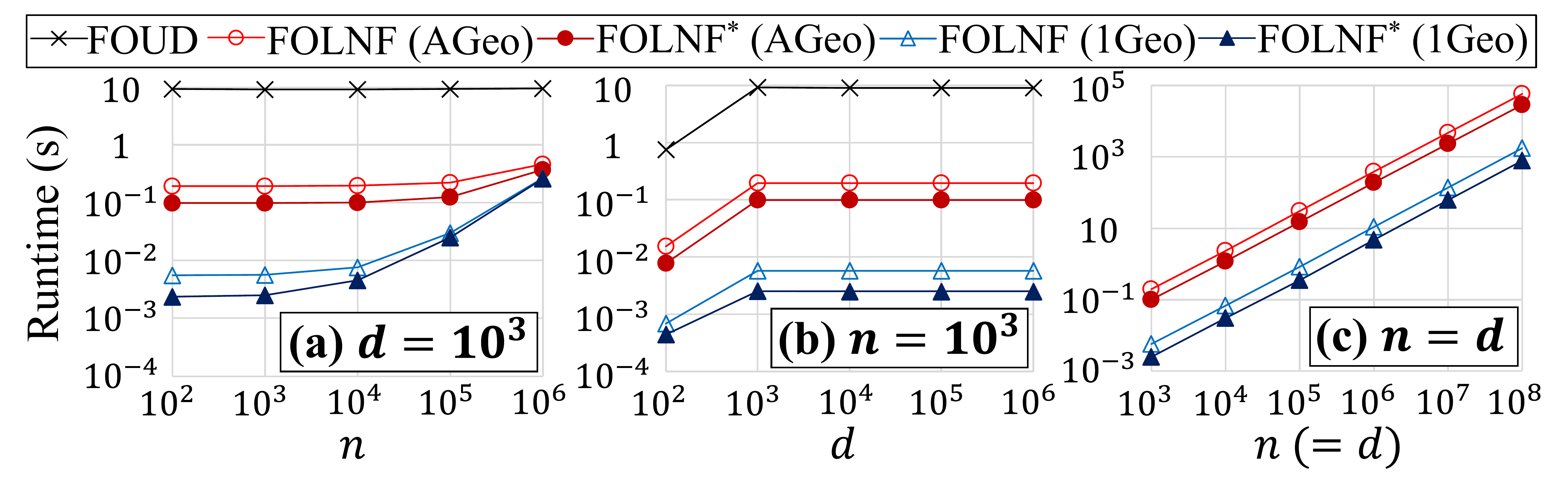}
  \caption{Total runtime of our algorithms ($\epsilon_E = 0.1$, $\epsilon_I = 1$). 
  In (b), we used the count-min sketch ($b=n$) when $n < d$.} 
  \label{fig:res_total_runtime}
\end{figure}

\smallskip{}
\noindent{\textbf{Runtime.}}~~\colorB{Then}, 
we measured the runtime of our algorithms. 
Figure~\ref{fig:res_each_runtime} shows the time for random sampling, dummy data addition, and shuffling when $(n,d)=(10^6,10^3)$ or $(10^3,10^6)$. 
Figure~\ref{fig:res_total_runtime} shows the total runtime when $n$ and $d$ are varied to various values. 
In Figure~\ref{fig:res_total_runtime}(c), we 
confirmed that the runtime of \FOLNF{} and \FOLNFast{} was almost linear in $\bn \log^2 \bn$, where $\bn$ is the number of shuffled values. 
Based on this property, we estimated the runtime of these algorithms for $n \geq 10^6$. 


Figure~\ref{fig:res_each_runtime} shows that the time for shuffling is dominant. 
This is consistent with our theoretical results that the runtime of random sampling, dummy data addition, and shuffling is $O(n \log d)$, $O((\lambda + \sum_{i=1}^d \kappa_i) \log d)$, and $O((\bn \log^2 \bn) \log d)$, respectively, where  $\bn = n + \lambda + \sum_{i=1}^d \kappa_i$ (Theorem~\ref{thm:Proposal_runtime}). 

Figure~\ref{fig:res_total_runtime}(a) shows that \FOLNF{} is more efficient than \FOUD{}, as \FOLNF{} adds fewer dummies than \FOUD{}. 
In addition, \FOLNFast{} and \OGeo{} are more efficient than \FOLNF{} and \AGeo{}, respectively, because they add fewer bots, as shown in Figure~\ref{fig:res_FOLNF}. 
Figure~\ref{fig:res_total_runtime}(b) shows that the runtime for large-domain data can be significantly reduced from $\tO(n+d)$ to $\tO(n)$ by using the count-min sketch. 
Furthermore, Figure~\ref{fig:res_total_runtime}(c) shows that our algorithms handle very large data. 
For example, 
the runtime of our \FOLNF{} (\AGeo{}), \FOLNFast{} (\AGeo{}), \FOLNF{} (\OGeo{}), and \FOLNFast{} (\OGeo{}) is $16$ hours, $8$ hours, $29$ minutes, and 
\colorB{$13$} 
minutes, respectively. 

\begin{figure}[t]
  \centering
  \includegraphics[width=0.99\linewidth]{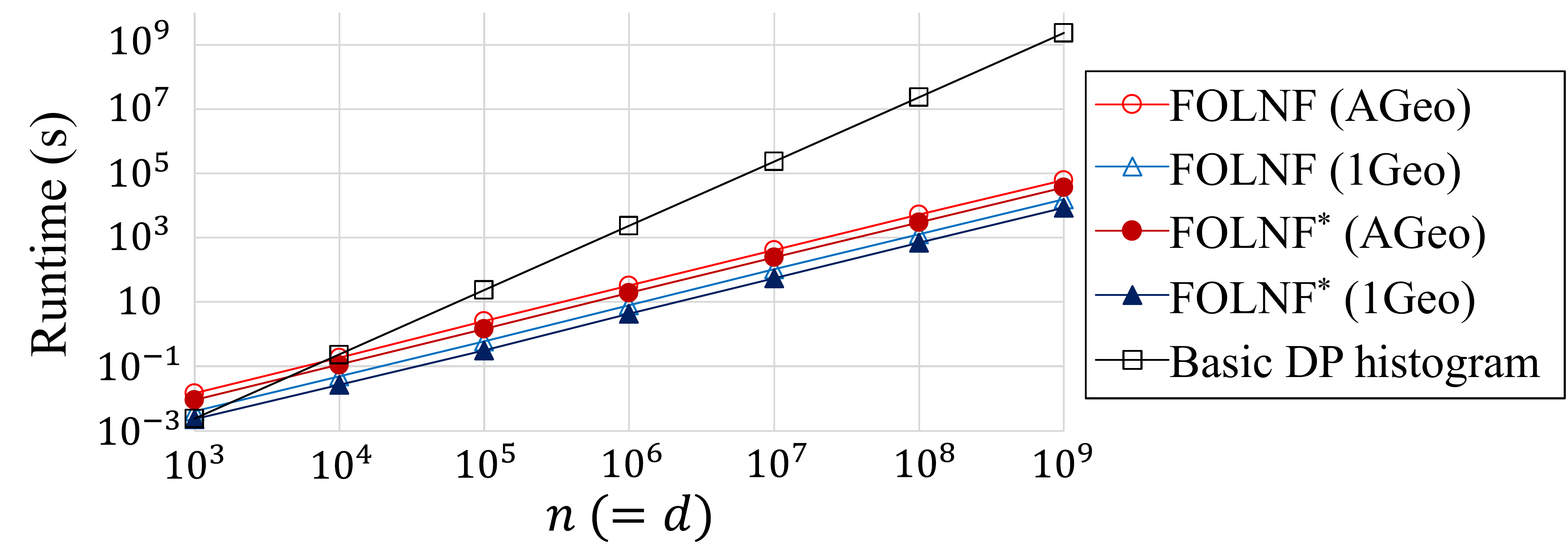}
  \caption{\colorB{Total runtime of our algorithms and the basic DP histogram ($\epsilon_E = 1$, $\epsilon_I = 5$).}} 
  \label{fig:res_histogram}
\end{figure}

\smallskip{}
\noindent{\textbf{\colorB{Comparison with a Central DP Algorithm.}}}~~\colorB{Finally, we compared the runtime of our algorithms with that of a central DP algorithm within the TEE. 
Specifically, we evaluated the basic DP histogram~\cite{Balcer_JPC19,Dwork_TCC06} because it has the same time complexity as the stability histogram~\cite{Bun_JMLR19,Lebeda_FORC25} when they are fully oblivious; see Appendix~\ref{sub:direct_histogram} for details. 
Figure~\ref{fig:res_histogram} shows the results. 
Here, we confirmed that the total runtime of the basic DP histogram (resp.~our algorithms) was almost linear in $nd$ (resp.~$\bn \log^2 \bn$, where $\bn$ is \#shuffled values). 
Based on this, we estimated the total runtime when $n=d=10^3$ to $10^9$.} 

\colorB{Figure~\ref{fig:res_histogram} shows that our algorithms are more efficient than the central DP algorithm when $n\geq10^4$. 
Note that this also holds when $d \gg n \geq 10^4$ and all algorithms (including the basic DP histogram) use the count-min with $b=n$. 
Figure~\ref{fig:res_total_runtime}(c) and Figure~\ref{fig:res_histogram} show that our algorithms have a lower runtime when $\epsilon_E = 1$, as the number of bots is smaller.} 

\colorB{For example, when $n=d=10^8$, the central DP algorithm requires about $270$ days ($=2.3 \times 10^7$ seconds). 
In contrast, our algorithms take less than one day when $n$ and $d$ are at most $10^9$ (e.g., 
$n=3 \times 10^8$ and $d=6 \times 10^8$ in Amazon~\cite{amazon_statistics}), demonstrating their efficiency.}

\colorB{We also note that when $n=d \leq 10^3$, the basic DP histogram is comparable to or more efficient than our algorithms. 
Thus, the former would be preferable in such small $n$ and $d$.}

\section{Conclusion}
\label{sec:conclusion}
We proposed a new privacy notion, FODP, and its algorithms for the augmented shuffle model with TEEs. 
Through theoretical analysis and extensive experiments, we 
showed that our algorithms achieve high accuracy, robustness, and efficiency (e.g., $16$ hours or less when $n=d=10^8$) under FODP. 
\colorB{For future work, we would like to study FODP for other tasks, such as the report noisy max~\cite{DP}, graph analysis~\cite{Fang_CCS25,Imola_CCS22}, and federated learning~\cite{Girgis_AISTATS21,Liu_AAAI21}.}

\section*{\colorB{Acknowledgments}}
\colorB{We thank Shun Takagi, Satoshi Hasegawa, and anonymous reviewers for their helpful comments. 
We also thank Tadanori Teruya for his comments on the implementation of ORShuffle used in Section~\ref{sec:experiments} and Appendix~\ref{sec:addendum}. 
This paper was edited for grammar using ChatGPT and Grammarly. 
This study was supported in part by JSPS KAKENHI 22H00521, 24H00714, 24K20775, JST K Program Grant Number JPMJKP24U5, JST NEXUS JPMJNX25C2, JST AIP Acceleration Research JPMJCR22U5, and JST CREST JPMJCR22M1.}

\appendix
\section*{Ethical Considerations}
Our work develops a new privacy notion, FODP, and algorithms for FODP in the augmented shuffle model with TEEs. 
Our algorithms can be used to calculate a frequency distribution on a single server while strongly preventing the leakage of personal data from the algorithms' outputs and their side-channel information, such as memory access patterns and control flows. 
Following the ethics guidelines, we conduct a stakeholder-based analysis. 

\smallskip{}
\noindent{\textbf{Stakeholder Analysis.}}~~Our stakeholders include 
\colorB{(1) users who join the protocol, (2) a service provider who manages the server and collects their personal data, and (3) a data analyst (or anyone) who obtains the published frequency distribution}. 

\smallskip{}
\noindent{\textbf{\colorB{Impacts}.}}~~Our algorithms provide high accuracy, privacy, 
robustness, and efficiency, thereby having a positive ethical impact on all of the stakeholders. 
\colorB{In addition, we do not encounter privacy issues that may arise from the use of private datasets, as we used only public datasets. 
However, there are some potential negative impacts. 
Specifically: 
\begin{itemize}
\item \textbf{Users.} Robustness means that privacy is not violated, even if the service provider (or data analyst) colludes with some users. 
Thus, high privacy and robustness reduce the risks of compromising user privacy. 
One potential negative impact is that privacy information might be leaked by timing attacks, as described in Section~\ref{sub:threat_model}. 
\item \textbf{Service Provider.} The high privacy and robustness reduce the risks of damaging his/her reputation. 
One potential negative impact is that using our algorithms may increase the service provider's effort and overhead. 
\item \textbf{Data Analyst.} The data analyst benefits from high accuracy and efficiency, as he/she wants to calculate an accurate frequency distribution in a timely manner. 
\end{itemize}}

\smallskip{}
\noindent{\textbf{\colorB{Mitigation.}}}~~\colorB{As discussed in Section~\ref{sub:threat_model}, a mitigation strategy for timing attacks is to make the runtime constant~\cite{Haeberlen_USENIX11,Jin_SP22}. 
In addition, we have published our code to decrease the effort and overhead for the service provider. 
Improving efficiency is an important future direction to further decrease overhead.}

\smallskip{}
\noindent{\textbf{\colorB{Decision}.}}~~\colorB{We decided to proceed with the research and publish it because the high accuracy, privacy, robustness, and efficiency of our algorithms have a positive impact on all stakeholders, as explained above. 
The potential negative impacts can also be mitigated by, e.g., making the runtime constant and improving efficiency.} 
We also believe that our work will serve as a basis for further research on more accurate, private, and efficient algorithms that have more positive ethical impacts on all stakeholders.

\section*{Open Science}
Following the open science policy, we have published our artifacts on the following website: 
\url{https://doi.org/10.5281/zenodo.20287113}. 
Our artifacts include: 
\begin{itemize}
    \item Source code of an implementation of our algorithms.
    \item Instructions on how to run our code to obtain our experimental results (README).
\end{itemize}

We have explained how to run our code for all experimental results. 
See README for more details.

\bibliographystyle{abbrv}
\bibliography{main_short}

\appendix
\section{Details of Existing Shuffle Protocols and Robustness against Collusion with Users}
\label{sec:robustness_collusion}

\subsection{Privacy Amplification Results}
\label{sub:pure_shuffle_collusion}

Below, we present state-of-the-art privacy amplification results for pure shuffle protocols: 
\begin{theorem}[Amplification-by-shuffling result~\cite{Feldman_SODA23}]
\label{thm:privacy_amplification}
Let 
$\epsilon_L \in \nnreals$ and $\delta \in [0,1]$. 
Let $\calR_L$ be a local randomizer. 
Let $\calR_S$ be a pure shuffle algorithm that takes $\bmx = (x_1, \ldots, x_n)$ as input 
and outputs $\calR_S(\bmx) = (\calR_L(x_{\pi(1)}), \ldots, \calR_L(x_{\pi(n)}))$, where $\pi$ is a random permutation over $[n]$. 
If $\calR_L$ provides $\epsilon_L$-LDP, then 
$\calR_S$ provides $(\epsilon, \delta)$-DP with $\epsilon = g(n,\epsilon_L,\delta)$, where 
$g(n,\epsilon_L,\delta) = 
\ln\left(1 + \frac{4(e^{\epsilon_L} - 1)\sqrt{2 \ln(4/\delta)}}{\sqrt{(e^{\epsilon_L} + 1)n}} + \frac{4}{n} \right)$
if $\epsilon_L \leq \ln(\frac{n}{8\ln(2/\delta)}-1)$ and $g(n,\epsilon_L,\delta)=\epsilon_L$ otherwise. 
\end{theorem}

\colorB{Theorem~\ref{thm:privacy_amplification} is typically used for single-message protocols in the pure shuffle model.} 
\cite{Balcer_ITC20,Cheu_SP22,Luo_CCS22} show different values of $g(n,\epsilon_L,\delta)$ 
for their multi-message protocols.

\subsection{Formal Definition of the Robustness against Collusion with Users}
\label{sub:robustness_collusion_formal}

Below, we formally define the robustness against collusion with users by using $\Omega$-neighboring databases~\cite{Beimel_CRYPTO08}: 

\begin{definition} [$\Omega$-neighboring databases~\cite{Beimel_CRYPTO08}] \label{def:omega_neighboring} 
Let $\Omega \subset [n]$. 
Two databases $\bmx = (x_1, \ldots, x_n) \in [d]^n$ and $\bmx' = (x'_1, \ldots, x'_n) \in [d]^n$ are \emph{$\Omega$-neighboring} if they differ on a single entry whose index $i$ is not in $\Omega$, i.e., 
$\exists i \notin \Omega$: $x_i \ne x'_i$ and $\forall j \ne i$: $x_j = x'_j$. 
\end{definition}
In Definition~\ref{def:omega_neighboring}, we consider an adversary who colludes with users $\{u_i | i \in \Omega\}$ and tries to infer input values of the remaining users (i.e., victims) $\{u_i | i \notin \Omega\}$. 

Consider a shuffle DP algorithm $\calR$ that achieves $(\epsilon_E,\delta_E)$-DP and $(\epsilon_I,\delta_I)$-FODP, as shown in Figure~\ref{fig:FODP}. 
Then, we can define the robustness against collusion with users as follows: 

\begin{definition} [Robustness against collusion with users] 
\label{def:robustness_to_collusion}
Let $\epsilon_E, \epsilon_I \in \nnreals$ and $\delta_E, \delta_I \in [0,1]$. 
Let $\calR$ be a shuffle algorithm with domain $[d]^n$ that provides $(\epsilon_E,\delta_E)$-DP and $(\epsilon_I,\delta_I)$-FODP. 
For $i \in [n]$, let $\psi_i$ be data sent from user $u_i$ to the shuffler in $\calR$. 
For $\Omega \subset [n]$, let $\calR_\Omega$ be an algorithm that takes a database $\bmx \in [d]^n$ as input and outputs $\calR_\Omega(\bmx) = (\calR(\bmx), (\psi_i)_{i\in\Omega})$. 
Then, $\calR$ is \emph{robust against collusion with users} if it satisfies the following two conditions for any $\Omega \subset [n]$ and any $\Omega$-neighboring databases $\bmx, \bmx' \in [d]^n$: 
\begin{itemize}
    \item For any $O \subseteq \mathrm{Range}(\calR_\Omega)$, $\calR_\Omega$ satisfies (\ref{eq:DP_inequality}) with $(\epsilon,\delta) = (\epsilon_E,\delta_E)$. 
    \item For any $(O,O_M,O_I) \subseteq \mathrm{Range}((\calR_\Omega, \calR^\calM, \calR^\calI))$, $(\calR_\Omega, \allowbreak \calR^\calM, \calR^\calI)$ satisfies (\ref{eq:FODP_inequality}) with $(\epsilon,\delta) = (\epsilon_I,\delta_I)$. 
\end{itemize}
\end{definition}
$\psi_i$ in Definition~\ref{def:robustness_to_collusion} depends on the protocol, e.g., $\psi_i = x_i$ in our protocols. 
Definition~\ref{def:robustness_to_collusion} states that the values of $(\epsilon_E,\delta_E)$ and $(\epsilon_I,\delta_I)$ in $\calR$ are not increased by collusion with users. 

\section{Details of Algorithms}
\label{sec:details_algorithms}
\subsection{Fully Oblivious Sampling from a Dummy-Count Distribution}
\label{sub:dummy-count_distribution}

\colorB{The function $\texttt{DummyCountGeneration}$ can be implemented using the CDF $F$ for a dummy-count distribution $\calD$ as follows. 
First, we compare $F(0),\ldots,F(z_{max}-1)$ to a random value $r \in [0,1)$, where $z_{max}$ is the maximum value of $z_i$, i.e., $z_{max} = \kappa$ in \FOLNF{}. 
Then, it outputs the first value whose CDF exceeds $r$ as $z_i$ ($z_i = z_{max}$ if no CDF exceeds $r$). 
This algorithm is fully oblivious to $z_i$, as memory and instruction traces are independent of $z_i$. 
Its time complexity is $O(z_{max})$.}

It is also possible to directly calculate $z_i$ from $r$ by arithmetic operations. 
Algorithm~\ref{alg:dummy_count_distribution_AGeo} shows such an algorithm in the asymmetric geometric distribution $\AGeo(\nu,q_l,q_r)$. 
Note that when $\nu=q_l=0$ (i.e., one-sided geometric distribution), $z_i$ can be easily calculated from $r$ as: $z_i \leftarrow \lfloor \log_{q_r} (1-r) \rfloor$. 
Algorithm~\ref{alg:dummy_count_distribution_AGeo} is also fully oblivious, as it consists of arithmetic operations. 
The time complexity of this algorithm is $O(1)$. 
We used Algorithm~\ref{alg:dummy_count_distribution_AGeo} in our experiments. 

\setlength{\algomargin}{5mm}
\begin{algorithm}[t]
  \SetAlgoLined
  \KwData{Asymmetric geometric distribution $\AGeo(\nu,q_l,q_r)$.
  }
  \KwResult{Random sample $z_i$ from $\AGeo(\nu,q_l,q_r)$.}
  $r \leftarrow \texttt{ORAND\_REAL}()$\;
  $\eta = \frac{q_l(1-q_l^\nu)}{1-q_l} + \frac{1}{1-q_r}$\;
  $B \leftarrow [r \geq 1 - \frac{q_r}{\eta(1-q_r)}]$\;
  $z_i^l \leftarrow \nu - \lceil \log_{q_l} (\eta(1-q_l)r + q_l^{\nu+1}) \rceil + 1$\;
  $z_i^r \leftarrow \lfloor \log_{q_r} \eta (1-q_r)(1-r) \rfloor + \nu$\;
  $z_i \leftarrow (1-B) z_i^l + B z_i^r$\;
  \KwRet{$z_i$}
  \caption{\texttt{DummyCountGeneration} via arithmetic operations when $\calD = \AGeo(\nu,q_l,q_r)$.}
  \label{alg:dummy_count_distribution_AGeo}
\end{algorithm}

\subsection{\colorB{Central DP Algorithms within the TEE}}
\label{sub:direct_histogram}

\noindent{\textbf{\colorB{Basic DP Histogram}.}}~~Algorithm~\ref{alg:direct_histogram} shows \colorB{a fully oblivious version of the basic DP histogram~\cite{Balcer_JPC19,Dwork_TCC06}, which calculates a histogram and adds DP noise to each histogram bin within the TEE.} 
We first calculate a histogram $\bmh \triangleq (h_1,\ldots,h_d) \in (\nnints)^d$ of input values $\bmx$ (lines 1-6). 
Then, we add the two-sided Geometric noise $sz$ to each bin of $\bmh$ and divide the result by $n$, where $s$ is a random value in $\{-1,1\}$ and $z$ is a sample from the one-sided Geometric distribution $\OGeo(2/\epsilon)$ (lines 7-12). 
Algorithm~\ref{alg:direct_histogram} provides $\epsilon$-DP~\cite{Balcer_JPC19}. 
It is also fully oblivious to a single secret, as memory and instruction traces are independent of $\bmx$. 
Thus, by Theorem~\ref{thm:DP_FO_FODP}, it provides $\epsilon$-FODP. 

\setlength{\algomargin}{5mm}
\begin{algorithm}[t]
  \SetAlgoLined
  \KwData{Input values $\bmx = (x_1, \ldots, x_n)$, 
  \#items $d \in \nats$.}
  \KwResult{Estimate $\bmhf = (\hf_i,\ldots,\hf_d)$.}
  \tcc{Histogram calculation}
  $\bmh \triangleq (h_1,\ldots,h_d) \leftarrow (0,\ldots,0)$\;
  \ForEach{$i \in [n]$}{
    \ForEach{$j \in [d]$}{
      $h_j \leftarrow h_j + [x_i = j]$\;
    }
  }
  \tcc{Adding the two-sided Geometric noise ($s \in \{-1,1\}$, $z \sim \OGeo(2/\epsilon)$)}
  \ForEach{$i \in [d]$}{
    $r \leftarrow \texttt{ORAND\_REAL}()$\;
    $z \leftarrow \lfloor \log_{2/\epsilon} (1-r) \rfloor$\;
    $s \leftarrow 2 \cdot \texttt{ORAND\_NATS}(2) - 3$\;
    $\hf_i \leftarrow \frac{1}{n}(h_i + s z)$\;
  }
  \KwRet{$\bmhf = (\hf_i,\ldots,\hf_d)$}
  \caption{\colorB{Basic DP histogram.}}
  \label{alg:direct_histogram}
\end{algorithm}

\smallskip{}
\noindent{\textbf{Runtime.}}~~Algorithm~\ref{alg:direct_histogram} accesses the entire memory (i.e., each element in $\bmh$) to provide full obliviousness when adding each input value to the histogram. 
In addition, the size of each element in $\bmh$ is $|h_j| = \log_2 d$. 
Thus, the runtime is upper bounded by $O(nd \log d)$\footnote{Note that when $\log_2 d$ is smaller than the word length in the word RAM model~\cite{Fredman_STOC90}, the runtime is $O(nd)$.} 
and is simplified as $\tO(nd)$.

\smallskip{}
\noindent{\textbf{\colorB{Stability Histogram.}}}~~\colorB{The runtime $\tO(nd)$ also applies to the stability histogram~\cite{Bun_JMLR19,Lebeda_FORC25}, which adds DP noise to only items whose frequencies exceed a threshold in a sparse histogram. 
Specifically, the stability histogram has the same issue as Algorithm~\ref{alg:direct_histogram}; i.e., it must access the entire memory to provide full obliviousness when adding each input value to the histogram $\bmh$ (lines 2-6 in Algorithm~\ref{alg:direct_histogram}). 
In other words, 
it requires the runtime $\tO(nd)$ to calculate a \textit{clean} histogram before adding DP noise. 
We also confirmed that the runtime for the histogram calculation (lines 2-6) is dominant in Algorithm~\ref{alg:direct_histogram}. 
Since the stability histogram needs $\bmh$, it also requires $\tO(nd)$ when running within the TEE; e.g., 
it requires about $270$ days when $n=d=10^8$, as shown in Figure~\ref{fig:res_histogram}.}

\section{Details of Theoretical Properties}
\label{sec:details_bounds}

\subsection{\colorB{Robustness of Our General Algorithm against Poisoning Attacks}}
\label{sub:robustness_poisoning}
\colorB{Following~\cite{Cao_USENIX21}, we consider a targeted attack that attempts to increase the estimates of some target items $\calT \subseteq [d]$. 
Without loss of generality, assume that users $u_1,\ldots,u_n$ are genuine and that the attacker injects $n'\in\nats$ fake users $u_{n+1},\ldots,u_{n+n'}$. 
The fake users can send arbitrary messages to the shuffler. 
For example, when honest users apply a local randomizer $\calR_L$ and send the noisy values $\calR_L(x_1),\ldots,\calR_L(x_n)$ to the shuffler, the fake users can change their noisy values to arbitrary values. 
This is called the \textit{output poisoning attack}~\cite{Li_USENIX23}.} 

\colorB{Let $\hat{f}'_i \in \reals$ be an estimate of $f_i$ after poisoning. 
Let $\Delta f_i = \hat{f}'_i - \hat{f}_i$ be the frequency gain for the $i$-th item. 
Let $\bmy' = (y'_1,\ldots,y'_{n'})$ be messages sent from $n'$ fake users.
Let $G(\bmy')$ be the attacker's overall gain defined as follows: 
$G(\bmy') = \sum_{i\in\calT} \E[\Delta f_i]$. 
Cao \textit{et al.}~\cite{Cao_USENIX21} propose an optimal attack called the MGA (Maximal Gain Attack) that maximizes $G(\bmy')$. 
Let $G_{MGA}$ be the MGA's overall gain, i.e., $G_{MGA} = \max_{\bmy'} G(\bmy')$. 
Then, we can show the robustness of our general algorithm:} 
\begin{theorem}\label{thm:G_MGA}
\colorB{Our general algorithm $\calR_{\calD,\calD',\beta,\lambda}$ (Algorithm~\ref{alg:proposal}) provides the following robustness guarantee:}
\begin{align}
\colorB{G_{MGA} = \textstyle{\frac{n'}{n+n'}(1-\sum_{i\in\calT} f_i)}.}
\label{eq:G_MGA_general}
\end{align}
\end{theorem}
\colorB{$G_{MGA}$ in (\ref{eq:G_MGA_general}) is exactly the same as the overall gain of the \textit{input poisoning attack}~\cite{Cao_USENIX21}, in which fake users change their input values and then follow the protocol (e.g., apply $\calR_L$). 
This is quite obvious because users do not add noise in our algorithm; i.e., output values are identical to input values. 
It is shown that pure shuffle protocols are vulnerable to the output poisoning attack, especially when $\epsilon$ is small~\cite{Cao_USENIX21,Murakami_SP25}. 
Our algorithm prevents the output poisoning attack, as the attacker can perform only the input poisoning attack.}

\subsection{Privacy Parameters in FOUD}
\label{sub:comparison_bounds_FOUD}

Below, we compare our bound on $(\epsilon,\delta)$-DP for \FOUD{} (Theorem~\ref{thm:FOUD_FODP}) with the bound directly derived from \cite{Wang_PVLDB20}: 
\begin{theorem}[$(\epsilon,\delta)$-DP of \FOUD{} derived from \cite{Wang_PVLDB20}]\label{thm:FOUD_FODP_Wang}
For $\epsilon \leq 1$, $\calR_\lambda^{\FOUD}$ provides $(\epsilon,\delta)$-DP, where $\epsilon = \sqrt{\frac{14\ln(2/\delta)d}{\lambda}}$. 
\end{theorem}

\begin{figure}[t]
  \centering
  \includegraphics[width=0.99\linewidth]{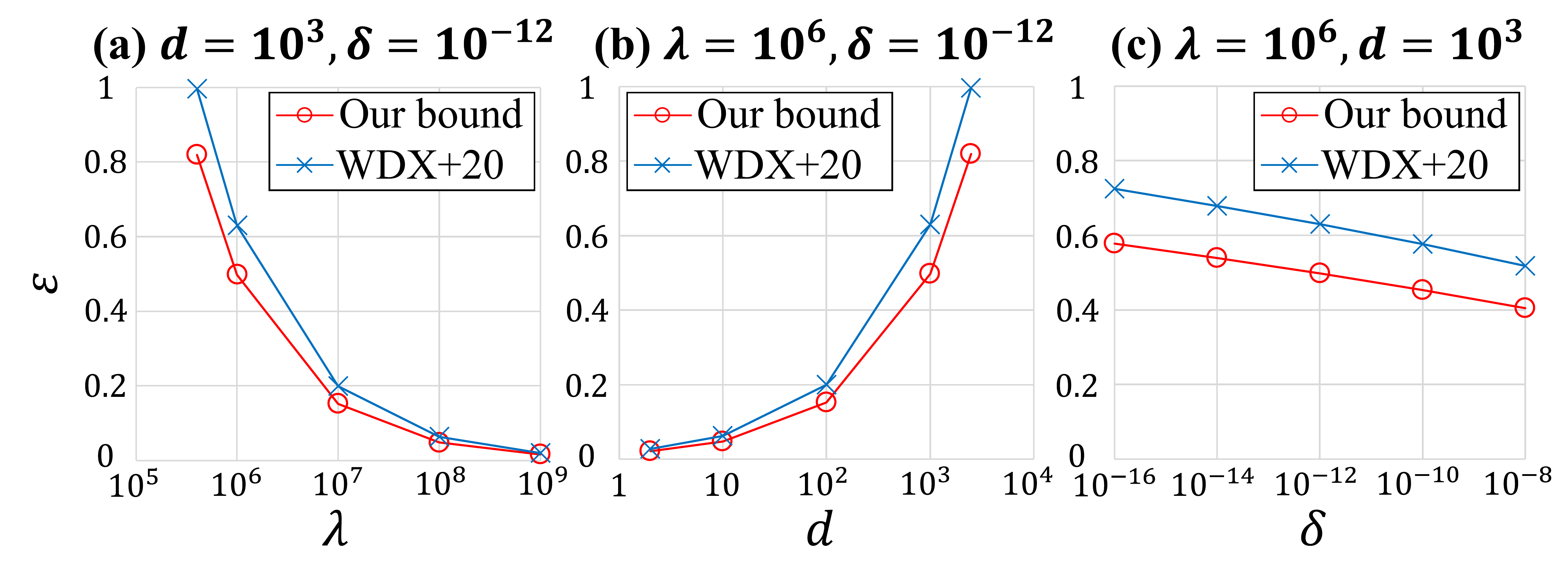}
  \caption{Two bounds on $\epsilon$ for \FOUD{}.} 
  \label{fig:FOUD_bound}
\end{figure}

Figure~\ref{fig:FOUD_bound} shows $\epsilon$ in Theorem~\ref{thm:FOUD_FODP} (our bound) and Theorem~\ref{thm:FOUD_FODP_Wang} (denoted by \textsf{WDX+20}). 
Our bound provides a tighter bound on $\epsilon$ than \textsf{WDX+20} for various values of $\lambda$, $d$, and $\delta$. 

\subsection{Privacy, Robustness, and Accuracy of Our Algorithms for Large-Domain Data}
\label{sub:comparison_bounds_Large}

\noindent{\textbf{Privacy and Robustness.}}~~We use the optimal composition theorem in \cite{Kairouz_ICML15} to show that our general algorithm $\calR_{\calD,\calD',\beta,\lambda}^{\Largesf}$ for large-domain data provides the following guarantees: 
\begin{theorem} \label{thm:general_large_privacy_robustness}
Let $\epsilon_E,\epsilon_I \in \nnreals$ and $\delta_E, \delta_I \in [0,1]$. 
If $\calR_{\calD,\calD',\beta,\lambda}$ (Algorithm~\ref{alg:proposal}) provides $(\epsilon_E,\delta_E)$-DP and $(\epsilon_I,\delta_I)$-FODP, then $\calR_{\calD,\calD',\beta,\lambda}^{\Largesf}$ (Algorithm~\ref{alg:proposal_large}) provides $((\tau - 2i)\epsilon_E, 1-(1-\delta_E)(1-\delta_i))$-DP and $((\tau - 2i)\epsilon_I, 1-(1-\delta_I)(1-\delta_i))$-FODP for any $i = \{0,1,\ldots,\lfloor \frac{\tau}{2}\rfloor\}$, where 
$\delta_i = \frac{1}{(1+e^\epsilon)^\tau} \left(\sum_{j=0}^{i-1} \binom{\tau}{j}(e^{(\tau-j)\epsilon} - e^{(\tau-2i+j)\epsilon})\right)$, 
and is robust against collusion with users. 
\end{theorem}

\noindent{\textbf{Accuracy.}}~~We denote our general algorithm $\calR_{\calD,\calD',\beta,\lambda}^{\Largesf}$ applied to \FOUD{}, \FOLNF{}, and \FOLNFast{} by 
$\calR_\lambda^{\FOUDLarge}$, $\calR_{\calD,\beta}^{\FOLNFLarge}$, and $\calR_{\calD,\calD',\beta}^{\FOLNFastLarge}$, respectively. 
Then, the following corollaries are derived from Theorem~\ref{thm:Proposal_large_accuracy}: 
\begin{corollary} \label{cor:Proposal_large_accuracy_FOUD}
$\calR_\lambda^{\FOUDLarge}$ 
achieves 
\begin{align}
\Pr[|\hf_i - f_i| \leq \gamma] \geq \textstyle{1 - ( \frac{2}{b\gamma} + d_1 )^\tau - \tau d_2}, 
\label{eq:Proposal_large_accuracy_FOUD}
\end{align}
for each $i\in[d]$, where 
$d_1 = e^{-\lambda \cdot D \left( \frac{1}{b}+\frac{n\gamma}{2\lambda} \parallel \frac{1}{b} \right)}$ if $\gamma < \frac{2\lambda(b-1)}{nb}$ and $d_1 = 0$ otherwise, 
$d_2 = e^{-\lambda \cdot D \left( \frac{1}{b}-\frac{n\gamma}{\lambda} \parallel \frac{1}{b} \right)}$ if $\gamma < \frac{\lambda}{nb}$ and $d_2 = 0$ otherwise, 
and $D(x \parallel y) = x \ln \frac{x}{y} + (1-x)\ln \frac{1-x}{1-y}$. 
\end{corollary}

\begin{corollary} \label{cor:Proposal_large_accuracy_FOLNF_AGeo}
$\calR_{\calD,\beta}^{\FOLNFLarge}$ and $\calR_{\calD,\calD',\beta}^{\FOLNFastLarge}$ with 
$\calD = \AGeo(\nu, q_l,\allowbreak q_r)$ in Definition~\ref{def:joint_ageo} and $\beta=1$ achieve 
\begin{align}
\Pr[|\hf_i - f_i| \leq \gamma] \geq \textstyle{1 -  ( \frac{2}{b \gamma} + d_3 )^\tau - \tau d_4}, 
\label{eq:Proposal_large_accuracy_FOLNF}
\end{align}
for each $i\in[d]$, where 
$d_3 = \frac{1}{\eta (1 - q_r)} \cdot q_r^{a_r - \nu}$, $d_4 = \frac{1}{\eta(1-q_l)} \cdot (1-q_l^{a_l +1})q_l^{\nu - a_l}$ if $\gamma < \frac{\mu}{n}$ and $d_4 = 0$ otherwise, 
$\eta = \frac{q_l(1-q_l^\nu)}{1-q_l} + \frac{1}{1-q_r}$, $a_r = \lceil \mu + \frac{n\gamma}{2} \rceil$, and $a_l = \lfloor \mu - n\gamma \rfloor$. 
\end{corollary}

Corollary~\ref{cor:Proposal_large_accuracy_FOLNF_AGeo} assumes $\beta=1$. 
When $\beta < 1$, the factors of random sampling (i.e., $Bin(nf_i,\beta)$ and $nf_i \beta$ in Theorem~\ref{thm:Proposal_large_accuracy}) are independent of $\tau$. 
Thus, in our experiments, we used the same optimization method for $\beta < 1$ (\OGeo{}) as for $\beta = 1$. 
We show that it works well \conference{\colorB{in~\cite{Murakami_arXiv26}}}\arxiv{in Appendix~\ref{sub:tau_b}}.

\noindent{\textbf{Comparison with the Bounds in \cite{Cheu_SP22,Ghazi_EUROCRYPT21}.}}~~Ghazi \textit{et al.}~\cite{Ghazi_EUROCRYPT21} analyze the additive error of a pure shuffle protocol, where each user adds a dummy value with probability $\varphi \in [0,1]$ for each entry of the count-min sketch (i.e., each of $\tau b$ hash values) and sends the dummy values along with her hash values. 
Shuffled values of this protocol are the same as those of 
$\calR_{\calD,\beta}^{\FOLNFLarge}$ 
with $\calD = Bin(n,\varphi)$ and $\beta=1$. 
Therefore, we show our bound in this case and compare it with the bound in \cite{Ghazi_EUROCRYPT21} (we discuss the bound in \cite{Cheu_SP22} at the end of this section): 

\begin{corollary}\label{cor:Proposal_large_accuracy_FOLNF_Bin}
$\calR_{\calD,\beta}^{\FOLNFLarge}$ 
with $\calD = Bin(n,\varphi)$ and $\beta=1$ 
achieves 
\begin{align}
\Pr[|\hf_i - f_i| \leq \gamma] \geq \textstyle{1 - ( \frac{2}{b\gamma} + d_5 )^\tau - \tau d_6}, 
\label{eq:Proposal_large_accuracy_FOLNF_Bin}
\end{align}
for each $i\in[d]$, where 
$d_5 = e^{-n \cdot D \left( \varphi+\frac{\gamma}{2} \parallel \varphi \right)}$ if $\gamma < 2(1-\varphi)$ and $d_5 = 0$ otherwise, $d_6 = e^{-n \cdot D \left( \varphi-\gamma \parallel \varphi \right)}$ if $\gamma < \varphi$ and $d_6 = 0$ otherwise, and $D(x \parallel y) = x \ln \frac{x}{y} + (1-x)\ln \frac{1-x}{1-y}$. 
\end{corollary}

\begin{theorem}[Additive error derived from \cite{Ghazi_EUROCRYPT21}]\label{thm:additive_error_Ghazi}
$\calR_{\calD,\beta}^{\FOLNFLarge}$ 
with $\calD = Bin(\lambda,\frac{1}{b}$) and $\beta=1$ achieves 
\begin{align}
\Pr[|\hf_i - f_i| \leq \gamma] \geq \textstyle{1 - (\frac{n}{b})^\tau - 2^{\log_2(2b\tau)- \gamma^2 n / (3\varphi)}}. 
\label{eq:Proposal_large_accuracy_FOLNF_Bin_Ghazi}
\end{align}
\end{theorem}
The privacy parameters in DP can also be derived from~\cite{Ghazi_EUROCRYPT21}. 
Specifically, it follows from Lemma~4.10 in~\cite{Ghazi_EUROCRYPT21} that $\calR_{\calD,\beta}^{\FOLNFLarge}$ 
with $\calD = Bin(n,\varphi)$ and $\beta=1$ provides $(\epsilon_E,\delta_E)$-DP for a single hash function ($\tau=1$), where $\epsilon_E = \sqrt{\frac{90 \ln(2/\delta_E)}{\varphi n}}$. 
By Theorem~\ref{thm:FOLNF_FODP}, $\calR_{\calD,\beta}^{\FOLNFLarge}$ also provides $(\epsilon_I,\delta_I)$-FODP, where $(\epsilon_I,\delta_I) = (\epsilon_E,\delta_E)$. 
Based on these results, we set $(n,\varphi)=(10^4,0.26)$ in Figure~\ref{fig:efficient_bound} so that $(\epsilon_I,\delta_I)=(\epsilon_E,\delta_E)=(1,10^{-12})$. 

The first and second terms $(=1-(\frac{n}{b})^\tau)$ in the right side of (\ref{eq:Proposal_large_accuracy_FOLNF_Bin_Ghazi}) represent the probability that no hash collision occurs among $n$ input values. 
Note that these terms are zero or less when $b \leq n$. 
In other words, the bound in \cite{Ghazi_EUROCRYPT21} cannot lower bound $\Pr[|\hf_i - f_i| \leq \gamma]$ in this case, as shown in Figure~\ref{fig:efficient_bound}. 

Finally, we note that the bound in \cite{Cheu_SP22} is also based on the probability that no hash collision occurs and therefore cannot lower bound $\Pr[|\hf_i - f_i| \leq \gamma]$. 
Specifically, $\Pr[|\hf_i - f_i| \leq \gamma]$ in their bound is less than or equal to $1 - d(\frac{n}{b})^\tau$. 
Since $d \geq 2$, $\Pr[|\hf_i - f_i| \leq \gamma]$ becomes 
negative 
when $b \leq n$. 

\arxiv{
\section{Additional Experiments}
\label{sec:additional}

\subsection{Runtime of Our Algorithms with the Codes in \cite{Sasy_code}}
\label{sub:runtime_Sasy_code}
Figure~\ref{fig:res_Sasy_runtime} shows the total runtime of our algorithms when we used the codes of ORShuffle and WaksShuffle in \cite{Sasy_code}. 
In Figure~\ref{fig:res_Sasy_runtime}(b) and (d), we did not show the runtime of \FOUD{}, because the number of shuffled values exceeded ten million and caused a stack overflow in their codes. 

\begin{figure}[t]
  \centering
  \includegraphics[width=0.99\linewidth]{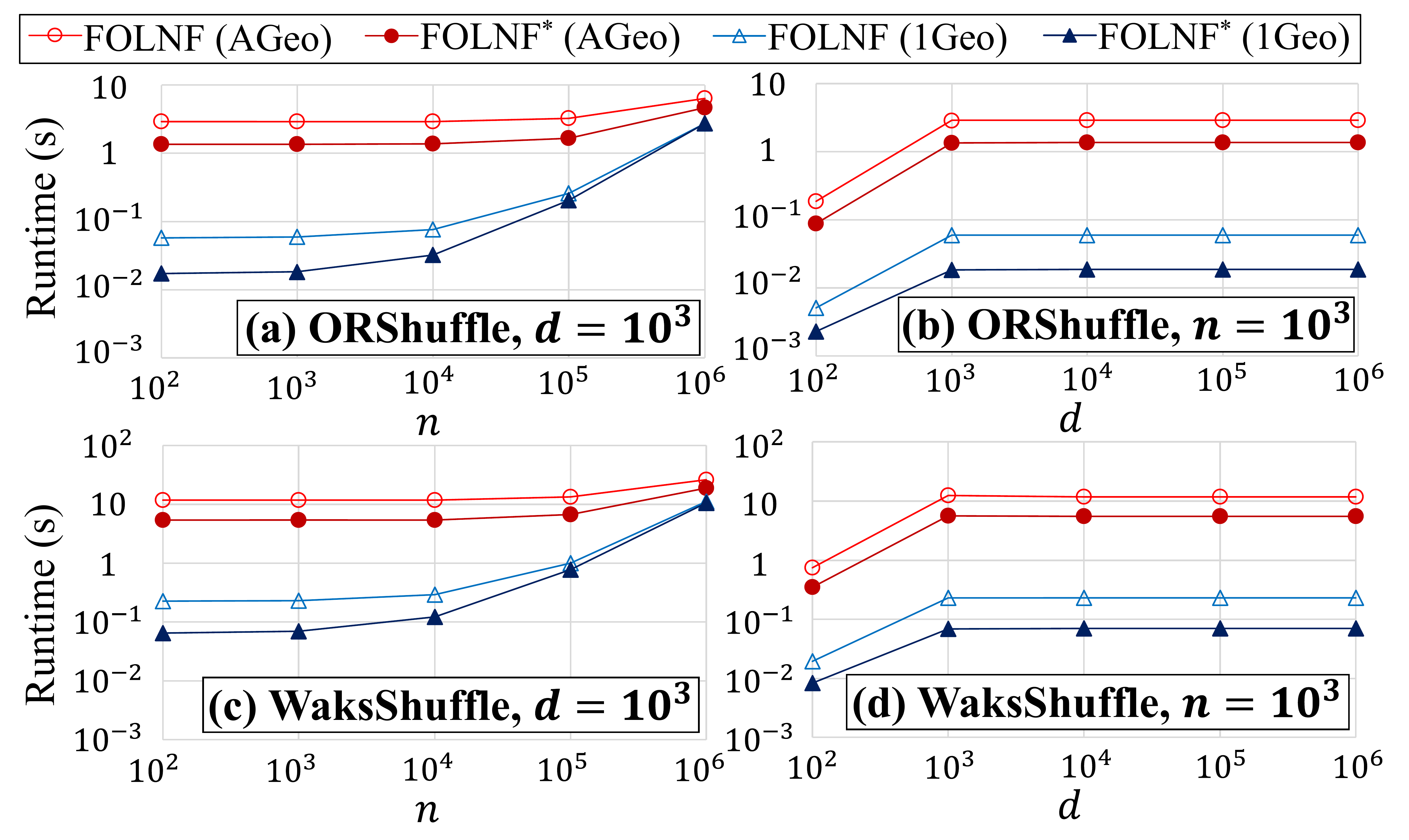}
  \caption{Total runtime of our algorithms with the codes of ORShuffle and WaksShuffle in \cite{Sasy_code} ($\epsilon_E = 0.1$). 
  In (b) and (d), we used the count-min sketch ($b=n$) when $n < d$.} 
  \label{fig:res_Sasy_runtime}
\end{figure}

Figure~\ref{fig:res_Sasy_runtime} shows a similar tendency to Figure~\ref{fig:res_total_runtime}. 
For example, \FOLNFast{} and \OGeo{} are more efficient than \FOLNF{} and \AGeo{}, respectively, and the count-min sketch significantly reduces the runtime for large-domain data. 
Figure~\ref{fig:res_Sasy_runtime} also shows that ORShuffle is more efficient than WaksShuffle, which is consistent with their upper bounds on the runtime \cite{Sasy_CCS22,Sasy_CCS23}. 
Note that the online runtime (i.e., the runtime after obtaining $n$ input values) of WaksShuffle is smaller than that of ORShuffle~\cite{Sasy_CCS23}. 
However, we evaluate the \textit{total} runtime, in which case ORShuffle is more efficient. 

\subsection{Communication Efficiency}
\label{sub:communication_efficiency}
We also compared our algorithms with \FME{} in terms of communication costs. 
Note that \FME{} uses multiple encryption using a public key encryption scheme. 
To conduct a fair comparison with \FME{}, 
we followed \cite{Murakami_NDSS26} and used the ECIES to encrypt a message. 
Then, we evaluated the expected total number $C_{tot} \in \nnreals$ of bits sent from one party (i.e., a user, shuffler, or data collector) to another. 

\begin{figure}[t]
  \centering
  \includegraphics[width=0.99\linewidth]{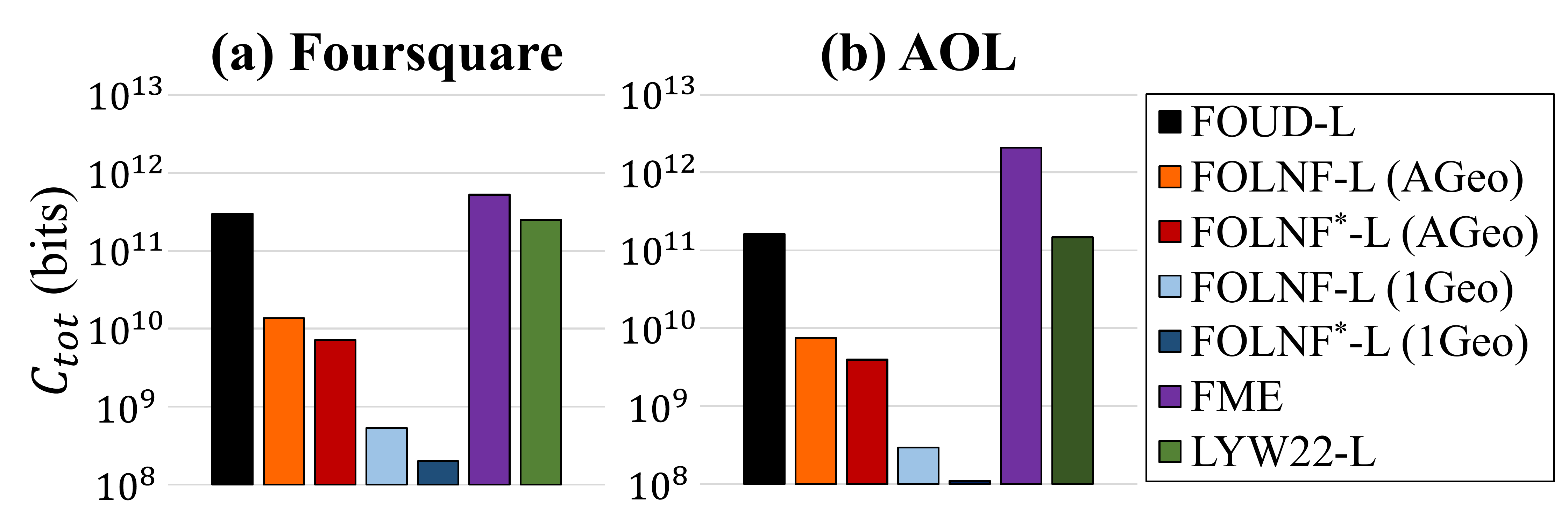}
  \caption{Total communication cost $C_{tot}$ ($\epsilon_E = 0.1$, $\epsilon_I = 1$).} 
  \label{fig:res_communication}
\end{figure}

Figure~\ref{fig:res_communication} shows $C_{tot}$ of each algorithm in \Foursquare{} and \AOL{} when $\epsilon_E = 0.1$ and $\epsilon_I = 1$. 
We observe that \FME{} is the least efficient. 
This stems from the fact that \FME{} uses multiple encryption and introduces two rounds of communication between the shuffler and the data collector. 
Our algorithms significantly reduce $C_{tot}$ of \FME{}; e.g., \FOLNFLarge{} and \FOLNFastLarge{} reduce $C_{tot}$ of \FME{} by two to four orders of magnitude. 
We also confirmed that our algorithms are much more efficient than \FME{} in other values of $\epsilon_E$ and $\epsilon_I$. 

\subsection{Changing $\tau$ and $b$}
\label{sub:tau_b}

Figures~\ref{fig:res_optimization_1geo} and \ref{fig:res_hash_range_Foursquare} show the MSE when varying $\tau$ in \OGeo{} and varying $b$ in \Foursquare{}. 
These figures show similar tendencies to Figures~\ref{fig:res_optimization} and \ref{fig:res_hash_range}; e.g., our optimization method works well, and the accuracy is close between $b=n$ and $b=10n$. 

\begin{figure}[t]
  \centering
  \includegraphics[width=0.99\linewidth]{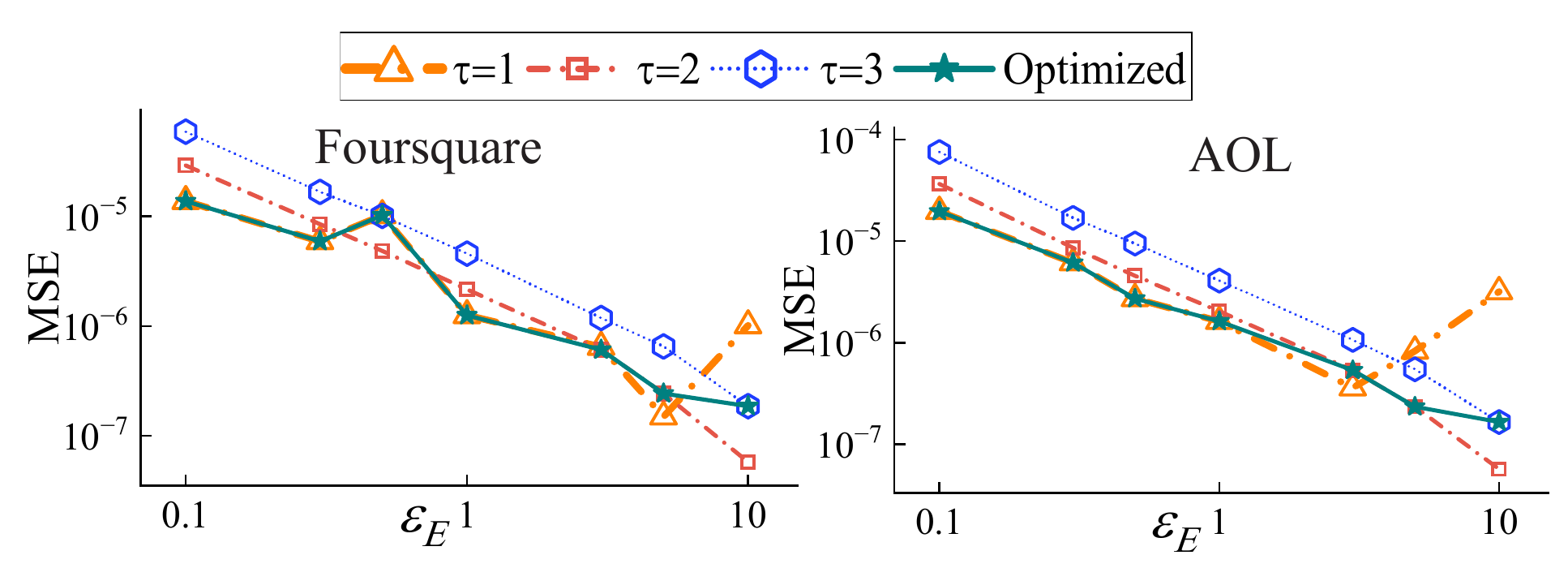}
  \caption{MSE for various values of $\tau$ 
  ($b=n$, \OGeo{}).} 
  \label{fig:res_optimization_1geo}
\end{figure}

\begin{figure}[t]
  \centering
  \includegraphics[width=0.95\linewidth]{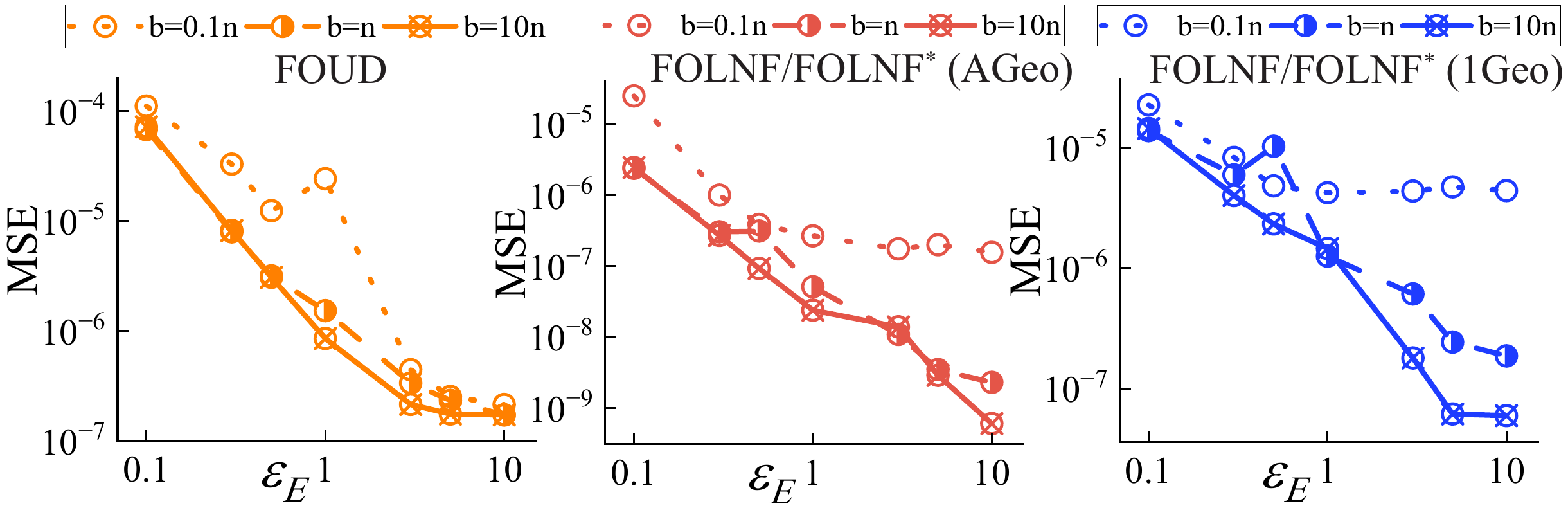}
  \caption{MSE vs. hash range $b$ ($\tau$: optimized, \Foursquare{}).} 
  \label{fig:res_hash_range_Foursquare}
\end{figure}
}

\arxiv{
\section{Proofs of Statements}
\label{sec:proofs}
\subsection{Proof of Theorem~\ref{thm:DP_FO_FODP}}

Consider two neighboring databases $\bmx$ and $\bmx'$ such that $x_i \neq x'_i$ and $x_j = x'_j$ for any $j \in [n]\setminus\{i\}$. 
Let $\bmx_{-i} = (x_1, \ldots, x_{i-1}, x_{i+1}, \ldots, x_n)$. 
Then, $\bmx_{-i} = \bmx'_{-i}$. 

The equation (\ref{eq:FO_equality}) states that the distribution of $(\calR^\calM(\bmx), \allowbreak \calR^\calI(\bmx))$ conditioned on $\calR(\bmx)=o$ can be simulated using the simulator $\calS$, i.e., for any $(o,(o_M,o_I)) \in \mathrm{Range}(\calR \times (\calR^\calM, \calR^\calI))$, 
\begin{align*}
&\Pr[(\calR^\calM(\bmx),\calR^\calI(\bmx)) = (o_M,o_I) | \calR(\bmx) = o] \nonumber\\
&= \Pr[\calS ((x_i)_{i\in [n] \setminus \Lambda}, (|x_i|)_{i \in \Lambda}) = (o_M,o_I) | \calR(\bmx) = o]. 
\end{align*}
This implies the following equation for any $(O,O_M,O_I) \subseteq \mathrm{Range}((\calR, \calR^\calM, \calR^\calI))$: 
\begin{align}
&\Pr[(\calR^\calM(\bmx),\calR^\calI(\bmx)) \in (O_M,O_I) | \calR(\bmx) \in O] \nonumber\\
&= \Pr[\calS ((x_i)_{i\in [n] \setminus \Lambda}, (|x_i|)_{i \in \Lambda}) \in (O_M,O_I) | \calR(\bmx) \in O]. 
\label{eq:FO_equality_O}
\end{align}

Assume that $\calR$ provides $(\epsilon,\delta)$-DP and is fully oblivious 
to a single secret. 
If $\Pr[\calR(\bmx) \in O] = 0$, (\ref{eq:FODP_inequality}) clearly holds. 
If $\Pr[\calR(\bmx') \in O] = 0$, (\ref{eq:FODP_inequality}) holds (as the left-hand side of (\ref{eq:FODP_inequality}) $\leq \Pr[\calR(\bmx) \in O] \leq \delta$ by (\ref{eq:DP_inequality})). 
If $\Pr[\calR(\bmx) \in O] \neq 0$ and $\Pr[\calR(\bmx') \in O] \neq 0$, we have 
\begin{align*}
&\Pr[(\calR(\bmx),\calR^\calM(\bmx),\calR^\calI(\bmx)) \in (O,O_M,O_I)] \\
&= \Pr[\calR(\bmx) \in O] \\ 
& \hspace{4mm} \cdot \Pr[(\calR^\calM(\bmx),\calR^\calI(\bmx)) \in (O_M,O_I) | \calR(\bmx) \in O] \\
&= \Pr[\calR(\bmx) \in O] \Pr[\calS (\bmx_{-i}, |x_i|) \in (O_M,O_I) | \calR(\bmx) \in O] \\
& \hspace{5mm} \text{(by (\ref{eq:FO_equality_O}) with $\Lambda = \{i\}$)} \\
&= \Pr[\calR(\bmx) \in O] \Pr[\calS (\bmx'_{-i}, |x_i|) \in (O_M,O_I) | \calR(\bmx') \in O] \\
& \hspace{5mm} \text{(as $\bmx_{-i} = \bmx'_{-i}$)} \\
&\leq e^\epsilon \Pr[\calR(\bmx') \in O] \\
& \hspace{4mm} \cdot \Pr[(\calR^\calM(\bmx'),\calR^\calI(\bmx')) \in (O_M,O_I) | \calR(\bmx') \in O] + \delta \\
& \hspace{5mm} \text{(by (\ref{eq:DP_inequality}) and (\ref{eq:FO_equality_O}) with $\Lambda = \{i\}$)} \\
& = e^\epsilon \Pr[(\calR(\bmx'),\calR^\calM(\bmx'),\calR^\calI(\bmx')) \in (O,O_M,O_I)] + \delta. 
\end{align*}
Thus, (\ref{eq:FODP_inequality}) always holds for any two neighboring databases $\bmx$ and $\bmx'$ and any $(O,O_M,O_I) \subseteq \mathrm{Range}((\calR, \calR^\calM, \calR^\calI))$, which means that $\calR$ provides $(\epsilon,\delta)$-FODP. 
\qed

\subsection{Proof of Lemma~\ref{lem:general_privacy}}

Random sampling (lines 1-6) and adding uniform dummies (lines 7-10) in $\calR_{\calD,\calD',\beta,\lambda}$ are fully oblivious to $\bmx$ and $r$ (i.e., memory and instruction traces do not depend on $\bmx$ and $r$), as $\texttt{ORAND\_REAL}$, $\texttt{OSWAP}$, and $\texttt{ORAND\_NATS}$ are fully oblivious (as described in Section~\ref{sub:FO}) and the numbers $n$ and $\lambda$ of iterations do not depend on $\bmx$ and $r$. 

$\texttt{DummyCountGeneration}$ (line 12) can be implemented via a CDF for $\calD$ or arithmetic operations and is fully oblivious to $z_i$ (see Appendix~\ref{sub:dummy-count_distribution} for details). 
$\texttt{BotCountGeneration}$ (line 13) consists of sampling a random value from $\calD'$ and arithmetic operations; e.g., the output is $\kappa_i = \kappa$ (resp.~$\kappa_i = z_i + \omega_i$, $\omega_i \sim \calD'$) in \FOLNF{} (resp.~\FOLNFast{}). 
As with sampling from $\calD$, sampling from $\calD'$ can be implemented via a CDF or arithmetic operations. 
Thus, $\texttt{BotCountGeneration}$ is also fully oblivious to $z_i$. 
The output $\kappa_i$ is also encrypted. 

It should be noted, however, that $\calR_{\calD,\calD',\beta,\lambda}$ allocates memory of size $\kappa_i$ for $(a_1,\ldots,a_{\kappa_i}) = (\bot,\ldots,\bot)$ (line 14). 
Thus, $\kappa_i$ is leaked from the size of the allocated memory. 
The subsequent $\texttt{OSWAP}$ (line 15-17) is fully oblivious to the dummy value $i$ and $z_i$. 
However, $\kappa_i$ is leaked from the instruction trace, as the process is repeated $\kappa_i$ times in this ``foreach'' loop. 
$\texttt{OSHUFFLE}$ (line 20) is fully oblivious to the shuffled values, as described in Section~\ref{sub:FO}. 

In summary, memory and instruction traces of $\calR_{\calD,\calD',\beta,\lambda}$ depend on $\bmx$ only through $(\kappa_1,\ldots,\kappa_d)$, which means that there exists a randomized post-processing algorithm $\phi$ such that (\ref{eq:lemma_post_processing}) holds for any $\bmx \in [d]^n$. 
\qed

\subsection{Proof of Lemma~\ref{lem:general_robustness}}
Intuitively, Lemma~\ref{lem:general_robustness} can be explained from the following two facts: (i) users do not add noise to their input values in $\calR_{\calD,\calD',\beta,\lambda}$, (ii) DP and FODP guarantee privacy even if the adversary obtains the input values of others. 

Formally, we can prove Lemma~\ref{lem:general_robustness} as follows. 
Let $\bmx = (x_1,\ldots,x_n)$ and $\bmx' = (x'_1,\ldots,x'_n)$ be neighboring databases such that 
$x_i \ne x'_i$ 
and $x_j = x'_j$ for $j\in[n]\setminus\{i\}$. 
Let $\Omega \subseteq [n]\setminus\{i\}$, and 
$\calR_\Omega(\bmx) = (\calR(\bmx), (x_k)_{k\in\Omega})$. 
Then, 
it suffices to show the following inequalities 
for any $O \subseteq \mathrm{Range}(\calR_\Omega)$ and any $(O,O_M,O_I) \subseteq \mathrm{Range}((\calR_\Omega, \calR^\calM, \calR^\calI))$: 
\begin{align}
&\Pr[\calR_\Omega(\bmx) \in O] \leq e^{\epsilon_E} \Pr[\calR_\Omega(\bmx') \in O] + \delta_E \label{eq:general_robustness_DP}\\
&\Pr[(\calR_\Omega(\bmx),\calR^\calM(\bmx),\calR^\calI(\bmx)) \in (O,O_M,O_I)] \nonumber\\
&\leq e^{\epsilon_I} \Pr[(\calR_\Omega(\bmx'),\calR^\calM(\bmx'),\calR^\calI(\bmx')) \in (O,O_M,O_I)] + \delta_I. \label{eq:general_robustness_FODP}
\end{align}
Since $\bmx$ and $\bmx'$ are fixed, the randomness in $\calR_\Omega$ arises solely from $\calR$. 
Thus, if (\ref{eq:DP_inequality}) and (\ref{eq:FODP_inequality}) hold, then (\ref{eq:general_robustness_DP}) and (\ref{eq:general_robustness_FODP}) also hold with the same privacy parameters $(\epsilon_E,\delta_E)$ and $(\epsilon_I,\delta_I)$. 
\qed

\subsection{Proof of Theorem~\ref{thm:general_accuracy}}
For any $i\in[d]$, the expectation of $\hf_i$ can be written as 
\begin{align*}
\E[\hf_i] 
= \textstyle{\frac{1}{\beta n}(\E[c_i] - \frac{\lambda}{d} - \mu)} = \textstyle{\frac{1}{\beta n}(\beta n f_i + \frac{\lambda}{d} + \mu - \frac{\lambda}{d} - \mu)} = f_i. 
\end{align*}
Thus, the estimate $\bmhf$ is unbiased, in which case the expected $l_2$ loss $\E[\sum_{i=1}^d (\hf_i - f_i)^2]$ is equal to the variance $\sum_{i=1}^d \V[\hf_i]$~\cite{mlpp}. 

For $i\in[d]$, let $y_i \in \nnints$ be the number of uniform dummies added to item $i$. 
Note that $\sum_{i=1}^d y_i = \lambda$. 
Then, by the law of total variance, and noting that $y_i$ and $z_i$ are independent, 
\begin{align*}
\V[c_i] 
&= \E[\V[c_i | y_i,z_i]] + \V[\E[c_i | y_i,z_i]] \\
&= \E[n f_i \beta (1 - \beta)] + \V[\beta n f_i + y_i + z_i] \\
&= n f_i \beta (1 - \beta) + \V[y_i] + \V[z_i] \\
&= \textstyle{n f_i \beta (1 - \beta) + \lambda \frac{1}{d} (1 - \frac{1}{d}) + \sigma^2}. 
\end{align*}
Thus, the expected $l_2$ loss can be written as follows: 
\begin{align*}
&\textstyle{\E[\sum_{i=1}^d (\hf_i - f_i)^2]} \\
&= \textstyle{\sum_{i=1}^d \V[\hf_i]}
= \textstyle{\frac{\sum_{i=1}^d \V[c_i]}{\beta^2 n^2}} 
= \textstyle{\frac{1-\beta}{\beta n} + \frac{\lambda(d-1)}{\beta^2 n^2 d} + \frac{\sigma^2 d}{\beta^2 n^2}}.
\end{align*}
\qed

\subsection{Proof of Theorem~\ref{thm:FOUD_FODP}}
\label{sub:thm:FOUD_FODP_proof}

\noindent{\textbf{DP.}}~~We 
prove 
DP of $\calR_\lambda^{\FOUD}$ 
based on the insight that $\calR_\lambda^{\FOUD}$ with $n=1$ is equivalent to the balls-into-bins mechanism in~\cite{Luo_CCS22} (with parameter $s=1$ and $p=0$ in their notations). 
Specifically, let $\bmx = (x_1,\ldots,x_n)$ and $\bmx' = (x'_1,\ldots,x'_n)$ be neighboring databases such that $x_i = k$, $x'_i = l$, $k \ne l$, and $x_j = x'_j$ for $j\in[n]\setminus\{i\}$. 
For $i\in[d]$, let $h_i$ (resp.~$h'_i$) $\in \nnints$ be the absolute frequency of item $i$ in $\bmx$ (resp.~$\bmx'$). 
Let $c_i \in \nnints$ be the absolute frequency of item $i$ in the shuffled values output by $\calR_\lambda^{\FOUD}$. 
Note that $\sum_{i=1}^d h_i = \sum_{i=1}^d h'_i = n$, $\sum_{i=1}^d c_i = n + \lambda$, $h_k = h'_k + 1$, and $h'_l = h_l + 1$. 
Let $\bmc = (c_1,\ldots,c_d)$. 
Let $\bar{\calR}_\lambda^{\FOUD}$ be a randomized algorithm with domain $[d]^n$ that outputs the histogram of the shuffled values output by $\calR_\lambda^{\FOUD}$. 
Then, 
\begin{align*}
\Pr[\bar{\calR}_\lambda^{\FOUD}(\bmx) = \bmc] 
&= \textstyle{\frac{\lambda !}{\prod_{i=1}^d (c_i - h_i)!} \frac{1}{d^\lambda}} \\
\Pr[\bar{\calR}_\lambda^{\FOUD}(\bmx') = \bmc] 
&= \textstyle{\frac{\lambda !}{\prod_{i=1}^d (c_i - h'_i)!} \frac{1}{d^\lambda}}, 
\end{align*}
and therefore, 
\begin{align*}
\textstyle{\frac{\Pr[\bar{\calR}_\lambda^{\FOUD}(\bmx) = \bmc] 
}{\Pr[\bar{\calR}_\lambda^{\FOUD}(\bmx') = \bmc] 
}} 
= \textstyle{\frac{c_k - h'_k}{c_l - h_l}}
= \textstyle{\frac{c_k - h_k + 1}{c_l - h_l}}.
\end{align*}
Assume that $\bmc$ is generated from $\bar{\calR}_\lambda^{\FOUD}(\bmx)$. Then, both $c_k - h_k$ and $c_l - h_l$ follow the binomial distribution $Bin(\lambda,\frac{1}{d})$. 
Thus, we consider random variables $X_1$ and $X_2$ such that $X_1, X_2 \sim Bin(\lambda,\frac{1}{d})$. 
By the multiplicative Chernoff bound~\cite{probability_computing}, it holds that for $\theta_1 \in \nnreals$ and $\theta_2 \in [0,1)$, 
\begin{align*}
\textstyle{\Pr[X_1 \geq \frac{(1+\theta_1)\lambda}{d}]} \leq \textstyle{e^{-\frac{\theta_1^2 \lambda}{(2+\theta_1)d}}},~ 
\textstyle{\Pr[X_2 \leq \frac{(1-\theta_2)\lambda}{d}]} \leq \textstyle{e^{-\frac{\theta_2^2 \lambda}{2d}}}. 
\end{align*}
In addition, when $X_1 < \frac{(1+\theta_1)\lambda}{d}$ and $X_2 > \frac{(1-\theta_2)\lambda}{d}$, we have 
\begin{align*}
\textstyle{\frac{X_1+1}{X_2}} < \textstyle{\frac{d + (1+\theta_1)\lambda}{(1-\theta_2)\lambda}} = e^{\epsilon}.
\end{align*}
Thus, by taking the union bound, 
\begin{align*}
\textstyle{\Pr\left[\frac{X_1+1}{X_2} \geq e^\epsilon\right]} 
&\leq \textstyle{\Pr[X_1 \geq \frac{(1+\theta_1)\lambda}{d}] + \Pr[X_2 \leq \frac{(1-\theta_2)\lambda}{d}]} \leq \delta.
\end{align*}
Therefore, we have 
\begin{align}
\textstyle{\Pr_{\bmc \sim \bar{\calR}_\lambda^{\FOUD}(\bmx)}\left[ \frac{\Pr[\bar{\calR}_\lambda^{\FOUD}(\bmx) = \bmc] 
}{\Pr[\bar{\calR}_\lambda^{\FOUD}(\bmx') = \bmc] 
} \geq e^{\epsilon} \right] \leq \delta}, 
\label{eq:FOUD_probabilistic_DP}
\end{align}
which implies 
$(\epsilon,\delta)$-DP~\cite{Balle_CRYPTO19,Ghazi_EUROCRYPT21}.

\smallskip{}
\noindent{\textbf{FODP.}}~~$\calR_\lambda^{\FOUD}$ sets $\kappa_i = 0$ for each item $i\in[d]$. 
Thus, the bot counts $(\kappa_1,\ldots,\kappa_d) = (0,\ldots,0)$ do not depend on $\bmx$. 
Then, by Lemma~\ref{lem:general_privacy}, memory and instruction traces do not depend on $\bmx$, which means that $\calR_\lambda^{\FOUD}$ is fully oblivious to a single secret. 
By Theorem~\ref{thm:DP_FO_FODP}, 
$\calR_\lambda^{\FOUD}$ also provides $(\epsilon,\delta)$-FODP. 

\smallskip{}
\noindent{\textbf{Robustness.}}~~The robustness of $\calR_\lambda^{\FOUD}$ against collusion with users is immediately derived from Lemma~\ref{lem:general_robustness}. 
\qed

\subsection{Proof of Theorem~\ref{thm:FOLNF_FODP}}

\noindent{\textbf{DP.}}~~After discarding $\bot$, the output of $\calR_{\calD,\beta}^{\FOLNF}$ is the same as that of the LNF protocol~\cite{Murakami_SP25}, except that $\calR_{\calD,\beta}^{\FOLNF}$ truncates 
$\calD$ at $\kappa$; i.e., the number $z_i$ of dummies is $z_i = \kappa$ if $F(\kappa) \leq r$, where $r \in [0,1)$ is a random value. 
Let $\calD^*$ be the dummy-count distribution over $\{0,1,\ldots,\kappa\}$ obtained by truncating $\calD$ at $\kappa$. 
The probability of $\kappa$ in $\calD^*$ is $1 - F(\kappa - 1) = \frac{\delta_0}{2}$. 
Therefore, if the binary input mechanism $\calM_{\calD,\beta}$ provides $(\frac{\epsilon}{2},\frac{\delta}{2})$-DP, then $\calM_{\calD^*,\beta}$ obtained by replacing $\calD$ in $\calM_{\calD,\beta}$ with $\calD^*$ provides $(\frac{\epsilon}{2},\frac{\delta + \delta_0}{2})$-DP (as the probability that DP is violated can be transformed into $\delta$ in DP~\cite{Kasivisiwanathan_JPC14}). 
The LNF protocol provides $(\epsilon,\delta)$-DP if the corresponding binary input mechanism with the same dummy-count distribution provides $(\frac{\epsilon}{2},\frac{\delta}{2})$-DP (Theorem 2 in~\cite{Murakami_SP25}). 
Thus, if $\calM_{\calD^*,\beta}$ provides $(\frac{\epsilon}{2},\frac{\delta + \delta_0}{2})$-DP, then $\calR_{\calD,\beta}^{\FOLNF}$ provides $(\epsilon,\delta + \delta_0)$-DP (as the dummy-count distribution is the same between the two). 

\smallskip{}
\noindent{\textbf{FODP.}}~~In $\calR_{\calD,\beta}^{\FOLNF}$, the bot counts $(\kappa_1,\ldots,\kappa_d) = (\kappa,\ldots,\kappa)$ do not depend on $\bmx$. 
Thus, $\calR_{\calD,\beta}^{\FOLNF}$ is fully oblivious to a single secret (by Lemma~\ref{lem:general_privacy}) and provides $(\epsilon,\delta + \delta_0)$-FODP (by Theorem~\ref{thm:DP_FO_FODP}). 

\smallskip{}
\noindent{\textbf{Robustness.}}~~The robustness is derived from Lemma~\ref{lem:general_robustness}. 
\qed

\subsection{Proof of Theorem~\ref{thm:FOLNFast_FODP}}
\label{sub:proof_thm_FONLFast_FODP}

\noindent{\textbf{DP.}}~~By 
Theorem 2 in~\cite{Murakami_SP25}, $\calR_{\calD,\calD',\beta}^{\FOLNFast}$ provides $(\epsilon,\delta)$-DP if $\calM_{\calD,\beta}$ provides $(\frac{\epsilon}{2},\frac{\delta}{2})$-DP.

\smallskip{}
\noindent{\textbf{FODP.}}~~Below, we show FODP of $\calR_{\calD,\calD',\beta}^{\FOLNFast}$ by directly showing the inequality (\ref{eq:FODP_inequality}). 
We denote the bot-count algorithm of $\calR_{\calD,\calD',\beta}^{\FOLNFast}$ by $\calR_{\calD,\calD',\beta}^{\FOLNFast \bot}$. 
Let $\bmx = (x_1,\ldots,x_n)$ and $\bmx' = (x'_1,\ldots,x'_n)$ be neighboring databases such that 
$x_i \ne x'_i$ 
and $x_j = x'_j$ for $j\in[n]\setminus\{i\}$. 
By Lemma~\ref{lem:general_privacy}, showing (\ref{eq:FODP_inequality}) with $(\epsilon,\delta) = (\epsilon^*,\delta^*)$ is reduced to showing the following inequality for any $(O,O_\bot) \subseteq \mathrm{Range}((\calR_{\calD,\calD',\beta}^{\FOLNFast}, \calR_{\calD,\calD',\beta}^{\FOLNFast \bot}))$: 
\begin{align}
&\Pr[(\calR_{\calD,\calD',\beta}^{\FOLNFast}(\bmx),\calR_{\calD,\calD',\beta}^{\FOLNFast \bot}(\bmx)) \in (O,O_\bot)] \nonumber\\
&\leq e^{\epsilon^*} \Pr[(\calR_{\calD,\calD',\beta}^{\FOLNFast}(\bmx'),\calR_{\calD,\calD',\beta}^{\FOLNFast \bot}(\bmx')) \in (O,O_\bot)] + \delta^*. 
\label{eq:FODP_inequality_FOLNFast}
\end{align}
Meanwhile, outputting $\calM^*_{\calD,\calD',\beta}(x) = (\calM_{\calD,\beta}(x), \calM^-_{\calD',\beta}(x))$ is equivalent to outputting 
\begin{align*}
(\calM_{\calD,\beta}(x), \calM_{\calD',\beta}(x) + \calM^-_{\calD',\beta}(x)) 
= (\alpha x + z, z + \omega) 
\triangleq 
\Phi. 
\end{align*}

Notice that 
$\Phi$ is a binary input version of the output of $(\calR_{\calD,\calD',\beta}^{\FOLNFast},\calR_{\calD,\calD',\beta}^{\FOLNFast \bot})$. 
This can be explained as follows. 
Given the database $\bmx$, let 
$(c_1,\ldots,c_d) \in (\nnints)^d$ 
be a histogram of shuffled values output by $\calR_{\calD,\calD',\beta}^{\FOLNFast}$. 
The output of $\calR_{\calD,\calD',\beta}^{\FOLNFast}$ leaks no information other than $(c_1,\ldots,c_d)$, as it is randomly shuffled. 
Let 
$\bmx_i$ (resp.~$\bmx'_i$) $\in \{0,1\}^d$ be a vector whose $x_i$-th (resp.~$x'_i$-th) element is $1$ and other elements are $0$. 
Then, 
\begin{align*}
(c_1,\ldots,c_d) = \textstyle{(\sum_{j=1}^n \alpha_j \bmx_j) + \bmz}, 
\end{align*}
where $\alpha_j \sim Ber(\beta)$ and $\bmz = (z_1,\ldots,z_d)$. 
Since $x_j = x'_j$ for $j\in[n]\setminus\{i\}$, $\sum_{j \ne i} \alpha_j \bmx_j$ follows the same distribution as $\sum_{j \ne i} \alpha_j \bmx'_j$. 
In addition, the output of $\calR_{\calD,\calD',\beta}^{\FOLNFast \bot}$ is 
$(z_1+\omega_1,\ldots,z_d+\omega_d)$. 
Thus, 
showing (\ref{eq:FODP_inequality_FOLNFast}) is reduced to showing that for any set $O_\Phi \subseteq (\nnints \times \nnints)^d$, 
\begin{align}
\Pr[(\Phi_1,\ldots,\Phi_d) \in O_\Phi] \leq e^{\epsilon^*} \Pr[(\Phi'_1,\ldots,\Phi'_d) \in O_\Phi]  + \delta^* 
\label{eq:FODP_inequality_FOLNFast2}
\end{align}
holds, where $\Phi_j = (\alpha_i \bmx_i[j] + z_j, z_j + \omega_j)$, $\Phi'_j = (\alpha_i \bmx'_i[j] + z_j, z_j + \omega_j)$, and $\bmx_i[j]$ (resp.~$\bmx'_i[j]$) is the $j$-th element of $\bmx_i$ (resp.~$\bmx'_i$). 
Recall that  $\alpha, \alpha_i \sim Ber(\beta)$, $z, z_j \sim \calD$, and $\omega, \omega_j \sim \calD'$. 
Therefore, $\Phi$ is a binary input version of $(\Phi_1,\ldots,\Phi_d)$. 

Based on this, we show that (\ref{eq:FODP_inequality_FOLNFast2}) holds if $\calM^*_{\calD,\calD',\beta}$ provides $(\frac{\epsilon^*}{2},\frac{\delta^*}{2})$-DP. 
If $j \notin \{x_i,x'_i\}$, then $\Phi_j$ follows the same distribution as $\Phi'_j$. 
If $j = x_i$, then $\Phi_j$ (resp.~$\Phi'_j$) follows the same distribution as $\calM^*_{\calD,\calD',\beta}(1)$ (resp.~$\calM^*_{\calD,\calD',\beta}(0)$). 
If $j = x'_i$, then $\Phi_j$ (resp.~$\Phi'_j$) follows the same distribution as $\calM^*_{\calD,\calD',\beta}(0)$ (resp.~$\calM^*_{\calD,\calD',\beta}(1)$). 
Thus, by group privacy~\cite{DP}, if $\calM^*_{\calD,\calD',\beta}$ provides $(\frac{\epsilon^*}{2},\frac{\delta^*}{2})$-DP, then (\ref{eq:FODP_inequality_FOLNFast2}) holds for any set $O_\Phi \subseteq (\nnints \times \nnints)^d$, which means that $\calR_{\calD,\calD',\beta}^{\FOLNFast}$ provides $(\epsilon^*,\delta^*)$-FODP. 

\smallskip{}
\noindent{\textbf{Robustness.}}~~Immediately derived from Lemma~\ref{lem:general_robustness}. 
\qed

\subsection{Proof of Theorem~\ref{thm:joint_ageo_DP}}
Let $x=0$ and $x'=1$. 
Then, by (\ref{eq:M_M_minus}) with $\alpha \sim Ber(\beta)$, $z \sim \AGeo(\nu, q_l, q_r)$, and $\omega \sim \AGeo(\nu', q'_l, q'_r)$, we have 
\begin{align*}
&\Pr[\calM^A_{\calD,\beta}(x) = k] = 
\begin{cases}
\frac{1}{\eta} q_l^{\nu - k}   &   \text{(if $k = 0, 1, \ldots, \nu$)}\\
\frac{1}{\eta} q_r^{k - \nu}   &   \text{(if $k = \nu, \nu+1, \ldots $)}
\end{cases}
\end{align*}
\begin{align*}
&\Pr[\calM^A_{\calD,\beta}(x') = k] \\
&= 
\begin{cases}
\frac{1-\beta}{\eta} q_l^{\nu} &    \text{(if $k = 0$)}\\
\frac{1 - \beta}{\eta} q_l^{\nu - k} + \frac{\beta}{\eta} q_l^{\nu - k + 1}   &   \text{(if $k = 1, \ldots, \nu$)}\\
\frac{1 - \beta}{\eta} q_r^{k - \nu} + \frac{\beta}{\eta} q_r^{k - \nu - 1}   &   \text{(if $k = \nu+1, \nu+2, \ldots $)}
\end{cases}\\
&\Pr[\calM^{A-}_{\calD',\beta}(x) = k'] = 
\begin{cases}
\frac{1}{\eta'} {q'}_l^{\nu' - k'}   &   \text{(if $k' = 0, 1, \ldots, \nu'$)}\\
\frac{1}{\eta'} {q'}_r^{k' - \nu'}   &   \text{(if $k' = \nu', \nu'+1, \ldots $)}
\end{cases}\\
&\Pr[\calM^{A-}_{\calD',\beta}(x') = k'] \\
&= 
\begin{cases}
\frac{\beta}{\eta'} {q'}_l^{\nu'} &    \text{(if $k' = -1$)}\\
\frac{1 - \beta}{\eta'} {q'}_l^{\nu' - k'} + \frac{\beta}{\eta'} {q'}_l^{\nu' - k' - 1}   &   \text{(if $k' = 0, \ldots, \nu'-1$)}\\
\frac{1 - \beta}{\eta'} {q'}_r^{k' - \nu'} + \frac{\beta}{\eta'} {q'}_r^{k' - \nu' + 1}   &   \text{(if $k' = \nu', \nu'+1, \ldots $)}.
\end{cases}
\end{align*}
Let $P_1 = \Pr[(\calM^A_{\calD,\beta}(x), \calM^{A-}_{\calD',\beta}(x)) = (k,k')]$ and $P_2 = \Pr[(\calM^A_{\calD,\beta}(x'), \calM^{A-}_{\calD',\beta}(x')) = (k,k')]$. 
Below, we calculate $P_1$ and $P_2$ in three cases: (i) when $k \geq 1$ and $k' \geq 0$, (ii) when $k = 0$ and $k' \geq 0$, and (iii) when $k' = -1$. 

\smallskip{}
\noindent{\textbf{Case 1: $k \geq 1$ and $k' \geq 0$.}}~~Noting that $\calM^A_{\calD,\beta}$ and $\calM^{A-}_{\calD,\beta}$ share $\alpha$ generated from $Ber(\beta)$ in (\ref{eq:M_M_minus}), we have 
\begin{align}
&P_1 = 
\begin{cases}
\frac{1}{\eta\eta'} q_l^{\nu - k} {q'}_l^{\nu' - k'}  &   \text{(if $1 \leq k \leq \nu$ and $1 \leq k' < \nu'$)}\\
\frac{1}{\eta\eta'} q_l^{\nu - k} {q'}_r^{k' - \nu'}  &   \text{(if $1 \leq k \leq \nu$ and $k' \geq \nu'$)}\\
\frac{1}{\eta\eta'} q_r^{k - \nu} {q'}_l^{\nu' - k'}  &   \text{(if $k > \nu$ and $1 \leq k' < \nu'$)}\\
\frac{1}{\eta\eta'} q_r^{k - \nu} {q'}_r^{k' - \nu'}  &   \text{(if $k > \nu$ and $k' \geq \nu'$)}
\end{cases}\nonumber
\end{align}
\begin{align}
&P_2 = 
\begin{cases}
\frac{1}{\eta\eta'} \{(1-\beta) q_l^{\nu - k} {q'}_l^{\nu' - k'} + \beta q_l^{\nu - k + 1} {q'}_l^{\nu' - k' - 1}\}\\
\hspace{30mm} \text{(if $1 \leq k \leq \nu$ and $1 \leq k' < \nu'$)}\\
\frac{1}{\eta\eta'} \{(1-\beta) q_l^{\nu - k} {q'}_r^{k' - \nu'} + \beta q_l^{\nu - k + 1} {q'}_r^{k' - \nu' + 1}\}\\
\hspace{30mm} \text{(if $1 \leq k \leq \nu$ and $k' \geq \nu'$)}\\
\frac{1}{\eta\eta'} \{(1-\beta) q_r^{k - \nu} {q'}_l^{\nu' - k'} + \beta q_r^{k - \nu - 1} {q'}_l^{\nu' - k' - 1}\}\\
\hspace{30mm} \text{(if $k > \nu$ and $1 \leq k' < \nu'$)}\\
\frac{1}{\eta\eta'} \{(1-\beta) q_r^{k - \nu} {q'}_r^{k' - \nu'} + \beta q_r^{k - \nu - 1} {q'}_r^{\nu' - k' + 1}\}\\
\hspace{30mm} \text{(if $k > \nu$ and $k' \geq \nu'$)}.
\end{cases}\nonumber
\end{align}
Therefore, 
\begin{align}
\frac{P_2}{P_1} = 
\begin{cases}
(1-\beta) + \beta q_l \frac{1}{{q'}_l}  &   \text{(if $1 \leq k \leq \nu$ and $1 \leq k' < \nu'$)}\\
(1-\beta) + \beta q_l {q'}_r &   \text{(if $1 \leq k \leq \nu$ and $k' \geq \nu'$)}\\
(1-\beta) + \beta \frac{1}{q_r} \frac{1}{{q'}_l}   &   \text{(if $k > \nu$ and $1 \leq k' < \nu'$)}\\
(1-\beta) + \beta \frac{1}{q_r} {q'}_r  &   \text{(if $k > \nu$ and $k' \geq \nu'$)}.
\end{cases}
\label{eq:P_2_P_1}
\end{align}
By Definition~\ref{def:joint_ageo}, 
\begin{align*}
\textstyle{q_l \frac{1}{{q'}_l} = q_l q_r \frac{1}{R(\epsilon^*)},~ \frac{1}{q_r} {q'}_r = \frac{1}{q_l q_r} L(\epsilon^*)}, 
\end{align*}
and $q_l q_r = L(\epsilon) R(\epsilon)$. 
Since $\epsilon^* \geq \epsilon$, we have $L(\epsilon) R(\epsilon) = \frac{e^{-\epsilon/2}-1+\beta}{e^{\epsilon/2}-1+\beta} \geq \frac{e^{-\epsilon^*/2}-1+\beta}{e^{\epsilon^*/2}-1+\beta} = L(\epsilon^*) R(\epsilon^*)$. 
Thus, $q_l q_r \allowbreak \in [L(\epsilon^*) R(\epsilon^*), 1]$, which means that $q_l \frac{1}{{q'}_l} \in [L(\epsilon^*), \frac{1}{R(\epsilon^*)}]$ and $\frac{1}{q_r} {q'}_r \in [L(\epsilon^*), \frac{1}{R(\epsilon^*)}]$. 
Therefore, we have $(1-\beta) + \beta q_l \frac{1}{{q'}_l} \in [e^{-\epsilon^*/2}, e^{\epsilon^*/2}]$ and $(1-\beta) + \beta \frac{1}{q_r} {q'}_r \in [e^{-\epsilon^*/2}, e^{\epsilon^*/2}]$. 
In addition, $(1-\beta) + \beta q_l {q'}_r = e^{-\epsilon^*/2}$ and 
$(1-\beta) + \beta \frac{1}{q_r} \frac{1}{{q'}_l} = e^{\epsilon^*/2}$. 
Thus, by (\ref{eq:P_2_P_1}), we have $\frac{P_2}{P_1} \in [e^{-\epsilon^*/2}, e^{\epsilon^*/2}]$. 

\smallskip{}
\noindent{\textbf{Case 2: $k = 0$ and $k' \geq 0$.}}~~In this case, 
$P_1$ and $P_2$ are written as $P_1 = \frac{1}{\eta\eta'} q_l^\nu {r'}^{\nu'-k'}$ and $P_2 = \frac{1-\beta}{\eta\eta'} q_l^\nu {r'}^{\nu'-k'}$, respectively, 
where 
\begin{align*}
{r'}^{\nu'-k'} = 
\begin{cases}
{q'}_l^{\nu' - k'}   &   \text{(if $k' = 0, 1, \ldots, \nu'-1$)}\\
{q'}_r^{k' - \nu'}   &   \text{(if $k' = \nu', \nu'+1, \ldots $)}.
\end{cases}
\end{align*}
Thus, we have 
\begin{align}
\textstyle{P_1 \leq e^{\epsilon^*/2} P_2 + \delta_{1,k'},~ P_2 \leq P_1}, 
\label{eq:P_1_P_2_case2}
\end{align}
where $\delta_{1,k'} = \frac{1}{\eta\eta'} q_l^\nu {r'}^{\nu'-k'} (1 - e^{\epsilon^*/2} (1-\beta))$. 

\smallskip{}
\noindent{\textbf{Case 3: $k' = -1$.}}~~In this case, $\alpha = 1$. 
Thus, $P_1 = 0$ and $P_2 = \frac{\beta}{\eta\eta'} r^{\nu - k} {q'}_l^{\nu'}$,
where 
\begin{align*}
r^{\nu - k} = 
\begin{cases}
0                   &   \text{(if $k = 0$)}\\
q_l^{\nu - k + 1}   &   \text{(if $k = 1, \ldots, \nu$)}\\
q_r^{k - \nu - 1}   &   \text{(if $k = \nu+1, \nu+2, \ldots $)}.
\end{cases}
\end{align*}
Thus, we have 
\begin{align}
P_1 \leq P_2,~ P_2 \leq e^{\epsilon^*/2} P_1 + \delta_{2,k}, 
\label{eq:P_1_P_2_case3}
\end{align}
where $\delta_{2,k} = \frac{1}{\eta\eta'} \beta r^{\nu - k} {q'}_l^{\nu'}$. 

\smallskip{}
\noindent{\textbf{Putting Together.}}~~For $K \subseteq \mathrm{Range}(\calM^A_{\calD,\beta})$ and $K' \subseteq \mathrm{Range}\allowbreak(\calM^{A-}_{\calD,\beta})$, let $P_1^{K,K'} = \Pr[(\calM^A_{\calD,\beta}(x), \calM^{A-}_{\calD',\beta}(x)) \in (K,K')]$ and $P_2^{K,K'} = \Pr[(\calM^A_{\calD,\beta}(x'), \calM^{A-}_{\calD',\beta}(x')) \in (K,K')]$. 
Let $\delta_1^{K,K'}, \delta_2^{K,K'} \in [0,1]$ be the minimum values satisfying 
\begin{align*}
P_1^{K,K'} \leq e^{\epsilon^*/2} P_2^{K,K'} + \delta_1^{K,K'},~ 
P_2^{K,K'} \leq e^{\epsilon^*/2} P_1^{K,K'} + \delta_2^{K,K'}. 
\end{align*}
Then, by (\ref{eq:P_1_P_2_case2}), (\ref{eq:P_1_P_2_case3}), and $\epsilon \leq \epsilon^*$, 
\begin{align*}
&\textstyle{\delta_1^{K,K'} \leq \sum_{k' \geq 0} \delta_{1,k'} = \frac{1}{\eta} q_l^\nu (1 - e^{\epsilon^*/2} (1-\beta)) \leq \frac{\delta}{2}}\\
&\textstyle{\delta_2^{K,K'} \leq \sum_{k \geq 0} \delta_{2,k} = \frac{1}{\eta'} \beta {q'}_l^{\nu'}}. 
\end{align*}
Therefore, $\calM^{A*}_{\calD,\calD',\beta} = (\calM^A_{\calD,\beta}, \calM^{A-}_{\calD',\beta})$ provides $(\frac{\epsilon^*}{2},\frac{\delta^*}{2})$-DP, where 
$\delta^* = \max\{\delta, \frac{2}{\eta'} \beta {q'}_l^{\nu'} \}$. 
In addition, $\calM^A_{\calD,\beta}$ provides $(\frac{\epsilon}{2},\frac{\delta}{2})$-DP, as explained in Section~\ref{sub:FOLNF_fixed}. 
\qed

\subsection{Proof of Theorem~\ref{thm:Proposal_runtime}}
Since the size of each input value $x_i$ ($i\in[n]$) is $O(\log d)$, the runtime of $\texttt{ORAND\_REAL}$ and $\texttt{OSWAP}$ can be expressed as $O(1)$ and $O(\log d)$, respectively. 
The runtime of $\texttt{ORAND\_NATS}(d)$ is $O(\log d)$, as it generates a $\lceil \log_2 d \rceil$-bit binary integer. 
The runtime of $\texttt{DummyCountGeneration}$ (line 12 in Algorithm~\ref{alg:proposal}) is $O(1)$ when it is implemented via arithmetic operations (see Appendix~\ref{sub:dummy-count_distribution}). 
Similarly, the runtime of $\texttt{BotCountGeneration}$ (line 13) is $O(1)$. 

Therefore, the runtime of random sampling (lines 1-6) and dummy data addition (lines 7-19) in $\calR_{\calD,\calD',\beta,\lambda}$ can be expressed as $O(n \log d)$ and $O((\lambda + \sum_{i=1}^d \kappa_i) \log d)$, respectively. 
Then, the runtime of $\calR_{\calD,\calD',\beta,\lambda}$ is $O(\bn \log d + \rho)$, where $\bn = n + \lambda + \sum_{i=1}^d \kappa_i$, and 
$\rho$ is the runtime of $\texttt{OSHUFFLE}$. 
\qed

\subsection{Proof of Theorem~\ref{thm:Proposal_large_accuracy}}
Let $X_{t,i}$ (resp.~$X'_{t,i}$) be a random variable that represents the number of users who have (resp.~do not have) item $i$ and increase the count $c_{t,h_t(i)}$. 
Note that $X'_{t,i}$ is the increase in $c_{t,h_t(i)}$ due to hash collisions. 
Each input value is sampled with probability $\beta$, and the hash collision occurs with probability $\frac{1}{b}$. 
Thus, $X_{t,i} \sim Bin(nf_i,\beta)$ and $X'_{t,i} \sim Bin(n(1-f_i), \frac{\beta}{b})$. 
For $t\in[\tau]$ and $j\in[b]$, let $\lambda_{t,j}$ (resp.~$z_{t,j}$) $\in \nnints$ be the number of uniform dummies (resp.~dummies based on $\calD$) for the $j$-th hash value in the $t$-th hash function.  
Then, 
\begin{align}
\textstyle{c_{t,h_t(i)} = X_{t,i} + X'_{t,i} + \lambda_{t,h_t(i)} + z_{t,h_t(i)}}. 
\label{eq:proposal_large_c_thti}
\end{align}
For $\gamma > 0$, 
\begin{align}
\textstyle{\Pr[|\hf_i - f_i| > \gamma] = \Pr[\hf_i > f_i + \gamma] + \Pr[\hf_i < f_i - \gamma]}. 
\label{eq:proposal_large_Pr_hf_i_fi}
\end{align}
By (\ref{eq:hf_i_count_min}), (\ref{eq:proposal_large_c_thti}), $\lambda_{t,h_t(i)} \sim Bin(\lambda,\frac{1}{b})$, and $z_{t,h_t(i)} \sim \calD$, we have 
\begin{align}
\Pr[\hf_i > f_i + \gamma] 
&= \textstyle{\Pr[\forall t\in[\tau]: c_{t,h_t(i)} - \frac{\lambda}{b} - \mu > n\beta(f_i + \gamma)]} \nonumber\\
&= \textstyle{(\Pr[X'_{t,i} + \xi > \xi_{avg} + n\beta\gamma])^\tau}, 
\label{eq:proposal_large_Pr_hf_i_fi2}
\end{align}
where $\xi \sim Bin(nf_i, \beta) + Bin(\lambda, \frac{1}{b}) + \calD$ and $\xi_{avg} = nf_i\beta+\frac{\lambda}{b}+\mu$. 
Note that $X'_{t,i} + \xi$ exceeds $\xi_{avg} + n\beta\gamma$ only if 
$X'_{t,i} > \frac{n\beta\gamma}{2}$ or 
$\xi > \xi_{avg} + \frac{n\beta\gamma}{2}$. 
Thus, by the union bound, 
\begin{align}
&\Pr[X'_{t,i} + \xi > \xi_{avg} + n\beta\gamma] \nonumber\\
&\leq \textstyle{\Pr[X'_{t,i} > \frac{n\beta\gamma}{2}] + \Pr[\xi > \xi_{avg} + \frac{n\beta\gamma}{2}]}. 
\label{eq:proposal_large_Pr_xi_X'ti}
\end{align}
Intuitively, (\ref{eq:proposal_large_Pr_hf_i_fi2}) and (\ref{eq:proposal_large_Pr_xi_X'ti}) decompose 
$\Pr[\hf_i > f_i + \gamma]$ in (\ref{eq:proposal_large_Pr_hf_i_fi})
into two factors: hash collision (the first term in (\ref{eq:proposal_large_Pr_xi_X'ti})) and DP noise (the second term in (\ref{eq:proposal_large_Pr_xi_X'ti})). 
We bound the first term in (\ref{eq:proposal_large_Pr_xi_X'ti}) using Markov's inequality~\cite{probability_computing}: 
\begin{align}
\textstyle{\Pr[X'_{t,i} > \frac{n\beta\gamma}{2}]} 
&\leq \textstyle{\E[X'_{t,i}]/(\frac{n\beta\gamma}{2})} ~~~~ \text{(by Markov's inequality)} \nonumber\\
&= \textstyle{(\frac{n(1-f_i)\beta}{b})/(\frac{n\beta\gamma}{2})} 
\leq \textstyle{\frac{2}{b\gamma}}.
\label{eq:proposal_large_Markov}
\end{align}
$\Pr[\hf_i < f_i - \gamma]$ in (\ref{eq:proposal_large_Pr_hf_i_fi}) is caused by DP noise and 
can be written as follows: 
\begin{align}
\Pr[\hf_i < f_i - \gamma] 
&= \textstyle{\Pr[\exists t\in[\tau]: c_{t,h_t(i)} - \frac{\lambda}{b} - \mu < n\beta(f_i - \gamma)]} \nonumber\\
&\leq \textstyle{\tau \cdot \Pr[\xi < \xi_{avg} - n\beta\gamma]} \nonumber\\
&\hspace{4mm}\text{(by the union bound and $X'_{t,i} \geq 0$)}.
\label{eq:proposal_large_Pr_hf_i_fi3}
\end{align}
By (\ref{eq:proposal_large_Pr_hf_i_fi}), (\ref{eq:proposal_large_Pr_hf_i_fi2}), (\ref{eq:proposal_large_Pr_xi_X'ti}), (\ref{eq:proposal_large_Markov}), and (\ref{eq:proposal_large_Pr_hf_i_fi3}), we have (\ref{eq:Proposal_large_accuracy}). 
\qed

\subsection{\colorB{Proof of Theorem~\ref{thm:G_MGA}}}
Since the proof of Theorem~\ref{thm:G_MGA} is exactly the same as the proof of Theorem 3 in~\cite{Murakami_SP25}, we omit the proof; see \cite{Murakami_SP25}. 

\subsection{\colorB{Proof of Theorem~\ref{thm:FOUD_FODP_Wang}}}
$\calR_\lambda^{\FOUD}$ outputs the same data as the UD protocol without LDP noise. 
Thus, $(\epsilon,\delta)$-DP of $\calR_\lambda^{\FOUD}$ can be immediately derived from 
\cite{Wang_PVLDB20} (Theorem 1 in \cite{Wang_PVLDB20} with $Bin(n,p)$ replaced with $Bin(\lambda,\frac{1}{d})$). 
Note that $\epsilon \leq 1$ is necessary, as \cite{Wang_PVLDB20} is based on \cite{Balle_CRYPTO19}, which assumes $\epsilon \leq 1$.
\qed

\subsection{\colorB{Proof of Theorem~\ref{thm:general_large_privacy_robustness}}}
$\calR_{\calD,\calD',\beta,\lambda}^{\Largesf}$ uses $\calR_{\calD,\calD',\beta,\lambda}$ for $\tau$ times. 
Thus, DP and FODP of $\calR_{\calD,\calD',\beta,\lambda}^{\Largesf}$ follow immediately from the optimal composition theorem (Theorem 3.3 in \cite{Kairouz_ICML15}). 
The robustness of $\calR_{\calD,\calD',\beta,\lambda}^{\Largesf}$ is derived from Lemma~\ref{lem:general_robustness}. 
\qed

\subsection{\colorB{Proof of Corollary~\ref{cor:Proposal_large_accuracy_FOUD}}}
$\calR_\lambda^{\FOUDLarge}$ sets $\beta=1$ and $z_i = \kappa_i = 0$ for $i\in[d]$. 
Thus, $\xi \sim nf_i + Bin(\lambda, \frac{1}{b})$ and $\xi_{avg} = nf_i+\frac{\lambda}{b}$. 
Let $\xi_0 \sim Bin(\lambda, \frac{1}{b})$. 
Then, by the additive Chernoff bound~\cite{Arratia_bMB89}, 
\begin{align*}
\textstyle{\Pr[\xi > \xi_{avg} + \frac{n\gamma}{2}]} 
&= \textstyle{\Pr[\frac{\xi_0}{\lambda} > \frac{1}{b} + \frac{n\gamma}{2\lambda}]} 
\leq d_1 \\
\textstyle{\Pr[\xi < \xi_{avg} - n\gamma]} 
&= \textstyle{\Pr[\frac{\xi_0}{\lambda} < \frac{1}{b} - \frac{n\gamma}{\lambda}]} 
\leq d_2. 
\end{align*}
(\ref{eq:Proposal_large_accuracy_FOUD}) is derived from these inequalities and (\ref{eq:Proposal_large_accuracy}).
\qed

\subsection{\colorB{Proof of Corollary~\ref{cor:Proposal_large_accuracy_FOLNF_AGeo}}}
In this case, $\beta=1$, $\lambda=0$, and $\calD = \AGeo(\nu, q_l, q_r)$. 
Thus, $\xi \sim nf_i + \AGeo(\nu, q_l, q_r)$ and $\xi_{avg} = nf_i+\mu$. 
Let $\xi_0 \sim \AGeo(\nu, q_l, q_r)$. 
Then, we can calculate the exact values of $\Pr[\xi > \xi_{avg} + \frac{n\gamma}{2}]$ and $\Pr[\xi < \xi_{avg} - n\gamma]$: 
\begin{align*}
\textstyle{\Pr[\xi > \xi_{avg} + \frac{n\gamma}{2}] = \Pr[\xi_0 > \mu + \frac{n\gamma}{2}] = \frac{1}{\eta}\sum_{k=a_r}^\infty q_r^{k-\nu}} = d_3 \\
\textstyle{\Pr[\xi < \xi_{avg} - n\gamma] = \Pr[\xi_0 < \mu - n\gamma] = \frac{1}{\eta}\sum_{k=0}^{a_l} q_l^{\nu-k}} = d_4, 
\end{align*}
from which (\ref{eq:Proposal_large_accuracy_FOLNF}) is derived.
\qed

\subsection{\colorB{Proof of Corollary~\ref{cor:Proposal_large_accuracy_FOLNF_Bin}}}
In this case, $\xi \sim nf_i + Bin(n, \varphi)$ and $\xi_{avg} = nf_i+n\varphi$. 
Let $\xi_0 \sim Bin(n, \varphi)$. 
By the Chernoff bound, 
\begin{align*}
\textstyle{\Pr[\xi > \xi_{avg} + \frac{n\gamma}{2}]} 
&= \textstyle{\Pr[\frac{\xi_0}{n} > \varphi + \frac{\gamma}{2}]} 
\leq d_5 \\
\textstyle{\Pr[\xi < \xi_{avg} - n\gamma]} 
&= \textstyle{\Pr[\frac{\xi_0}{n} < \varphi - \gamma]} 
\leq d_6. 
\end{align*}
(\ref{eq:Proposal_large_accuracy_FOLNF_Bin}) follows from these inequalities and (\ref{eq:Proposal_large_accuracy}).
\qed

\subsection{\colorB{Proof of Theorem~\ref{thm:additive_error_Ghazi}}}
Immediately derived from Lemma~4.8 (with $k=1$) in \cite{Ghazi_EUROCRYPT21}. 
\qed
}

\section{Addendum to the Runtime}
\label{sec:addendum}
After this paper was published at USENIX Security 2026, we found an error in the implementation of ORShuffle used in Section~\ref{sec:experiments}. 
Therefore, we fixed the error and re-evaluated the runtime of our algorithms (i.e., Figures~\ref{fig:res_each_runtime}, \ref{fig:res_total_runtime}, and \ref{fig:res_histogram}) using the updated code. 
We have published the updated code on the following website: \\\\
\url{https://doi.org/10.5281/zenodo.22119784}\\ 

\begin{figure}[p]
  \centering
  \includegraphics[width=0.99\linewidth]{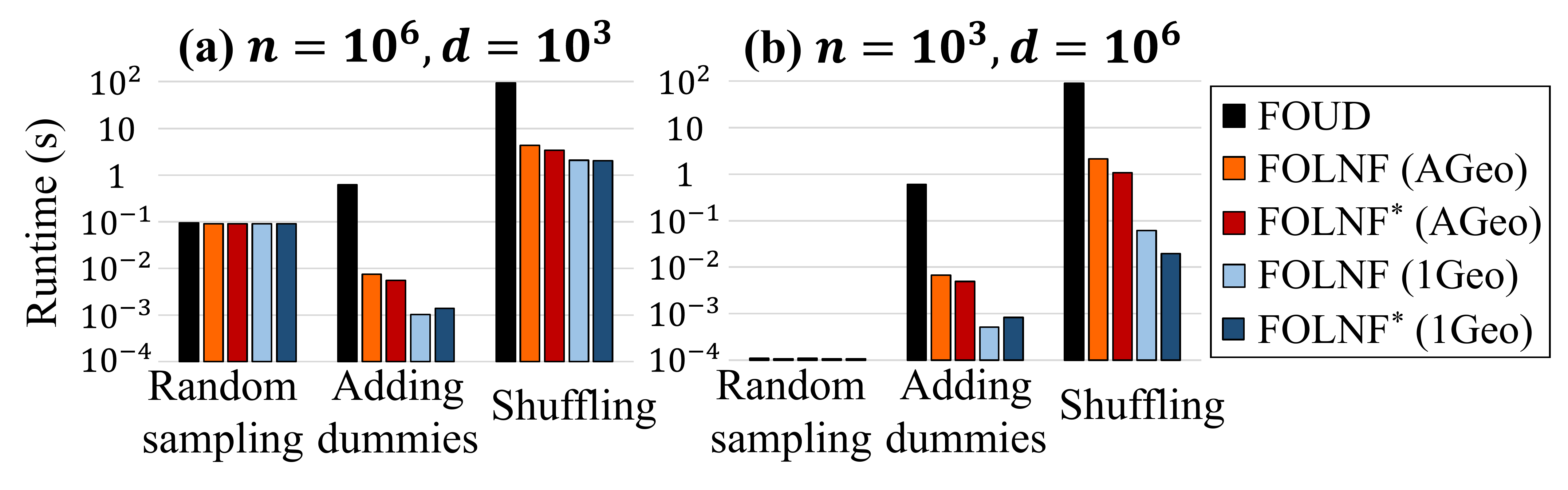}
  \caption{Time for random sampling, dummy data addition, and shuffling in our algorithms ($\epsilon_E = 0.1$, $\epsilon_I = 1$). 
  In (b), we used the count-min sketch ($b=n$).} 
  \label{fig:addendum_res_each_runtime}
\end{figure}

\begin{figure}[t]
  \centering
  \includegraphics[width=0.99\linewidth]{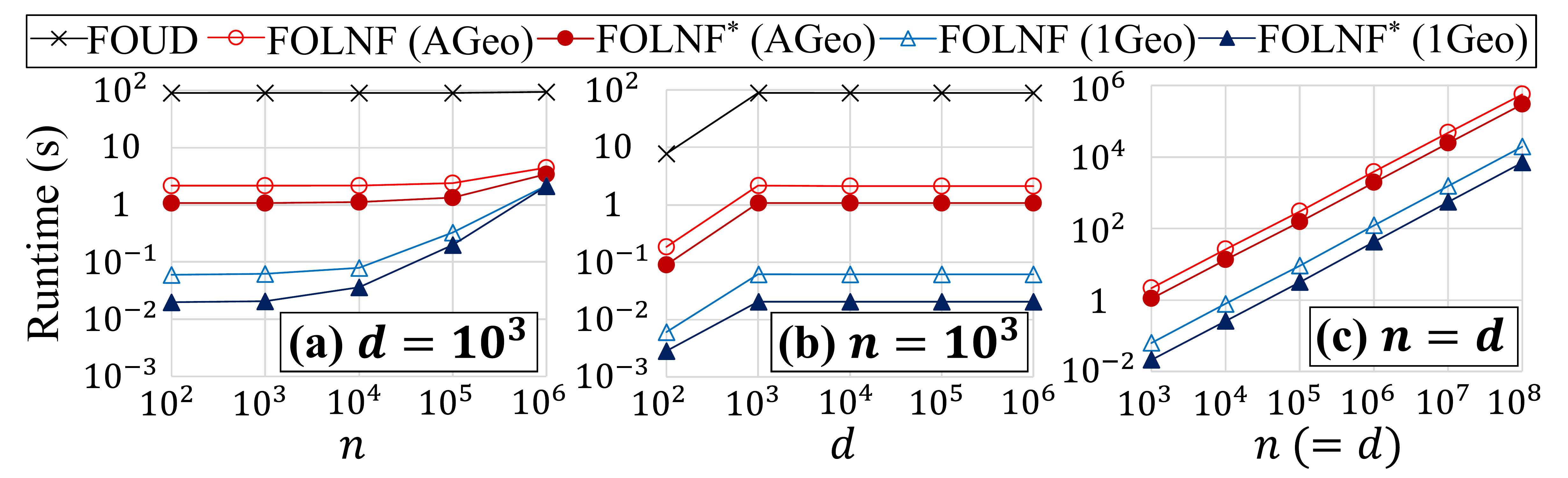}
  \caption{Total runtime of our algorithms ($\epsilon_E = 0.1$, $\epsilon_I = 1$). 
  In (b), we used the count-min sketch ($b=n$) when $n < d$.} 
  \label{fig:addendum_res_total_runtime}
\end{figure}

\begin{figure}[t]
  \centering
  \includegraphics[width=0.99\linewidth]{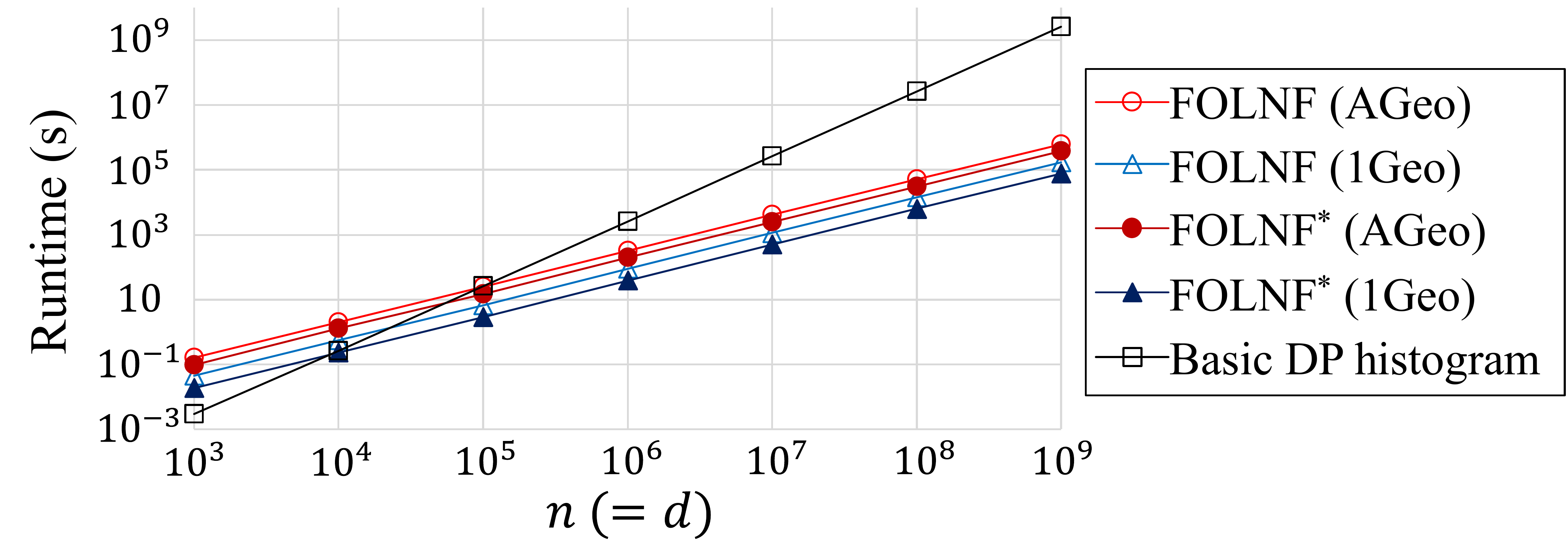}
  \caption{Total runtime of our algorithms and the basic DP histogram ($\epsilon_E = 1$, $\epsilon_I = 5$).} 
  \label{fig:addendum_res_histogram}
\end{figure}

\begin{figure}[t]
  \centering
  \includegraphics[width=0.99\linewidth]{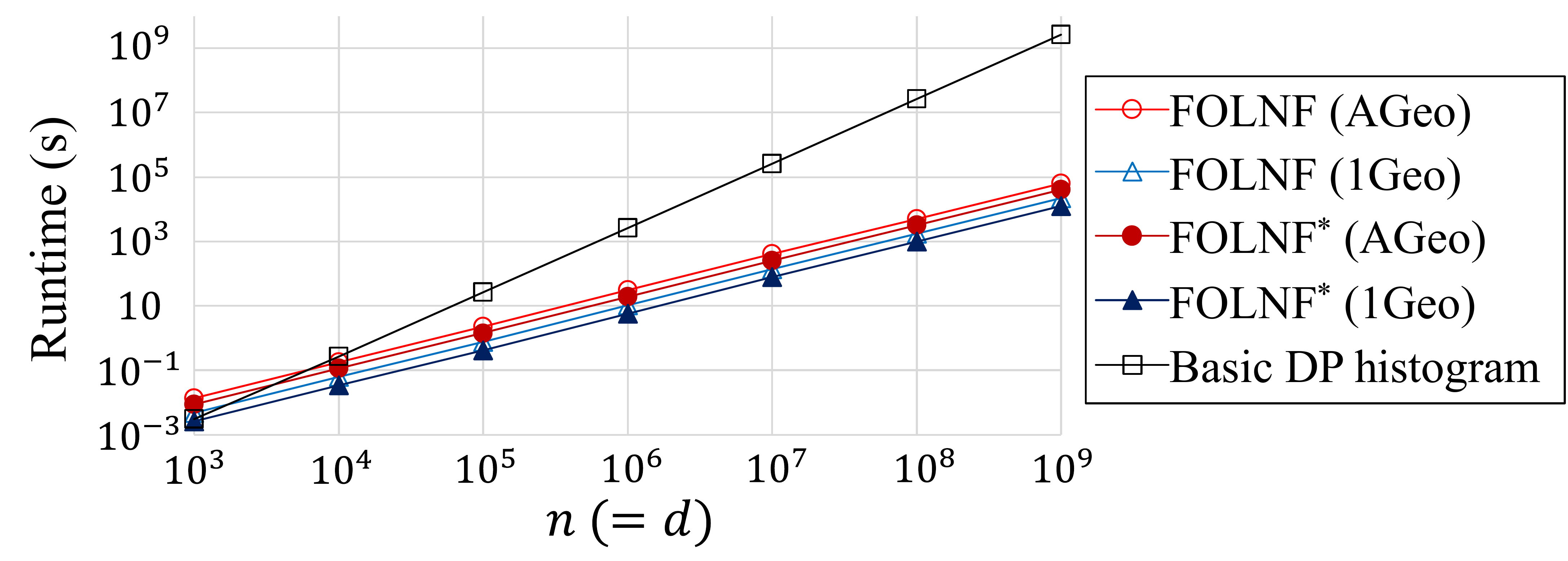}
  \caption{Total runtime of our algorithms with the count-min sketch ($b=0.1n$) and the basic DP histogram ($\epsilon_E = 1$, $\epsilon_I = 5$).} 
  \label{fig:addendum_res_histogram_count-min}
\end{figure}

Figures~\ref{fig:addendum_res_each_runtime}, \ref{fig:addendum_res_total_runtime}, and \ref{fig:addendum_res_histogram} show the results. 
In Figure~\ref{fig:addendum_res_histogram_count-min}, we also show the total runtime of our algorithms with the count-min sketch ($b=0.1n$). 
Figures~\ref{fig:addendum_res_each_runtime}, \ref{fig:addendum_res_total_runtime}, and \ref{fig:addendum_res_histogram} show that the total runtime of our algorithms has increased (when compared to Figures~\ref{fig:res_each_runtime}, \ref{fig:res_total_runtime}, and \ref{fig:res_histogram}) and is close to that of our algorithms with the code of ORShuffle in \cite{Sasy_code} (when compared to Figure~\ref{fig:res_Sasy_runtime}). 
However, our algorithms still handle very large data and are more efficient than the central DP algorithm when $n$ and $d$ are large. 
For example, Figure~\ref{fig:addendum_res_histogram} shows that when $n=d=10^8$, our algorithms ($\epsilon_E = 1$) require less than $16$ hours. 
Specifically, the runtime of our \FOLNF{} (\AGeo{}), \FOLNFast{} (\AGeo{}), \FOLNF{} (\OGeo{}), and \FOLNFast{} (\OGeo{}) is about $14$ hours, $9$ hours, $4$ hours, and $2$ hours, respectively. 
This also means that the runtime of \FOLNF{} (\AGeo{}) (resp.~\FOLNF{} (\OGeo{})) can be reduced from $14$ hours to $9$ hours (resp.~from $4$ hours to $2$ hours) by increasing $\epsilon_I$ from $1$ to $5$. 
Figure~\ref{fig:addendum_res_histogram_count-min} shows that even when $n=d=10^9$, our algorithms can be executed in less than one day by using the count-min sketch. 



\end{document}